\documentclass[useAMS,usenatbib]{mn2e}

\usepackage{graphicx}
\usepackage[caption=false]{subfig}
\newcommand{\ie}{\textit{i.e. }}

\newcommand{\eg}{\textit{e.g. }}

\newcommand{\zphot}{\ensuremath{z_{phot}}}
\newcommand{\zspec}{\ensuremath{z_{spec}}}
\newcommand{\ugriz}{\ensuremath{u^*g'r'i'z'}}
\newcommand{\photz}{photo-$z$}

\title[The Cosmic Web and galaxy evolution around the most luminous X-ray cluster: RX\,J1347.5-1145]{The Cosmic Web and galaxy evolution around the most luminous X-ray cluster: RX\,J1347.5-1145}
\author[M. Verdugo et al.]{M. Verdugo$^{1}$\thanks{E-mail: mverdugo@mpe.mpg.de}, M. Lerchster$^{{1}{,2}}$, H. B\"ohringer$^{1}$,    H. Hildebrandt$^{{3}{,4}{,5}}$,    B. L. Ziegler$^{{6},{7}}$ ,\newauthor  T. Erben$^{3}$, A. Finoguenov$^{1}$,  and G. Chon$^{1}$ \\
$^{1}$Max Planck Institut f\"ur Extraterrestrische Physik, Giessenbachstrasse 1 85748, Garching bei M\"unchen, Germany \\
$^{2}$University Observatory Munich,  Ludwigs-Maximillians University Munich, Scheinerstrasse 1, 81679 Munich, Germany \\
$^{3}$Argelander-Institut f\"ur Astronomie, University of Bonn,  Auf dem H\"ugel 71, 53121 Bonn, Germany \\
$^{4}$Sterrewacht Leiden, Leiden University, Niels Bohrweg 2, 2333 CA Leiden, The Netherlands  \\
$^{5}$University of British Columbia, Department of Physics and Astronomy, 6224 Agricultural Road, Vancouver, B.C. V6T 1Z1, Canada \\
$^{6}$European Southern Observatory, Karl-Schwarzschild-Strasse 2, 85748 Garching bei M\"unchen, Germany \\
$^{7}$University of Vienna, Department of Astronomy, T\"urkenschanzstraße 17, 1180 Vienna, Austria \\
}

\begin{document}

\date{\today}

\pagerange{\pageref{firstpage}--\pageref{lastpage}} \pubyear{2002}

\maketitle

\label{firstpage}

\begin{abstract}
In this paper we study the large scale structures and their galaxy content around the most X-ray luminous cluster known,
RX\,J1347.5-1145 at $z=0.45$. We make use of \ugriz\ CFHT MEGACAM photometry together with VIMOS 
VLT spectroscopy to identify structures around the RXJ1347 on a scale of $\sim$20$\times$20\,Mpc$^2$. 
We construct maps of the galaxy distribution and the fraction of blue galaxies. We study 
the photometric galaxy properties as a function of environment, traced by the galaxy density. 
We identify group candidates based on galaxy overdensities and study their galaxy content. We also use available 
GALEX NUV imaging to identify strong unobscured star forming galaxies.
We find that the large scale structure around RXJ1347 extends in the NE-SW direction for 
at least 20\,Mpc, in which most of the group candidates are located, some of which show X-ray emission
in archival XMM-$Newton$ observations. As other studies, we find that the fraction of blue galaxies ($F_{blue}$)
is a function of galaxy number density, but the bulk of the trend is due to galaxies belonging
to massive systems. The fraction of the UV-bright galaxies is also function of environment, but their relative 
numbers compared to the blue population seems to be constant regardless of the environment. These UV emitters also  
have similar properties at all galaxy densities, indicating that the transition between galaxy types occurs in short 
time-scales. Candidate galaxy groups show a large variation in their galaxy content and  $F_{blue}$ in those groups display 
little dependence with galaxy number density. This may indicate possible differences
in their evolutionary status or the processes that are acting in groups are different than in clusters. 
The large scale structure around rich clusters are dynamic places for galaxy evolution. In the case of RXJ1347 
the transformation may start within infalling groups to finish with the removal of the cold gas once galaxies 
are accreted in massive systems. 

\end{abstract}


\section{Introduction}

Galaxy properties, such as spectral type and morphology, are known to correlate with environment: The central regions
of galaxy clusters are mainly composed by bulge-dominated, passive evolving, red galaxies, whereas the low density field is 
preferentially inhabited by disk-like, blue, active star-forming types (\eg {\citealt{Dressler1980,Dressler1985}). 
This behaviour has been  studied in the local and distant universe out to large cluster centric distances 
(\citealt{balogh99,Christlein2005,Verdugo2008,Poggianti2008,vonderLinden2009,Bauer2011}).
Some of these studies have shown that the decline of the star formation activity starts  too far away from the clusters core 
to be caused only by cluster specific processes and some form of preprocessing would be necessary to account 
for the difference in galaxy populations.

Going to higher redshifts, it is observed that the fraction of blue galaxies in distant clusters increases with redshift 
(\citealt{BO78}). At first order this result indicates an increase of the star formation activity within clusters, which
has been recently confirmed by studies looking at the mid-infrared signatures of star formation 
(\citealt{Saintonge2008,Haines2009a,Finn2010}). As the overall star formation density was higher in the distant universe,
an obvious explanation would link this effect  to a higher rate of infall of active galaxies into clusters from their surroundings, 
as predicted by the hierarchical built-up of structures in modern cosmologies (\citealt{Ellingson2001}).

However, systematic studies of the cluster population have been often handicapped by the large cluster-to-cluster variation 
at all epochs, which displays little or no correlation with total mass (\eg\ \citealt{Popesso2007,Poggianti2008},
see however \citealt{Hansen2009}). Detailed analysis of clusters with similar global properties show that they can 
contain a rather different galaxy population (\eg \citealt{Moran2007a, Braglia2009}). It may be possible that 
the cluster galaxy content depends on more subtle aspects of cluster nature, such as, assembly history, 
substructure and the surrounding large scale structure. 

Recently a number of studies have begun to investigate the galaxy populations embedded 
in the filamentary structure around clusters. By far, the  most complete sample was published by 
\citet{Porter2008}, where a large catalogue of filaments drawn from the 2dFGRS was studied. 
They found a sharp increase of the star formation activity in filaments joining clusters at 
approximately 2-3\,Mpc from the nearest cluster centre. 
Similarly \citet{Mahajan2010} report an increase of the star formation 
activity among dwarf blue galaxies in the infall regions of the Coma supercluster.

At moderate redshifts, \citet{Braglia2007} have found a number of bright, blue 
star-forming galaxies in filaments around two distant X-ray luminous $z\sim0.3$ clusters
drawn from the REFLEX-DXL sample (\citealt{Zhang2006}). Using the Spitzer satellite, \citet{Fadda2008}
discovered a number of starbursts in a  ``cluster-feeding'' filament around a $z\sim0.2$ cluster. Similarly, 
\citet{Koyama2008} report an increase of the number density of 15$\mu$m sources detected with the  AKARI space mission in 
medium density environments in a cluster at $z\sim0.8$.  \citet{Marcillac2007} detected several dusty star-bursts 
in the infall regions of another rich $z\sim0.8$ cluster. \citet{Haines2009} also find a number of obscured 
star-forming galaxies in the Abell 1758 ($z=0.28$) cluster complex, coinciding with filaments and infalling groups. 
They also find that one of the subclusters (A1758N) is more active than the central one, reflecting probably different 
dynamical histories.

On the other hand, a detailed analysis of the Abell 901/902 supercluster ($z\sim0.15$) by \citet{Gallazzi2009} 
and \citet{Wolf2009}  shows that  there is indeed an increase of obscured star formation activity at intermediate densities, 
however, it appears to be rather mild, as most of the galaxies have star formation rates lower or similar to normal blue
star-forming galaxies. 

In a complementary analysis, \citet{Li2009} observed that the fraction of blue galaxies increases faster with redshift 
in groups associated with CNOC1 clusters (\citealt{Yee1996}) than in the clusters themselves. Finally, 
\citet{Tanaka2007,Tanaka2009a} have shown evidence of newly formed red-galaxies with residual star formation in the large 
scale structure around clusters, indicating  that these systems may be very effective in transforming galaxies through cosmic 
times.

In this study we follow similar ideas. We surveyed an area of $\sim$20$\times$20\,Mpc around the centre 
of the very rich cluster RX\,J1347.5-1145 at $z=0.45$ (RXJ1347 hereafter). This cluster is the most 
X-ray luminous one  (\citealt{Schindler1997}) found in the REFLEX cluster survey (\citealt{Bohringer2004}) and has been 
investigated by different techniques, 
including strong  and weak lensing 
(\citealt{Bradac2005,Bradac2008b,Halkola2008,Lu2010,Medezinski2010}), 
X-ray 
(\citealt{Ettori2001,Gitti2007,Ota2008}),
Radio 
(\citealt{Gitti2007a}), 
Sunyaev-Zeldovich effect 
(\citealt{Komatsu2001,Kitayama2004,Mason2009})
and optical spectroscopy 
(\citealt{Cohen2002,Lu2010}). 

Most of these studies have arrived to the conclusion that RXJ1347 is likely one 
of the most massive objects in the universe. In the context of the hierarchical growth of structures
predicted by $\Lambda$CDM cosmologies, such massive cluster should sit at the centre of a complex network of 
filaments and subclumps as our first results indicate, making it an ideal target for our investigation.

Throughout this paper, we will use a cosmology of $H_0=70$\,km~s$^{-1}$Mpc$^{-1}$, $\Omega_m=0.3$ and $\Omega_\Lambda=0.7$.

\section{Observations}
  \label{S:obs}

\subsection{Optical data}

Our study makes use of available data taken with the  {\tt MegaPrime} 
camera mounted at the 3.6\,m Canada-France-Hawaii Telescope (CFHT)
using the \ugriz\ filter set.  The field of view of the instrument covers 1\,deg$^2$, 
roughly 20$\times$20\,Mpc$^2$ at $z=0.45$. The total usable field  in our case 
is somewhat smaller (0.82\,deg$^2$)  due to the imperfect overlap of the target fields taken with the different filters.
Table\,\ref{T:obs} contains a summary of the observations with this instrument. 
Fig.\,\ref{F:observations} shows the total field of view of the observations.

The MEGACAM data were retrieved from the Elixir system\footnote{http://www.cfht.hawaii.edu/Instruments/Elixir/home.html}
 in a preprocessed form  from the The Canadian Astronomy Data Centre (CADC\footnote{http://cadcwww.dao.nrc.ca/cadc/}) 
archive and further processed as described
in \citet{Erben2009} and \citet{Hildebrandt2009}. The final data products consist of astrometrically
and photometrically calibrated co-added science and noise-map data with the characteristics summarised in Table\,\ref{T:obs}.
Photometric zero-points for our data were provided by Elixir. We checked the calibration against colour-colour 
diagrams for model stars from the \citet{Pickles1998} library. No offsets between the observed and theoretical
stellar locus were appreciable, which would indicate  a bias in the calibration. Further corrections are described in 
Section\,\ref{SS:photz}

A note about the filters. The $u^*$-band at $z=0.45$ cover a rest-frame wavelength range of 
2400 to 2800\,\AA\  and would fall within the GALEX NUV band at $z=0$. The reddest band (the $z$ filter) 
correspond approximately to the  $r$-band at rest-frame. Finally, the 4000\,\AA\ break at $z=0.45$ is well braketed 
by the $g$ and $r$-bands.

Photometry was performed with {\tt SExtractor} (\citealt{Bertin1996}) in dual mode, using the deepest image 
(\ie\  $r'$-band) as a detection frame. Images have been convolved with a gaussian filter to match the worst seeing image. 
Galaxy colours have been extracted in a 2.5\,\arcsec\ aperture, which is equivalent to 14.4\,kpc at $z=0.45$. This is 
a good trade off between the sampling of the PSF wings without increasing the noise for fainter objects.

The photometric catalogue contains almost 60000 sources, including 6890 stars according to the {\tt SExtractor
CLASS\_STAR} parameter ({\tt CLASS\_STAR}$\geq$0.95), which is reliable down to $r\sim23$. Those stellar sources 
were eliminated from the catalogue. Stars fainter than this limit were eliminated using an additional approach (
see Section\,\ref{S:selection}). 

Galactic extinction for each filter was calculated using the dust maps of \citet{schlegel98}.

Comparisons of number counts with other photometric surveys of greater depth like the CFHTLS-Deep 
(\citealt{Ilbert2006,Coupon2009})  indicate that our photometric catalogue is complete down to $r'=23.5$\,AB.  
Fainter sources are thus removed from the catalogue.

\begin{table}
\caption{Sumary of the CFHT/MEGACAM observations.}
\begin{minipage}[t]{\columnwidth}
\begin{tabular}{lcccc}
\hline\hline\\
Program / PI & Filter & Exp. Time & M$_{limit}$\footnote{S/N=5 in an aperture of 1''. Magnitudes are in the AB system.} & seeing \\
             &        &     [s]   &  [mag]  & ["]   \\
\hline \\

2006BH34/Ebeling & $u^*$   & 4260  & 25.09 & 0.95 \\
2005AC10/Hoekstra & $g'$   & 4200  & 26.44 & 0.82 \\
2005AC10/Hoekstra & $r'$   & 6000  & 25.94 & 0.77 \\
2005AC12/Balogh   & $i'$   & 1600  & 25.74 & 1.03 \\
2005AC12/Balogh   & $z'$   & 3150  & 24.80 & 1.14 \\
\end{tabular}
\end{minipage}
\label{T:obs}
\end{table}

\subsection{Spectroscopy}

Spectroscopy in the field was performed with VIMOS at the ESO VLT using the low resolution blue grism 
(LR-Blue, PID: 169.A-0595, PI: B\"ohringer) 
and  the medium resolution one 
(MR, PID: 381.A-0823, PI: Verdugo).

The first program covered the central regions  with an overlapping pattern 
to  properly  fill the gaps between the VIMOS CCDs (see Fig.\,\ref{F:observations}).  
The individual target selection  was based  on the galaxies I-band magnitudes.
 
The second program targeted the large scale structures first identified using colour selected galaxies using the MEGACAM
photometry. The individual object selection was colour based, excluding objects redder than the cluster red-sequence. 
In practice, the galaxy distribution and mask design constraints do not allow the selection of 
all candidate objects with the desirable photometric properties, leaving empty regions where we placed additional slits, 
so that up to 40\% of the galaxies were selected randomly.

Data reduction was carried out using  the {\tt VIPGI} pipeline (\citealt{Scodeggio2005}) and redshifts have been obtained 
using the EZ software (\citealt{Garilli2010}) and custom made tools.  In total we secured 745 redshifts. 
The distribution of galaxies with spectroscopic redshifts is presented in the upper panel of Fig.\,\ref{F:histoZ}. 
Our spectroscopic campaign over the full 1\,deg$^2$ field 
was recently completed and results will be presented 
in a forthcoming paper. 

We have also added 62 out of the 79 redshifts previously published  by \citet{Cohen2002} using Keck spectroscopy at 
the very central regions of the cluster, making a total of 807 redshifts. The remaining 17 redshifts are also common 
to our data set.

\begin{figure}
\includegraphics[width=0.99\columnwidth,clip,viewport=55 170 485 620]{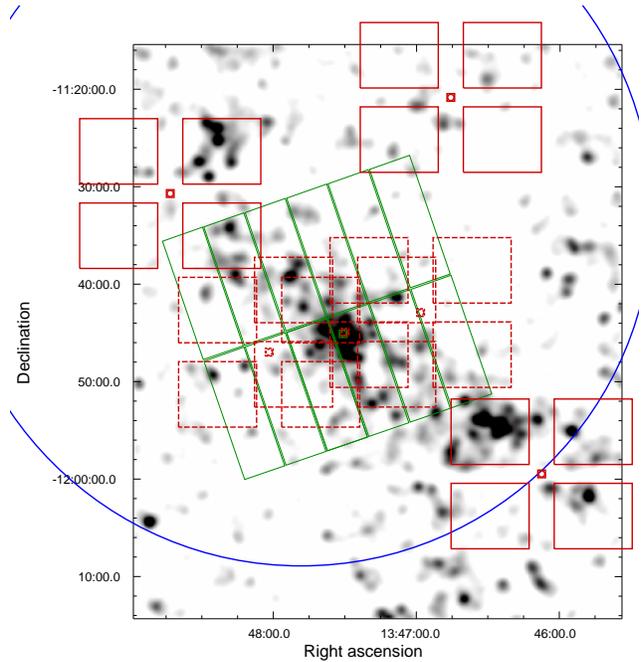}
\caption{Layout of the observations used in this paper, superimposed to the  density map of cluster members 
candidates (Section\,\ref{S:env}). 
Dashed red rectangles represent the  VIMOS LR spectroscopy and 
solid ones are VIMOS MR masks. The large semi-circle shows the region covered by the GALEX. 
The tilted green mosaic represent the XMM-$Newton$ EPIC-$pn$ observations.}
\label{F:observations} 
\end{figure}

\subsection{GALEX observations}
\label{SS:Galex}

Observations with the Galaxy Evolution Explorer (GALEX) were found in the MAST archive under program GI3\_103 
(PI: Hicks, see also \citealt{Hicks2010} for more details). 
Both  NUV ($\lambda_{eff}=2267$\,\AA) and FUV ($\lambda_{eff}=1516$\,\AA)  bands 
were exposed for 9120\,s. At $z\sim0.45$ the NUV filter is similar to the rest-frame FUV filter and thus a good indicator 
of the unobscured star formation activity. 

The limit magnitude of the NUV observations (NUV=24 AB at 3-$\sigma$) is deep enough to detect the strongest 
star-forming galaxies at $z\sim0.45$, however the coverage of our field is imperfect (see Fig.\,\ref{F:observations}). 
FUV observations are much shallower and will not be used in this work.

From the catalogues produced by the GALEX pipeline (\citealt{Morrissey2007}) we  extracted all sources with at 
least 3-$\sigma$ detection significance. Objects with distances larger than 0.58 degrees from the field centre 
were eliminated to avoid  artefacts present at the field edges. 

We matched the NUV sources with our optical photometric catalogue using a search radius of 3.5\arcsec. Increasing the search 
radius, due to the broader GALEX PSF, does not produce more matches and contamination appears beyond 5\,\arcsec. 
Matched objects were also visually inspected against the $u$-band image to check  the purity of the catalogue. In total, 
we matched 1040 objects with  $0.37<\zphot<0.52$, our cluster galaxy selection window (see Section\,\ref{S:selection}).

\begin{figure}
\centering
\includegraphics[width=0.85\columnwidth,clip,viewport=20 180 380 580]{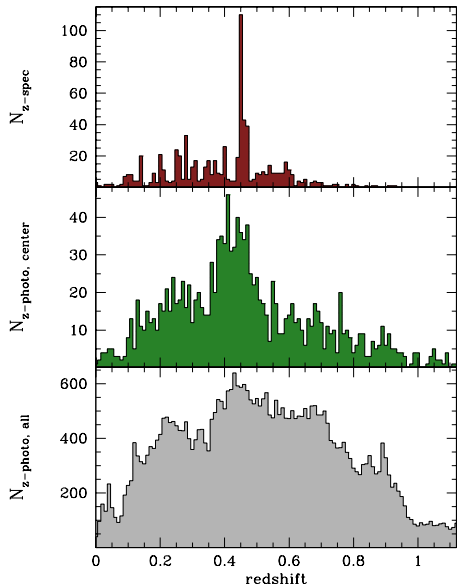}
\caption{Upper panel: Distribution of  spectroscopic redshifts from our 
VIMOS observations.  Mid panel: Photometric redshift distribution for all galaxies in the central 10\arcmin.
Lower panel: Histogram of photometric redshifts for all galaxies with $r\leq23.5$\,mag.   }
\label{F:histoZ}
\end{figure}

\subsection{X-ray data}
\label{SS:Xray}

RXJ1347 has been observed with the XMM-$Newton$ X-ray telescope (ObsID:0112960101, PI: M. Turner)
for 38\,ks total time. Results of these observations have been reported elsewhere by \citet{Gitti2004} and \citet{Gitti2007}.

The XMM-$Newton$ observations were focused on the central cluster and thus only a fraction of the field is covered by 
them (see Fig.\,\ref{F:observations}). Nevertheless, they represent an opportunity to detect the X-ray emission associated 
to the large scale structures around RXJ1347.

We have processed the imaging using custom made software. After flare cleaning, following \citet{Zhang2004}, we 
retained 24, 31 and 30\,ks of clean time for EPIC-{\it pn},  {\tt MOS1} and  {\tt MOS2}, respectively. The background has 
been estimated using the regions of the observation free of cluster emission 
and point-sources.  We have iterated on the definition of the background zone, as in \citealt{Bielby2010}, by also 
removing the zone of excess  X-ray emission, associated with the location of the large scale structure 
(see Section\,\ref{SS:LSS} and Fig.\,\ref{F:MapDens}).

In this paper we use the wavelet+PSF restoration image technique of \citet{Finoguenov2010}. The objective is to 
report the properties of the optically  selected groups (see Section\,\ref{SS:groups}), using the residual flux image after 
background and point source removal (including the PSF wings).

\section{Cluster member selection}
\label{S:selection}

\subsection{Photometric redshifts}
\label{SS:photz}

The only way to establish cluster membership  with high confidence is by using information from 
spectroscopic observations. Unfortunately, obtaining a complete sample of galaxies is very 
time-consuming and difficult to perform down to faint magnitudes. For that reason, in this work
 we use the photometric  information from our multicolour imaging.

We use the code {\tt LePhare} (Arnouts \& Ilbert, \citealt{Ilbert2006}) to obtain
photometric redshifts ($\zphot$) for all sources down to $r\sim23.5$\,mag using only the optical data. Because 
of the similar characteristics of our data, we have followed the procedures of \citet{Ilbert2006} 
to obtain photometric redshifts. In particular, we have used the same improved galaxy templates 
that they used for estimating redshifts for the CFHT Legacy Survey (CFHTLS). Since the GALEX data 
does not cover the whole field, it has not been used to obtain photometric redshifts. 

We have also used spectroscopic redshifts to optimise the templates and to compensate the 
systematic errors in the photometry. The results from  {\tt LePhare} have been compared with 
photometric redshifts obtained using  {\tt BPZ} (\citealt{Benitez2000}), see also 
\citet{Hildebrandt2009}, and {\tt PHOTO-z} (\citealt{Bender2001}, see also \citealt{Brimioulle2008}). 
The results of the three codes agree within the statistical uncertainties.

\begin{figure}
\centering
\includegraphics[width=0.85\columnwidth,clip,viewport=20 299 350 690]{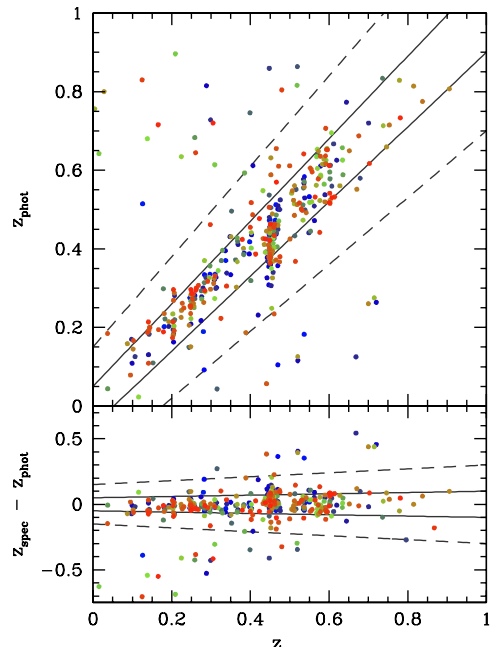}
\caption{Comparison between photometric and the 807 spectroscopic redshifts. The solid lines
mark where $| z_{spec} - z_{phot} |< 0.05(1+z)$ and the dashed ones where $| z_{spec} - z_{phot} | < 0.15(1+z)$.
Outside of this latter limit photometric redshifts are considered catastrophic. The colour of the points 
indicate the best fitting template with redder colours for earlier galaxy types. The mean scatter of  $z_{phot}$
is $\sigma_z = 0.058$. }

\label{F:ZvsZ}
\end{figure}

In Fig.\,\ref{F:histoZ} we plot the distribution of photometric redshifts for the whole field and the
central 10\arcmin. A clear peak in the redshift distribution centred at $z\sim 0.45$ can be 
discerned. As a comparison we plot the distribution of the  spectroscopic redshifts 
in the upper panel. 

{\tt LePhare} also provides information on the object type (via SED classification). 
This helped us to eliminate stars from the photometric catalogue below the limit  of the 
{\tt CLASS\_STAR SExtractor} parameter, as stellar templates have been also included in the 
analysis. As galaxies  dominate the number counts at faint magnitudes, we expect that the star 
contamination is very low in our sample. 

To assess the quality of the photometric redshifts we plot in Fig.\,\ref{F:ZvsZ} a comparison  
between \zspec\ and \zphot\ as a function of \zspec. The regions marked by lines show 
$| \zphot - \zspec |< 0.05(1+\zspec)$ and $| \zspec - \zphot |< 0.15(1+\zspec)$ respectively.  
Photometric redshifts with $| \zspec - \zphot |> 0.15(1+\zspec)$ are considered catastrophic outliers.

A number of statistical tests have been proposed in the literature  
(\eg\ \citealt{Ilbert2006,Pello2009}) to probe the quality of the photometric redshifts:

\begin{itemize}
\item The fraction of catastrophic outliers is defined  as $| \zspec - \zphot |> 0.15(1+\zspec)$. 
The result is 9.12\%. 

\item The normalised median absolute deviation 
$\sigma_{z,NMAD}=1.48\times$median$(| z_{spec} - z_{phot} |/(1+z))$. The result is 0.037.

\item The systematic deviation between \zphot\ and \zspec. 
$\langle\Delta(z)\rangle = \sum(\zspec-\zphot)/N $, with  $\Delta_z= | z_{spec} - z_{phot} |$. 
Catastrophic outliers are excluded.   The result is 0.0051.

\item The standard deviation of $\Delta(z)$:   $ \sigma_z = \sqrt{ (\Delta(z) - \langle \Delta(z) \rangle )^2 / (N-1) } $. 
The results is $ \sigma_z = 0.058$.  This value represent the scatter between spectroscopic and photometric redshifts.

\end{itemize}

\begin{figure}
\centering
\includegraphics[width=0.9\columnwidth,clip,viewport=20 150 580 600]{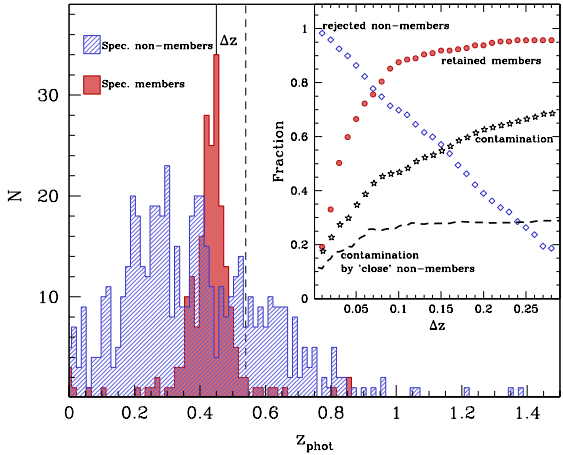}
\caption{Photometric redshift distribution for spectroscopic confirmed members (red filled histogram) 
and non-members (blue hashed one). The inset shows the fraction of retained members (red circles) and rejected 
non-members (blue diamonds) as a function of an increasing $\zphot$ window threshold. 
The contamination due to all galaxies (black stars) is also displayed. The contamination by ``close non-members'' 
(dashed line)  refers to non-cluster galaxies with $0.38<\zspec<0.52$, \ie\ the same redshift range we finally chose 
to select our cluster galaxies.}
\label{F:Select}
\end{figure}

These statistical tests show that the precision of  our results is comparable to those obtained with 
similar data (see \eg\ \citealt{Ilbert2006}) and the photometric redshifts display no bias compared 
to the spectroscopic ones. 
Still, they contain large uncertainties compared with spectroscopic redshifts  and, therefore,  
any selection will introduce contamination by field interlopers and loss of cluster
members.

\subsection{Optimal redshift window and statistical background subtraction}
\label{SS:thres}

As a first step, we select galaxies with photometric redshifts within an optimal range. This is done by dividing 
the spectroscopic sample in cluster members ($0.43<z_{spec}<0.48$)\footnote{This larger range is necessary because
of a possible associated structure at $z=0.47$, see Section\,\ref{S:LSS}} and non-members.
Then we compare the fraction of retained members and rejected non-members as a function of \emph{increasing} \zphot\ window 
threshold   ($\Delta z=|z_{cl}-z_{thres}|$). The contamination due to non-members is also calculated. The results of this 
procedure are presented  in Fig.\,\ref{F:Select}, where we plot the \emph{photometric redshift} distribution 
for spectroscopic cluster members and non-members. In the inset, we plot the fraction of retained members, rejected 
non-members and contamination as a function of \zphot\ window size.

This procedure is inspired in the selection scheme of \citet{Pello2009}. At difference of us, they used 
the full \zphot\ probability function. In our case the probability functions calculated by {\tt Lephare}  are, in general, 
too narrow and likely not reliable, despite that their peak values are statistically accurate as
shown in Fig.\,\ref{F:ZvsZ}.  

The optimal threshold is $\Delta z=0.07$, where  $\sim$80\% of the spectroscopic members
are retained and a similar fraction of non-members are rejected. The contamination by non-members is $\sim$45\%, 
however more than a half of this contamination (roughly 55\%,) come from ``close'' non-members, \ie\ field galaxies with 
spectroscopic redshifts within the thresholds ($z_{spec}= 0.45 \pm 0.07$) which are difficult to distinguish  by \photz\ 
techniques.

This final catalogue of candidate cluster members contains 5895 sources down  to $r=23.5$ AB.

To correct for the contamination by field interlopers, we perform an statistical background subtraction following 
\citet{Pimbblet2002}.  For this, we use two areas of the field where the galaxy density is the lowest 
(see Fig.\,\ref{F:MapDens})  to represent the field population. The total area used for this purpose is 0.285 deg$^2$.

We construct two colour-magnitude diagrams with the galaxy distribution binned in 0.1\,mag in the $(g-r)$ colour 
and 0.25 in the $r$ magnitude. One diagram correspond to the whole cluster area and another only to the field. Both are 
normalised by their respective total areas.

\begin{figure}
\centering
\includegraphics[width=0.8\columnwidth,clip,viewport=0 10 560 390]{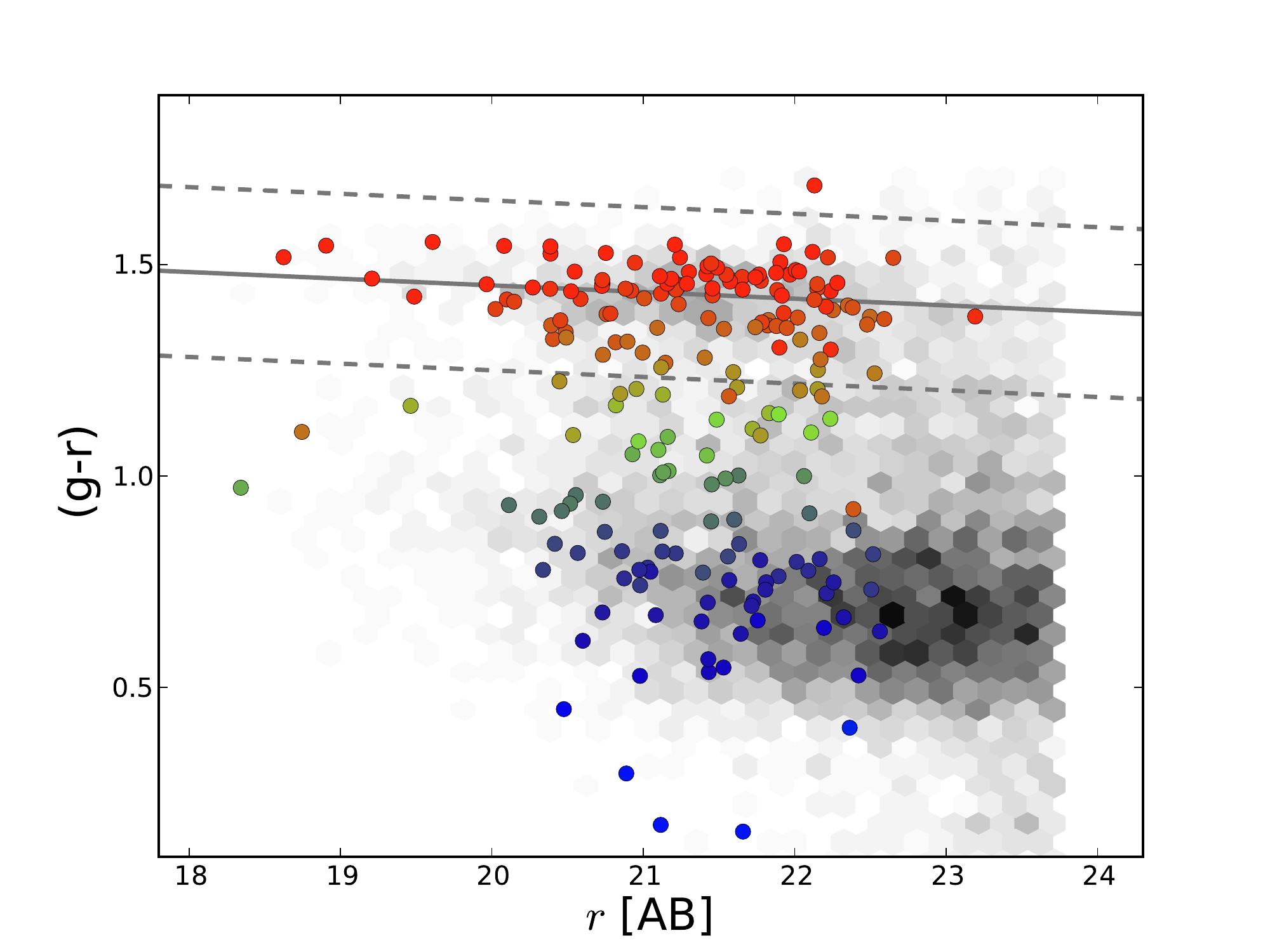}
\caption{Observed colour-magnitude relation for spectroscopic members of RXJ1347. Each galaxy has been colour 
coded according to its best-fit template with redder colours for earlier types. The  solid line marks the 
line that best describes the red-sequence fit to early type galaxies. The dashed lines are the upper and
lower 3-$\sigma$ limits. The colour distribution of all \photz\ selected members  is shown in 
grey scale in the background.}
\label{F:specCMD}
\end{figure}

Afterwards, the normalised field colour-magnitude diagram is divided by the cluster one 
(which also contains a field signal).  The result can be interpreted as the probability that a galaxy 
belongs to the field as a function of colour and  magnitude . In other words,

\begin{equation}
P(\mathrm{Field})_{col,mag} = \frac{ N\mathrm({Field})_{col,mag}}{N\mathrm({Cluster+Field})_{col,mag}}
\end{equation}

Note that this procedure is also valid in the presence of incompleteness as the field  and cluster signal 
should have a similar level of incompleteness.
'
Based on the probability map, we construct 100 Monte Carlo realisations of the parent catalogue. This is done
by generating a random number between 0.0 and 1.0 for each galaxy and comparing it with its field probability
according to its colour and magnitude. If this number is larger than $P(\mathrm{Field})_{col,mag}$, the galaxy is 
attributed to the field population and thus eliminated from the cluster catalogue.
Each resulting catalogue contains between 4000 and 4400 galaxies, which should represent the true cluster population. 
The different quantities presented in this paper are calculated separately for each individual catalogue and 
then averaged. Error bars are then the standard deviation of these quantities.

\subsection{The effect of the (in)accuracy of photometric redshifts}

This study is based in a complete catalogue of cluster galaxies selected using moderately 
accurate and unbiased photometric redshifts. We have been careful in the selection of an 
optimal redshift bin that maximises completeness and minimises contamination. 
Still, the large redshift bin  of $0.38<\zphot<0.52$ is equivalent to a transverse distance of $\sim$450\,Mpc (co-moving), 
much larger than extent of the large scale structure around RXJ1347. Superposition of unassociated systems is 
therefore expected and this should be kept in mind when interpreting the results.

The use of the statistical background subtraction, with their 100 Monte Carlo generated catalogues, should
correct for the \emph{mean} contamination of the unassociated field. As different quantities are calculated for
the individual 100 catalogues and averaged afterwards, the associated error bars should also include the 
uncertainty due to field contamination for a particular position of the field. 

Larger than expected background fluctuations are of course difficult to assess and correct. 
We expect that superposition affect more strongly
low density areas than high density ones like the cluster cores.

\begin{figure}
\includegraphics[width=0.8\columnwidth,clip,viewport=0 10 560 390]{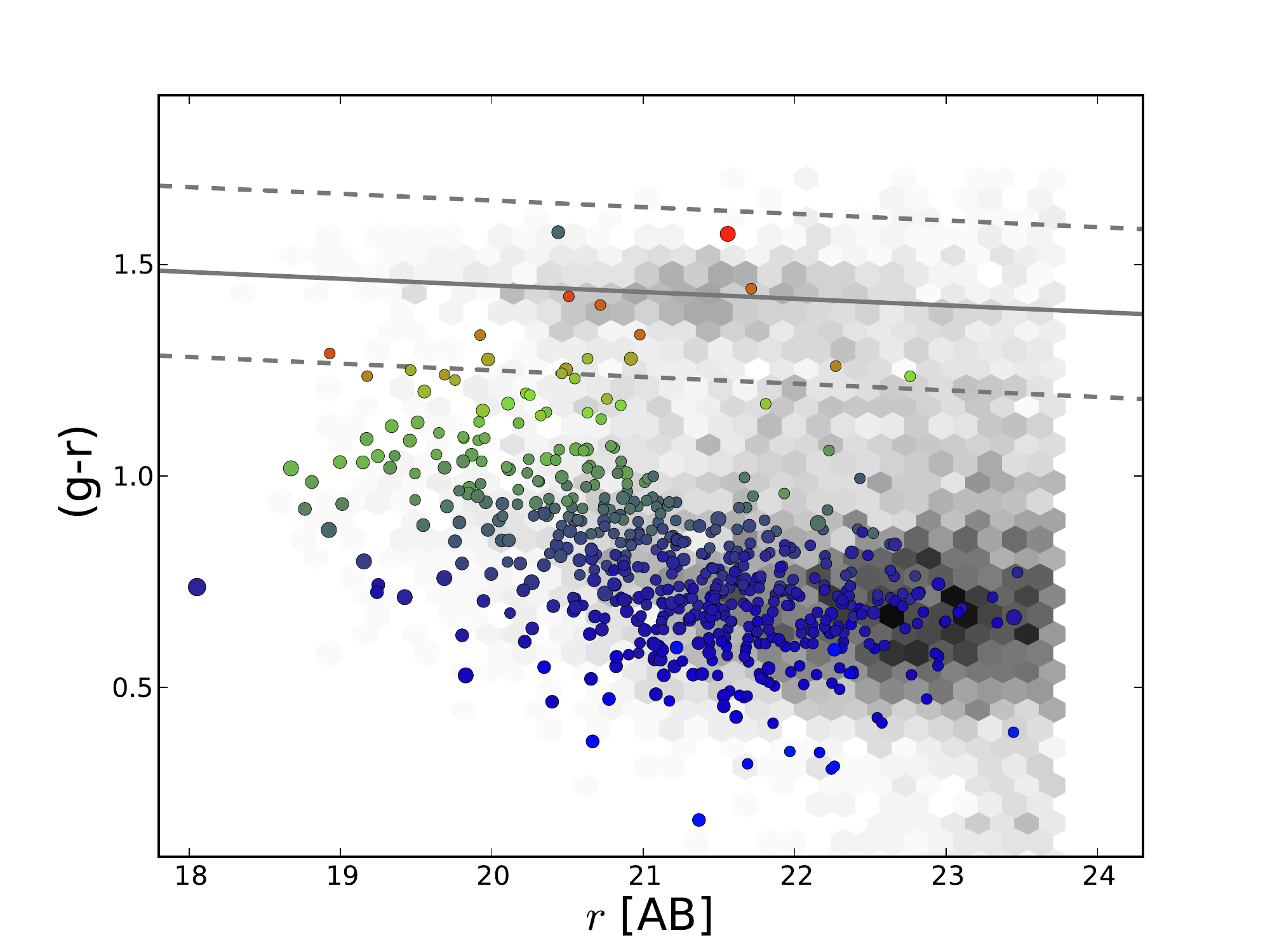}
\caption{Colour-magnitude diagram for NUV GALEX detections with \zphot\ similar to the cluster.
They have been colour-coded according to the best-fit template. 
The sizes of the symbols are a function of the NUV magnitude. As a comparison, we plot in the background the colour-magnitude diagram for all candidate cluster galaxies from the \photz\ selection. }
\label{F:CMDgalex}
\end{figure}

\section{Galaxy colours and luminosity function}
\label{S:Colors}

Old, passive, early type galaxies are typically located in a narrow region within the appropriate 
colour-magnitude diagram, usually well described by a straight line leading to the so-called the red-sequence
(\citealt{Baum1959,Gladders1998}).

To isolate galaxies belonging to the red-sequence, we ran {\tt LePhare} on the spectroscopic catalogue keeping 
this time the redshift fixed. This reduces the degrees of freedom for the $\chi^2$ template fitting, 
yielding, in principle, more accurate estimates of the galaxy spectroscopic types. From the resulting catalogue 
we selected galaxies at $z_{spec}= 0.45\pm 0.03$ with templates compatible with early types. We fitted a line to 
the colour-magnitude relation using a simple least-square algorithm with 3-$\sigma$ clipping in five iterations. Under 
these conditions this method is accurate enough for our purposes. The results can be seen in Fig.\,\ref{F:specCMD}.
The red-sequence is  well approximated by the following relation: $(g-r)=-0.016\times r +1.77$ with a scatter
of $\sigma=0.067$\,mag.

Based on the previous fit, we define red galaxies as all galaxies with colours redder than the lower 
3-$\sigma$ limit (\ie\ 0.201\,mag) and blue otherwise.  This limit will be used to calculate the fraction of blue 
galaxies in the following sections. Note that this definition is identical to the original scheme proposed  
by \citet{BO78}.

The colour-magnitude diagram for all spectroscopic members is plotted in Fig.\,\ref{F:specCMD}.
We also plot the distribution of all \photz\ selected members. The bimodality of galaxy colours reported by several authors 
(\eg\ \citealt{Baldry2006,Loh2008})  is  also clearly discernible here.

In Fig.\,\ref{F:CMDgalex} we plot the distribution of GALEX NUV sources in the optical colour-magnitude diagram. 
Practically all detected sources correspond to blue galaxies. More luminous NUV sources tend also to be brighter 
and bluer in the optical. 
As a comparison we plot also the distribution of all candidate cluster galaxies. It is evident that GALEX sources
represent  only a small subset the galaxy population, probably the  strongest, unobscured starbursts.

\begin{figure}
\centering
\includegraphics[width=0.85\columnwidth,clip,viewport=0 5 560 390]{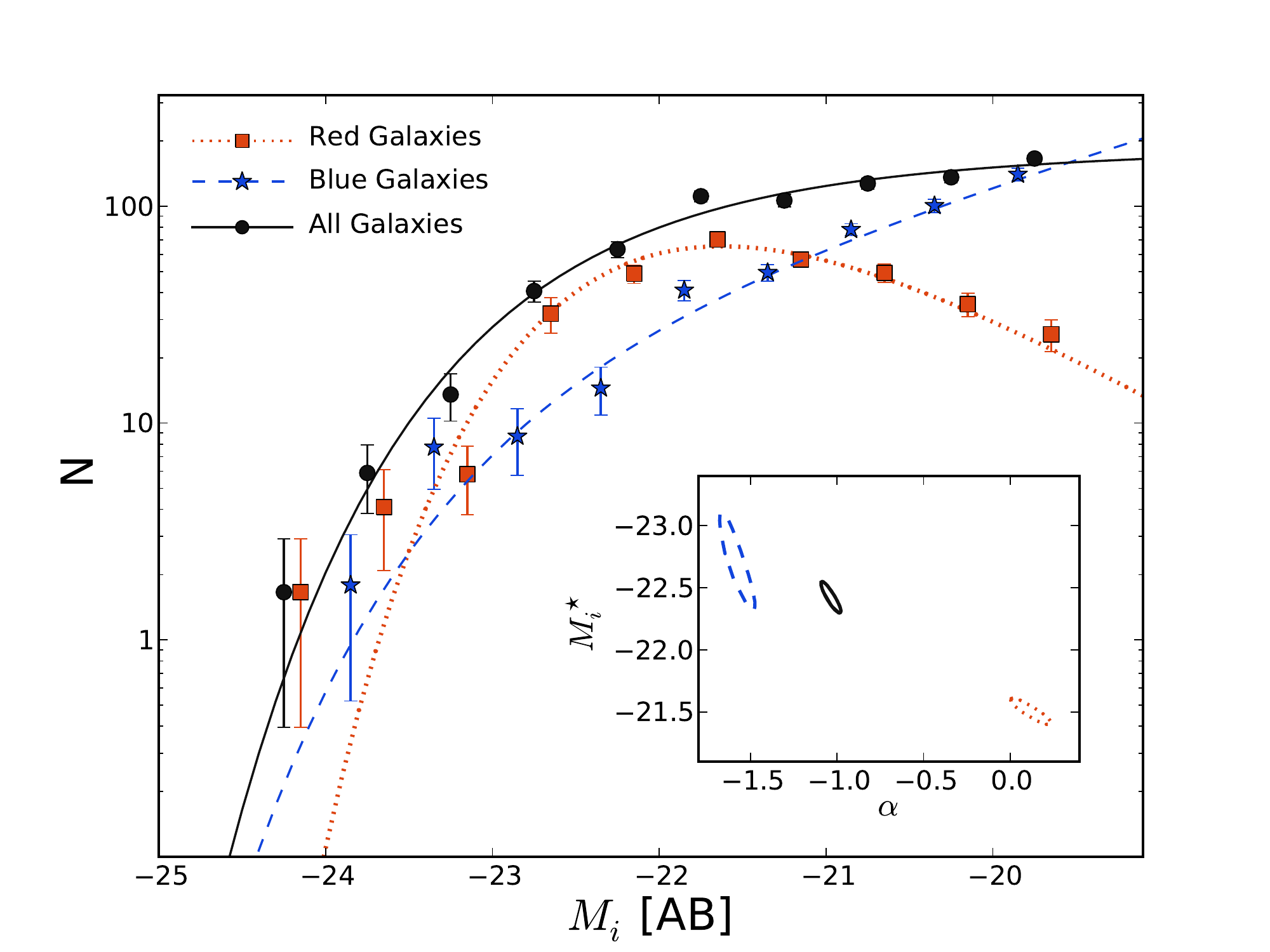}
\caption{Rest frame $i$-bad luminosity function for all, red and blue galaxies within $2 R_{200}$ (\ie\ 4.6\,Mpc) from 
the centre of RXJ1347 with the best-fit Schechter function overplotted. The inset show the 1-$\sigma$ uncertainty 
contours for every fit.}
\label{F:LFall}
\end{figure}

We also calculate the rest-frame $i$-band luminosity for all, red and blue galaxies within $2 R_{200}$ (4.6\,Mpc) from  
the centre of RXJ1347. Absolute magnitudes were obtained using the observed magnitudes and colours and 
applying typical k-corrections obtained with the online tool of 
\citealt{Chilingarian2010}\footnote{http://kcor.sai.msu.ru/UVtoNIR.html}. The frequency of galaxies 
per 0.5 magnitude bin was fitted with a Schechter function (\citealt{Schechter1976}) of the form:

\begin{eqnarray}\nonumber
 \phi(M)dM & = 0.4 \ln 10 \times \phi^\star \times 10^{-0.4(M-M^\star)(\alpha+1)} \\
  & \times \exp \left [ -10^{-0.4(M-M^\star)}\right ] dm
\label{eq:lfeq}
\end{eqnarray}

where $\phi^\star$ is the normalisation factor, $M^\star$ is the characteristic magnitude and $\alpha$ describes the faint 
end slope. The fitting was performed with a $\chi^2$-minimisation algorithm. The luminosity functions and the best fit 
Schechter 
parametrisations can be seen in Fig\,\ref{F:LFall}. 
The results for the key parameters 
$M^\star$ and  $\alpha$ are the following:  
\begin{itemize}
\item All galaxies: $M^\star= -22.4\pm 0.1$\,mag, $\alpha=-1.04\pm 0.06$, % chi2=2.28
\item Red galaxies: $M^\star= -21.5\pm 0.1$\,mag, $\alpha=0.11\pm 0.12 $,   %, $\chi^2=1.28$
\item Blue galaxies: $M^\star= -22.7\pm 0.4$\,mag, $\alpha=-1.58\pm 0.10$.  % chi2 1.34
\end{itemize}

The values for $M^\star$ are in general agreement with the stacked luminosity function 
for $0.4 < z < 0.8 $ clusters (see \citealt{Rudnick2009} and references therein), although the luminosity 
function for red galaxies in RXJ1347 displays a shallower faint-end slope than the (less massive) clusters studied 
by the previous authors.

%%%%%%%%%%%%%%%%%%%%%%%%%%%%%%%%%%%%%%%%%

\section{The cluster environment and the large scale structures around RXJ1347}
\label{S:LSS}

\subsection{Galaxy number density and map of the galaxy distribution}
\label{SS:Density}
.

We use the projected galaxy number density (\citealt{Dressler1980}) as a measure 
of the environment.  For each galaxy  we calculate the area of the circle of radius $r$ that 
encloses the 10th neighbour,  so the density is defined as:

\begin{equation}
 \Sigma_{10} = \frac{11}{\pi r_{10}^2}
\end{equation}

This quantity was calculated for each of the 100 Montecarlo realisations of 
the \photz\ cluster catalogue (Section\,\ref{S:selection}). 

 We preferred to use the 10th neighbour instead of the more usual fifth as projection 
is an issue in a photo-z selected catalogue. Using more galaxies reduces, in principle,
the effects of shot noise, limiting these uncertainties. 

This method can also be generalised to calculate the density for each point of a grid to produce
a map of the galaxy density distribution (see Fig.\,\ref{F:MapDens}). This procedure was called 
``cluster tomography'' by \citet{Pello2009} and can be considered analogous to the method used by 
\citet{Haines2006a} to investigate the environment in the  Shapley supercluster.  
It allows us to sample with adequate resolution the high density regions without increasing the noise in 
the low density ones. It can, therefore, be regarded as an adaptive smoothing method, with the advantage that it conserves the number density of galaxies at each position of the grid.

The first contour was defined at the mean density plus 1-$\sigma$ of the CFHTLS-Deep fields (\citealt{Coupon2009}). 
 Taking mean and standard deviation of the four fields, we obtain a value of $\sim$11$\pm$4 galaxies/Mpc$^2$ at $z=0.45$,  using the same magnitude ($r<23.5$) and photometric redshift cuts ($0.38<\zphot<0.52$). 
Note, however, that the underdense areas in our field (marked with dashed rectangles in Fig.\,\ref{F:MapDens}).
have an average density of $\sim$9$\pm$3 galaxies/Mpc$^2$, determined from  statistics of the 
100 Monte Carlo realisations from the parent catalogue. This makes the first contour 2-$\sigma$ over the 
background. 

This map helps to visualise the complex dynamical situation around RXJ1347. The overdensity extends approximately 
in diagonal across the field $\sim$20\,Mpc. Some filaments and several overdense subclumps can be seen, mainly, 
along this large scale structure. Some of the overdensities coincide with previously identified cluster candidates.

Besides the main cluster, two other galaxy concentrations are prominent. One towards the South-East, coincident with 
the cluster LCDS0825 (\citealt{Gonzalez2001}) and another towards the North-East without previous identification, which we
will call in this paper as the NE-Clump.

\begin{figure*}
\begin{minipage}{0.72\textwidth}
%\sidecaption
\includegraphics[width=0.95\textwidth,clip,viewport=80 20 510 415]{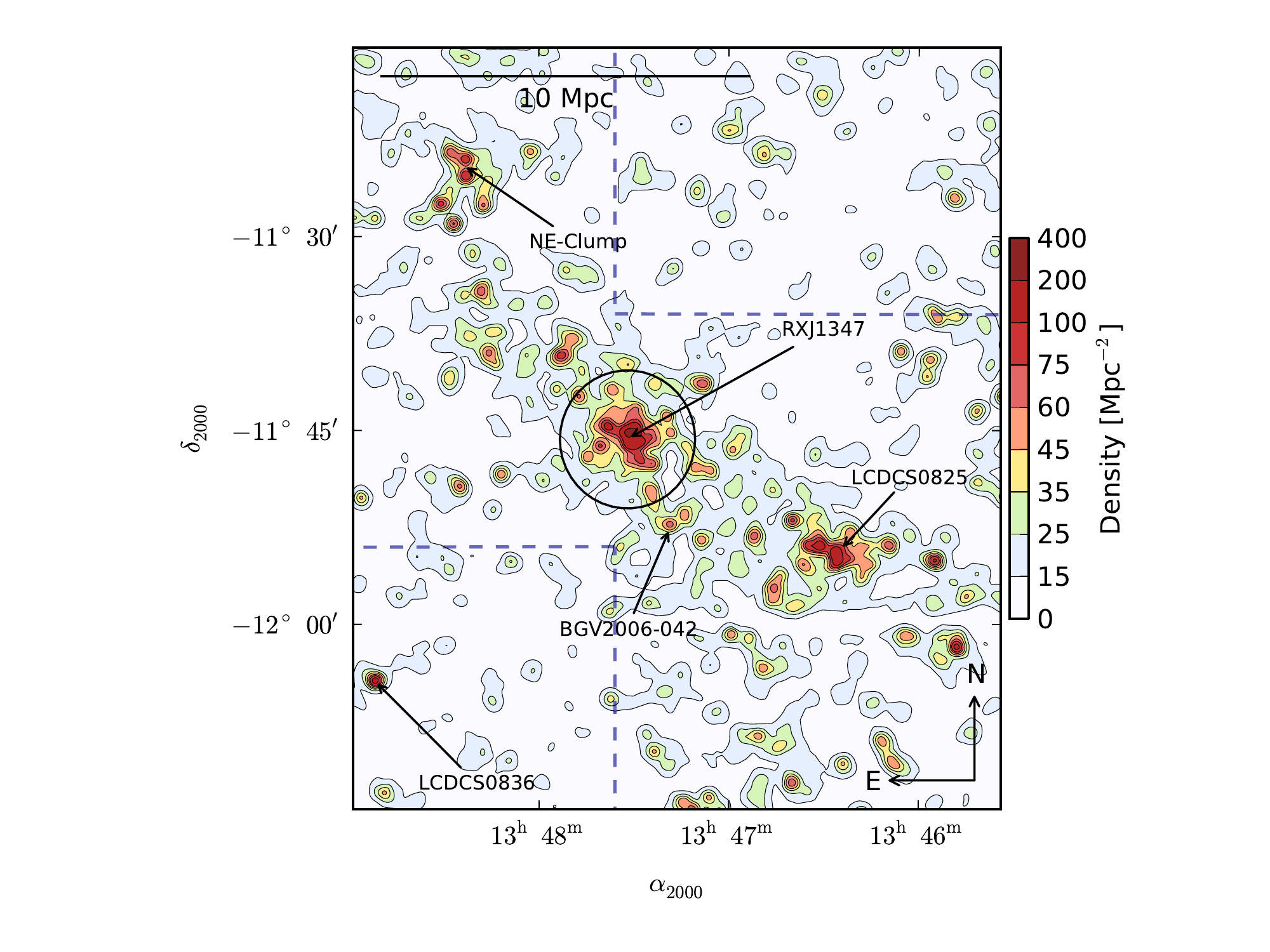}
\end{minipage}
\begin{minipage}{0.25\textwidth}
\caption{Map of the galaxy number density in Mpc$^{-2}$ produced by the nearest neighbour counting technique.
This map is the combination of 100 individual background subtracted maps. 
The first contour starts at the mean of the general field at $z\sim0.45$ plus 1-$\sigma$, obtained from 
CFHTLS (\citealt{Coupon2009}) and 2-$\sigma$ over the low density regions marked with dashed rectangles.
These regions were used for the statistical background subtraction (see Section\,\ref{S:selection}).
Note the overdensity associated with the central cluster runs in diagonal across the field, 
where a complex network of filaments and subclumps can be noticed. 
Previously detected clusters candidates associated with overdensities at $z\sim0.45$ have been marked, 
including the central cluster. The circle around RXJ1347 mark $R_{200}=2.3$\,Mpc from \citet{Gitti2007}. }
\label{F:MapDens}
\end{minipage}
\end{figure*}
\begin{figure*}
%\sidecaption
\begin{minipage}{0.72\textwidth}  
\includegraphics[width=0.95\textwidth,clip,viewport=80 20 510 415]{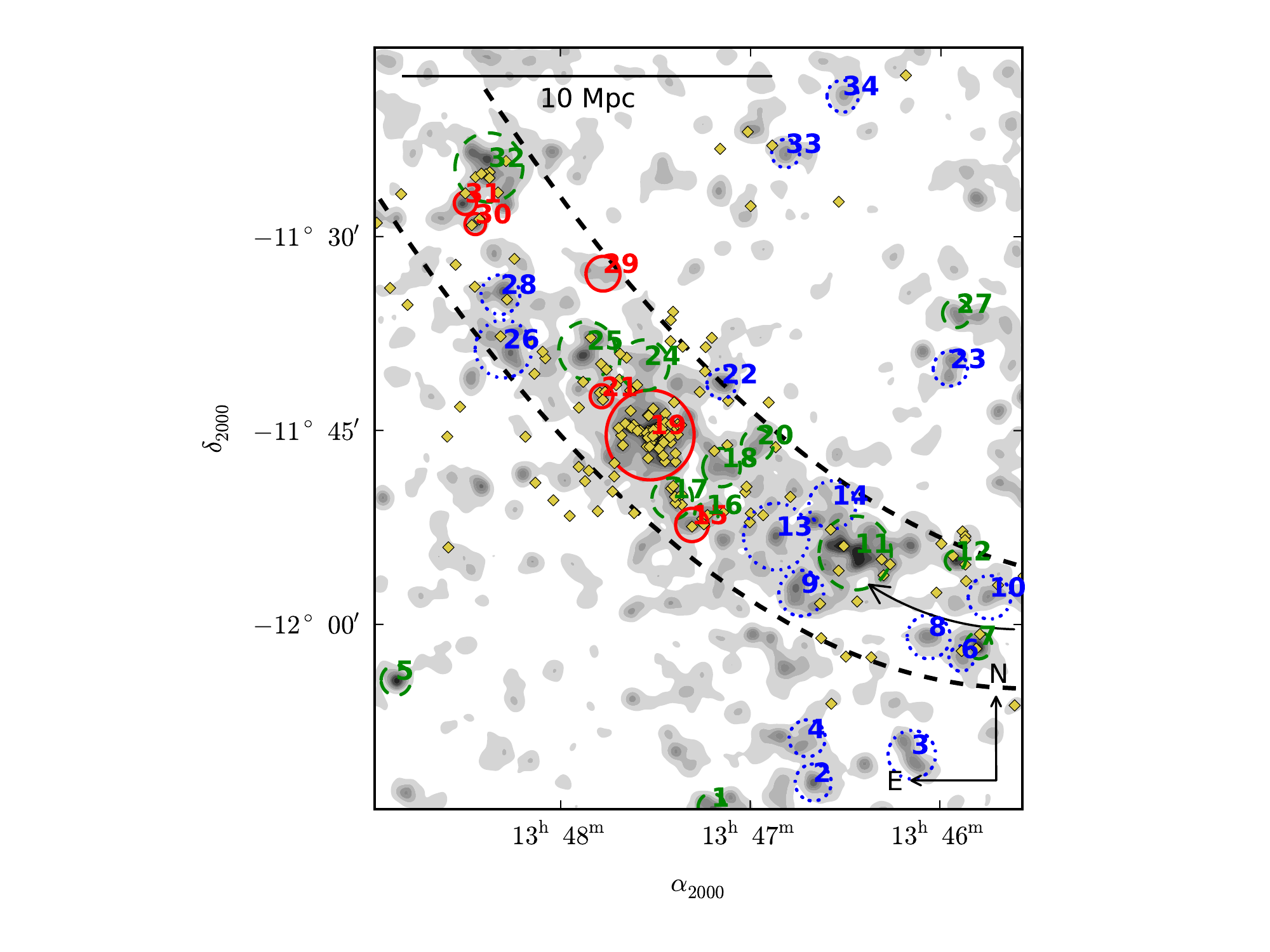}
\end{minipage}
\begin{minipage}{0.25\textwidth}
\caption{Distribution of the group candidates detected by the Voronoi-Delaunay algorithm over the density field. 
Contours are the same as in Fig.\,\ref{F:MapDens}. 
The numbers are group IDs as listed in Table\,\ref{T:tablegroups}. 
Groups have been colour coded according their galaxy content: 
Blue dotted circles represent groups with $F_{blue}>0.75$, green dashed circles are groups with  $0.75>F_{blue}>0.5$ 
and red solid circles are groups with $F_{blue}<0.5$. The large area delimited by  dashed lines marks  
large scale structure associated with the cluster. The curved arrow indicates that distances along the LSS are measured 
from the lower-right corner (South-West) towards the upper-right one (North-East). Dots are galaxies with 
\emph{spectroscopic} redshifts similar to the cluster. A zoom in of the central regions is shown in Fig.\,\ref{F:MapLx}.
}
\label{F:MapGroups}
\end{minipage}
\end{figure*}

\begin{figure*}
%\sidecaption
\begin{minipage}{0.72\textwidth}
\includegraphics[width=0.95\textwidth,clip,viewport=80 20 510 415]{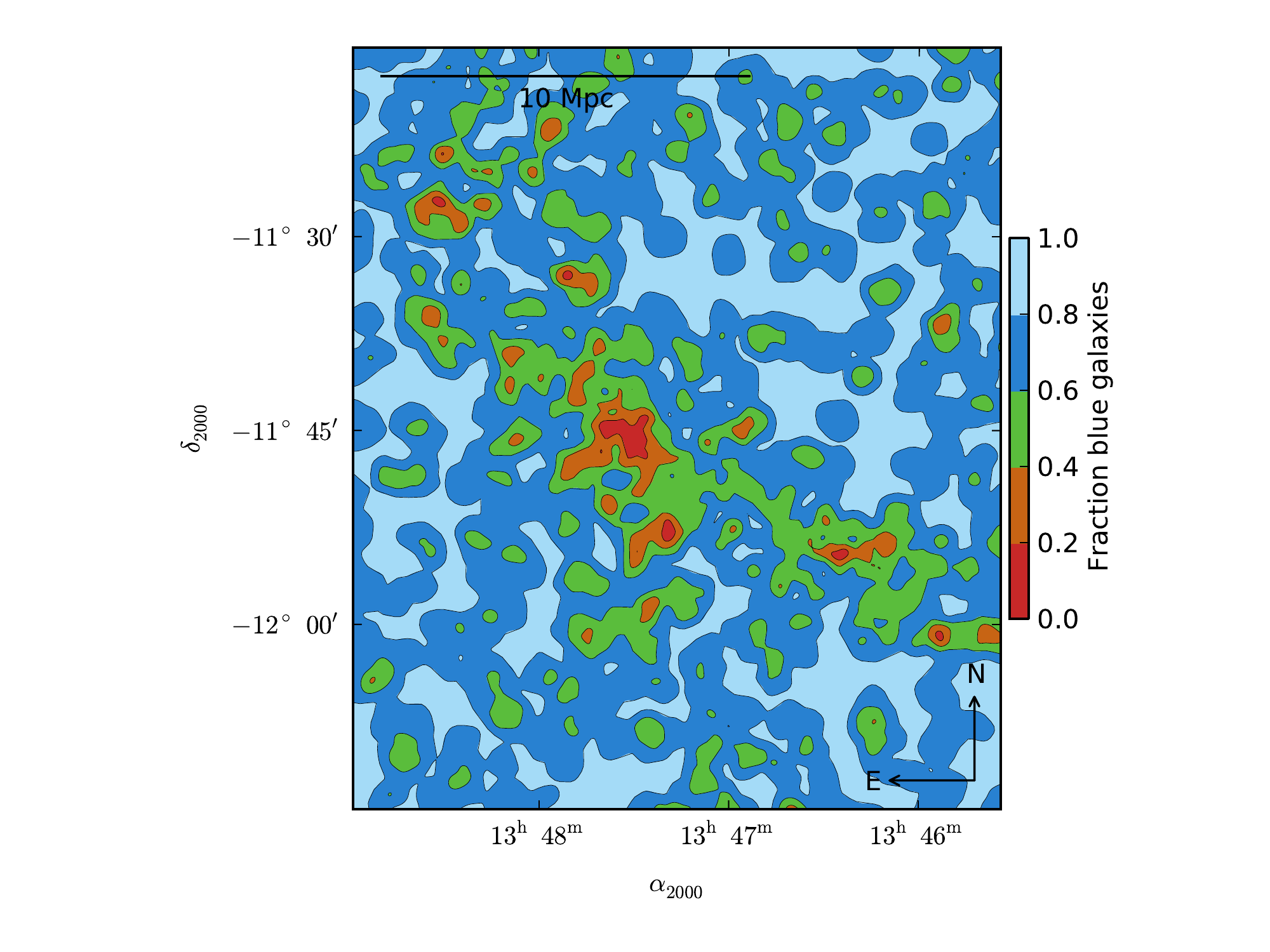}
\end{minipage}
\begin{minipage}{0.25\textwidth}
\caption{Map of the fraction of blue galaxies across the field. Comparing with Fig.\,\ref{F:MapDens}, the close relation
between galaxy mix and environment can be noted: Denser regions host larger fractions of red galaxies.This plot
illustrates however the complexity of this relation.  }
\label{F:MapFrac}
\end{minipage}
\end{figure*}

\begin{figure*}
\begin{minipage}{0.72\textwidth}
%\sidecaption
\includegraphics[width=0.95\textwidth,clip,viewport=80 20 510 415]{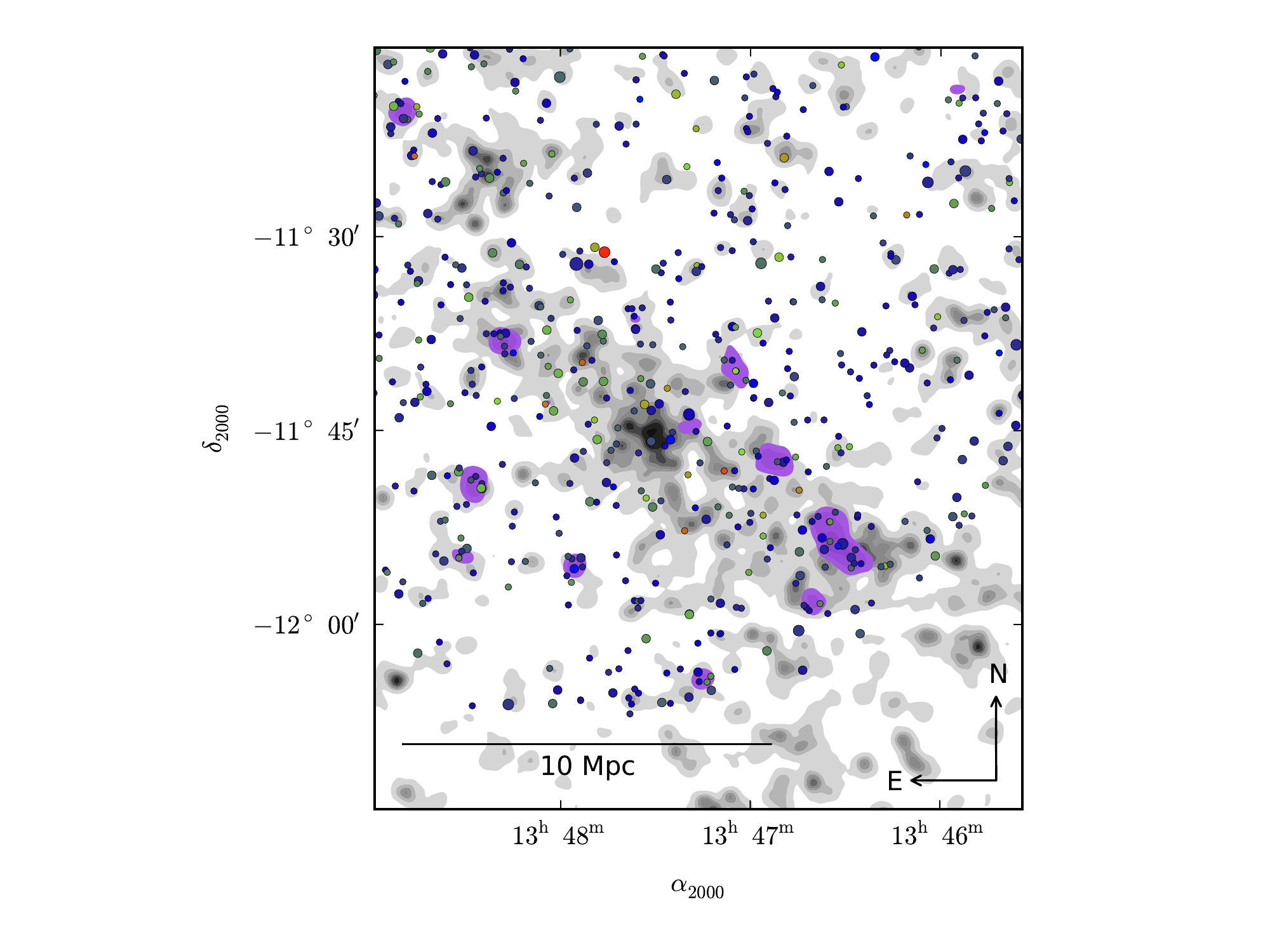}
\end{minipage}
\begin{minipage}{0.25\textwidth}
\caption{Distribution of GALEX NUV sources over the smoothed density field at $z=0.45$. Contours 
are the same as in Fig.\,\ref{F:MapDens}. Points have been colour-coded according the best-fit spectral template 
based on the optical photometry. Larger symbols indicate sources with higher NUV fluxes. The purple areas indicate regions
where the concentration of NUV sources exceeds 3-$\sigma$ of the field average. }
\label{F:MapGalex}
\end{minipage}
\end{figure*}

\subsection{Groups of galaxies}
\label{SS:groups}

Candidate galaxy groups were detected using the Voronoi-Delaunay tessellation technique (VTT) of \citet{Ramella2001} 
applied to the cluster photometric catalogue. The main advantage of this method is that it does not assume any 
particular physical properties of groups as other techniques. This allows to select galaxy concentrations
with different galaxy content and morphologies.  It is also very efficient in
detecting galaxy overdensities in inhomogeneous backgrounds, according to the simulations of SDSS fields by
\citet{Kim2002}. They also find that the efficiency of the VTT is greatly enhanced if galaxies are pre-selected 
in  colour space, because of the improved background contrast and smaller contamination. Variations
of this technique has been successfully used to produce galaxy cluster candidates catalogues either 
in  redshift space (\eg\ \citealt{Marinoni2002}) or by photometric selection (\eg\ \citealt{Geach2011a}).

The VTT of \citet{Ramella2001} is based in splitting the 2-D distribution of galaxies 
in independent cells, each containing only one galaxy. Galaxy groups candidates are selected 
from adjacent cells that satisfy a certain density threshold over the cumulative 
Kiang distribution (\citealt{Kiang1966}) of randomly positioned points. 
Once a significant number of cells have been associated, the overdensity is expanded circularly to include more 
galaxies until the density falls under the threshold. The radius of these circles ($R_{group}$) will be interpreted 
as the group physical size.

We select overdensities that are significant at the 90\% confidence level, \ie\ they form part of the top 90\% of 
the density distribution.   This level is higher than the one originally proposed by \citet{Ramella2001}, 
which was 80\% c.l.  We reject overdensities with  probabilities of being random fluctuation larger 
20\%, as we have already eliminated most of the contamination.  We note, however, that our results are 
resistant to the variation these choices.

The group detection was performed in the parent catalogue but individual quantities are calculated for each of the 
100 Monte Carlo realisations of the cluster catalogue. We do not claim that all of these groups are necessarily physical 
associations as this would require spectroscopic confirmation. They also may be unassociated to the cluster, 
however they allow us to probe particular environments. In total, we detect 34 group candidates.

Due to the lack of reliable mass measurements and possibly differences in their galaxy content, group
candidates will be characterised by two parameters: 

\begin{enumerate}
\item  The mean density of galaxies, calculated by counting the number of galaxies within $R_{group}$. Note 
that this number differs from $\Sigma_{10}$ indicator introduced in the previous section.

\item The total rest-frame $i$-band luminosity, obtained from summing the individual galaxies luminosities within $R_{group}$,
as a proxy of the total stellar content.
\end{enumerate}

The errors of both quantities are estimated from the statistics of the 100 MonteCarlo catalogues. 
Comparison of different density estimates are provided in Appendix\,\ref{A:DenComp}.

In Fig.\,\ref{F:MapGroups}, we plot the distribution of the candidate groups over the density map of the field. 
Most of the groups  are located along the large scale structure, but some are also quite isolated. The different
colours indicate their different  galaxy content, for red solid circles marking groups dominated by early types galaxies, 
green ones with an intermediate galaxy mix and blue ones, dominated by blue galaxies (see Section\,\ref{SS:groupevol} for 
more details). Properties of the group are summarised in Table\,\ref{T:tablegroups}.

\subsection{The Large Scale Structure}
\label{SS:LSS}

The overdensity associated to the cluster extends diagonally across 
the field for about 20\,Mpc.  As we also
want to study the effects of the filamentary structure on galaxy evolution, we isolate  
galaxies belonging to this structure. This is done by modelling it by two polynomial functions that 
encompass most of the groups found in the large  scale structure (see Fig.\,\ref{F:MapGroups}).
Distance is measured from the South-West corner towards the North-East with RXJ1347 roughly at
the middle.

\section{Search for X-ray emission for optically identified systems}
\label{SS:Xraygroups}

\subsection{Rosat All Sky Survey}

\begin{figure}
\includegraphics[width=0.98\columnwidth,clip,viewport=50 20 515 415]{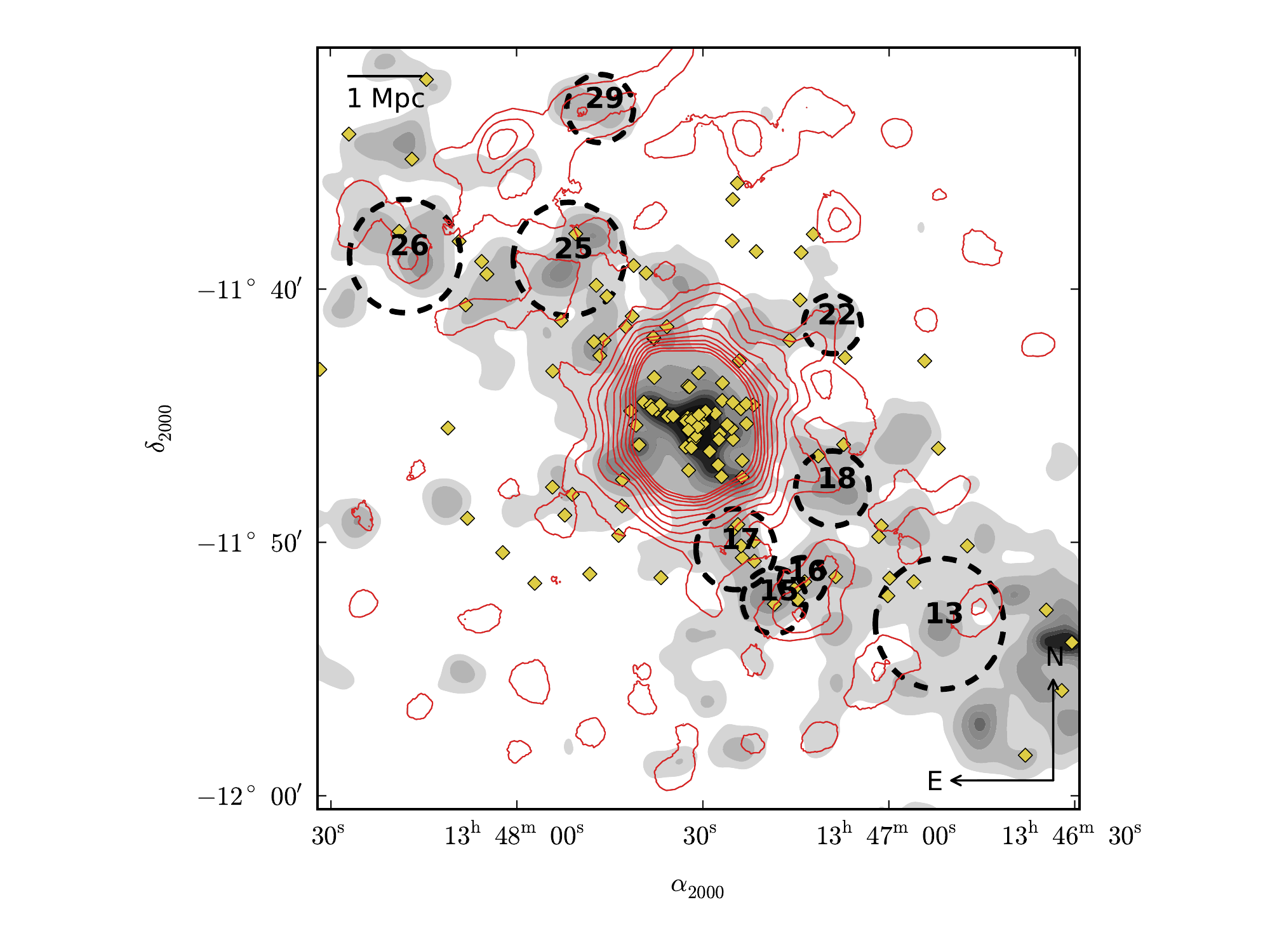}
\caption{XMM-$Newton$ X-ray significance contours (red) superimposed to the smooth galaxy density field (gray scale). 
The first X-ray contour starts at 1-$\sigma$ per resolution element (32\arcsec$\times$32\arcsec)
and are linearly spaced till 10-$\sigma$. The circles show the position of the optical group candidates where 
we  performed the measurements. The numbers mark the group IDs. Small dots show the position of spectroscopic member galaxies}
\label{F:MapLx}
\end{figure}

We  performed an analysis of ROSAT All Sky Survey (RASS)  data around RXJ1347 to see whether 
the two richer structures besides the central cluster (\ie\ LCDS0825 and the NE-Clump)  have  detectable 
X-ray emission.  We measured the flux within the aperture given by the cluster detection algorithm
and keeping the position fixed, \ie\ a forced detection. To measure the flux, we use two methods. 
1) A wavelet technique as in \citet{Finoguenov2010}, 2) the growth curve 
analysis of \citet{Bohringer2000}. 

For LCDCS0825, we find a 2-$\sigma$  detection with an upper limit for 
$F_{X,0.1-2.4\mathrm{\,keV}}, < 1.15\times 10^{-13}$\,ergs\,s$^{-1}$\,cm$^{-2}$,
which implies a luminosity of 
$L_{X,0.1-2.4\mathrm{\,keV}}<1.2\times 10^{44}$\,ergs\,s$^{-1}$. We take the $L_X-M$ relation of \citet{Leauthaud2010}: 

\begin{equation}
 \frac{\langle M_{200} E(z) \rangle}{M_0} = A \left( \frac{\langle L_X E(z)^{-1} \rangle}{L_{X,0}}  \right)^\alpha
\end{equation}

where $M_0=10^{13.7} h_{70}^{-1} M_\odot$, $L_{X,0}=10^{42.7}  h_{70}^{-1}$\,erg\,s$^{-1}$, $\log_{10}(A) = 0.068 \pm 0.063$ 
and $\alpha = 0.66 \pm 0.14$. This sets an upper limit of $M<3.1 \times 10^{14 } M_\odot$, consistent 
(within the errors) with the  reported weak lensing mass of \citet{Lu2010} (see more in Section\,\ref{SS:lcdcs0825}). 
However, the low significance prevent us to set further constrains.

The North-East Clump (ID=32) has a 1$\sigma$ detection in the RASS imaging. The upper limit for its X-ray flux is: 
$F_{X,0.1-2.4\mathrm{\,keV}} < 8.5\times 10^{14}$\,ergs\,s$^{-1}$\,cm$^{-2}$ and the luminosity 
$L_{X,0.1-2.4\mathrm{\,keV}}< 8.2 \times 10^{43}$\,ergs\,s$^{-1}$. Using the same scaling relation, the upper mass 
limit is  $M<1.4 \times 10^{14} M_\odot$.

\subsection{XMM-$Newton$}

XMM-$Newton$ observations cover the central regions of the field (see Fig.\,\ref{F:observations}), tracing part 
of the large scale structure associated with RXJ1347. A number of optically selected groups candidates are located in 
the region and we try to recover any X-ray emission from their warm-hot intergalactic medium.

Like in the previous case, we performed a forced detection, using the information from optical group detection algorithm.
Point sources were removed and the background estimation (in the 0.5--2\,keV range)  
was done using areas outside the cluster large scale structure. 

In total we were able to measure the flux with over 1-$\sigma$  significance for nine groups.  The properties are summarised in Table\,\ref{T:xmmgroups}. The flux errors are propagated to the luminosities and mass estimates. 
The 20\% systematic uncertainty for the mass estimates (due to the relevant scaling relation) is not listed in the errors.

The X-ray significance map is shown in Fig.\,\ref{F:MapLx}, with contours starting at 1-$\sigma$ per resolution element
(32\arcsec$\times$32\arcsec). The X-ray emission for some of the groups might the residual of
the  central cluster emission (groups 17, 18 and 22). 

\section{Clusters belonging to large scale structures around RXJ1347}
\label{S:Clusters}

Some structures in the studied field have been previously recognised. Some of them
match with detected overdensities at $z\sim0.45$ and thus are possibly associated with the RXJ1347. We have marked their
 position in Fig.\,\ref{F:MapDens}. In the following we provide an individual description for some of them.

\subsection{RX\,J1347.5-1145}
\label{SS:rxj1347}

This cluster was serendipitously  discovered by \citet{Schindler1995} as
part of the REFLEX Cluster Survey (\citealt{Bohringer2001}) using the ROSAT All Sky Survey (RASS).
It also showed as the first strong lensing system in the REFLEX identification optical imaging. 

RXJ1347 is also know as LCDCS\,0829 in the Las Campanas Distant Cluster Survey (\citealt{Gonzalez2001}).
Subsequent X-ray analysis with XMM-Newton have confirmed that it is indeed a massive structure with a X-ray luminosity 
in excess of $L_X = 6.0 \times 10^{45}$\,erg/s in the 2--10\,keV range, a temperature of $kT\sim10$\,keV and a mass 
estimate of $2.0 \pm 0.4 \times 10^{15} M_\odot$  within the central 1.7\,Mpc (\citealt{Gitti2004}). The same authors 
also find a  massive cooling flow  with a nominal accretion rate of $\sim$1900 M$_\odot$/yr.

Initial discrepancies in the mass derived using X-ray, strong and weak lensing analyses, were solved by more recent studies using a combination of techniques as in \citet{Bradac2005,Bradac2008b}. Furthermore, the initial low dynamical mass estimate ($\sim4\times10^{14}$\,$M_\odot$)  of \citet{Cohen2002}  appears to be solved in \citet{Lu2010}  by 
using a larger spectroscopic sample, yielding a mass well over $10^{15}$\,M$_\odot$ in concordance with the X-ray results.

In Fig.\,\ref{F:histoZgroups} we show the redshift distribution of the cluster members from our spectroscopy. 
A more detailed analysis of the cluster dynamics (including mass estimates) will be presented in a forthcoming
paper using the newer spectra.

\begin{figure}
\centering
\includegraphics[width=0.6\columnwidth,clip,viewport=40 180 330 570]{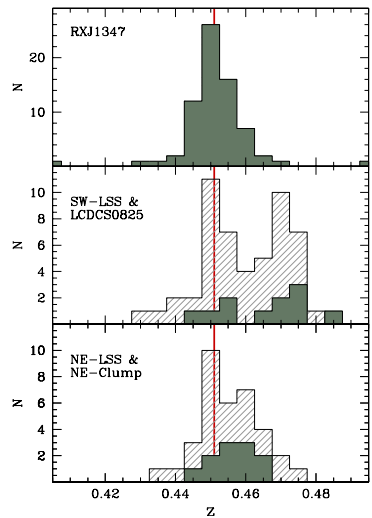}
\caption{Spectroscopic redshift distribution for some structures in the field:  Solid filled histograms 
show the redshift distribution for RXJ1347, LCDS0825 and the NE-Clump respectively. The hashed histograms display 
the redshift distributions for the associated large scale structures. The vertical line marks $z=0.45$}
\label{F:histoZgroups}
\end{figure}

For the moment, we would like to add the optical mass estimate using the richness measurements 
following \cite{Reyes2008}. The centre is taken as the peak of the X-ray emission from \citet{Schindler1997}.
As a first step we determine the richness $N_{200}$, which is the number of red-sequence galaxies within 
1\,Mpc\,h$^{-1}$ brighter than $0.4L^\star$ (\ie\ $M_i=-20.51$, see Fig.\,\ref{F:LFall}). From the 
statistical background subtracted catalogues, we count $N_{200}=86.9 \pm 3.1 $ galaxies.

Then we calculate the radius $R_{200}^{gal}=0.156\times N_{200}^{0.6}h^{-1}_{70}$\,Mpc, where the galaxy density is 200 
the mean of the Universe\footnote{This must not be confused with the $r_{200}$ parameter which refers to the matter density} yielding a result of $R_{200}^{gal}=2.27$\,Mpc. We iterate the richness estimator with this new radius
obtaining a richness measure of $N_{200}=154 \pm 5$ galaxies, which we  finally use to estimate the mass using the formula:

\begin{equation}
 M_{200}^{N_{200}}=(2.03 \pm 0.11 )\left (   \frac{N_{200}}{20}    \right)^{1.16\pm 0.09} \times 10^{14} M_\odot
\label{eq:m200gal}
\end{equation}

obtaining $M_{200}^{N_{200}} = 2.17^{+0.8}_{-0.6} \times 10^{15} M_\odot$ (random and systematics errors included), 
which is consistent with the estimates using other means. 

We  calculate the mass-to-light ratio by summing the rest-frame $i$-band luminosities of all galaxies within 
1.7\,Mpc. The result is $L=6.25\pm 0.2\times 10^{12} L_\odot$. Comparing with the reported 
X-ray determined mass of \citet{Gitti2004}, we obtain a mass-to-light ratio of $M/L=320\pm 64$\,$M_\odot/L_\odot$ which 
is typical for rich clusters (\citealt{Girardi2002}).

\begin{figure*}
\centering
\subfloat[RXJ1347]{\includegraphics[width=0.33\textwidth,clip,viewport=0 5 570 395]{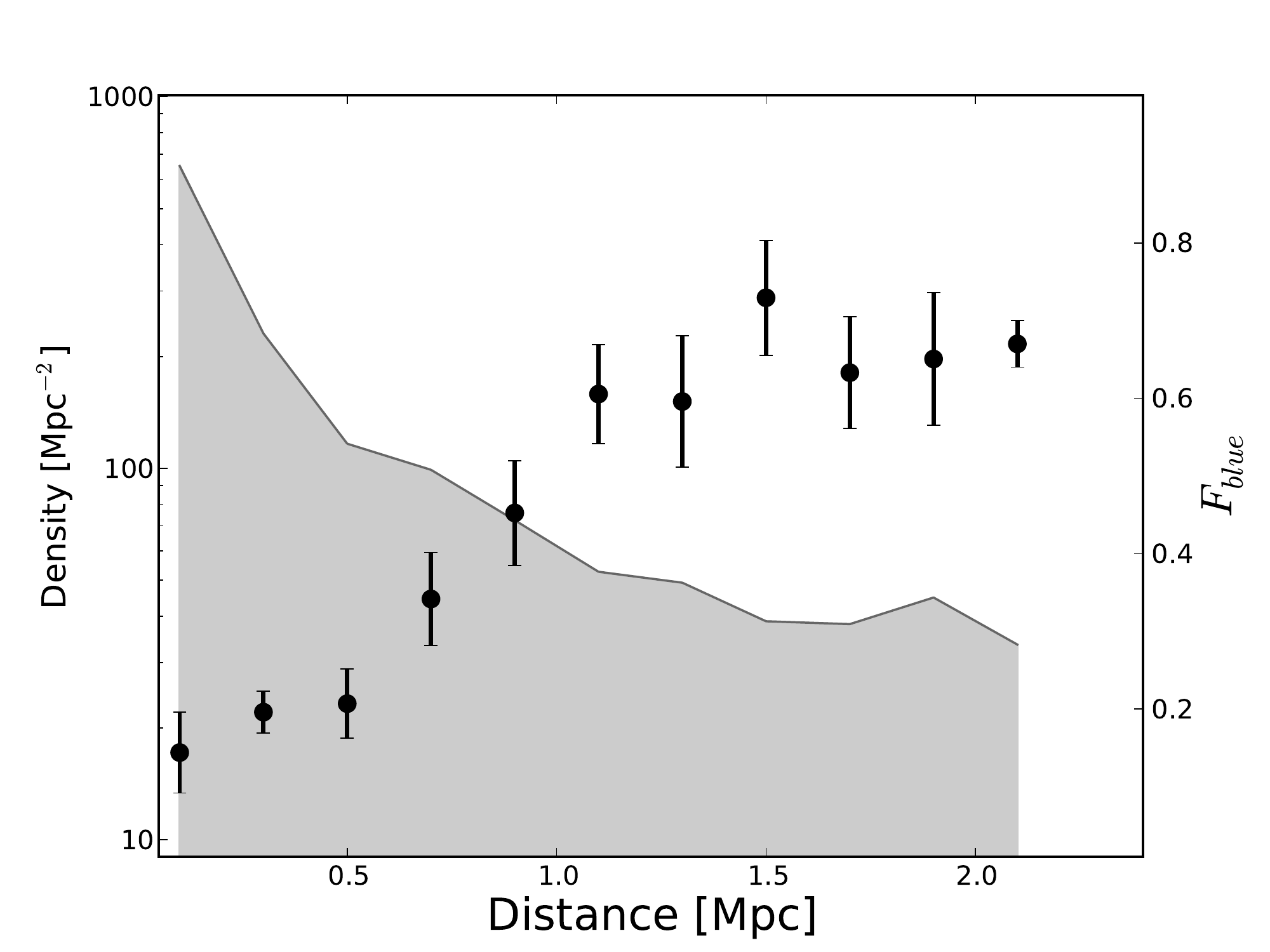}}
\subfloat[LCDCS0825]{\includegraphics[width=0.33\textwidth,clip,viewport=0 5 570 395]{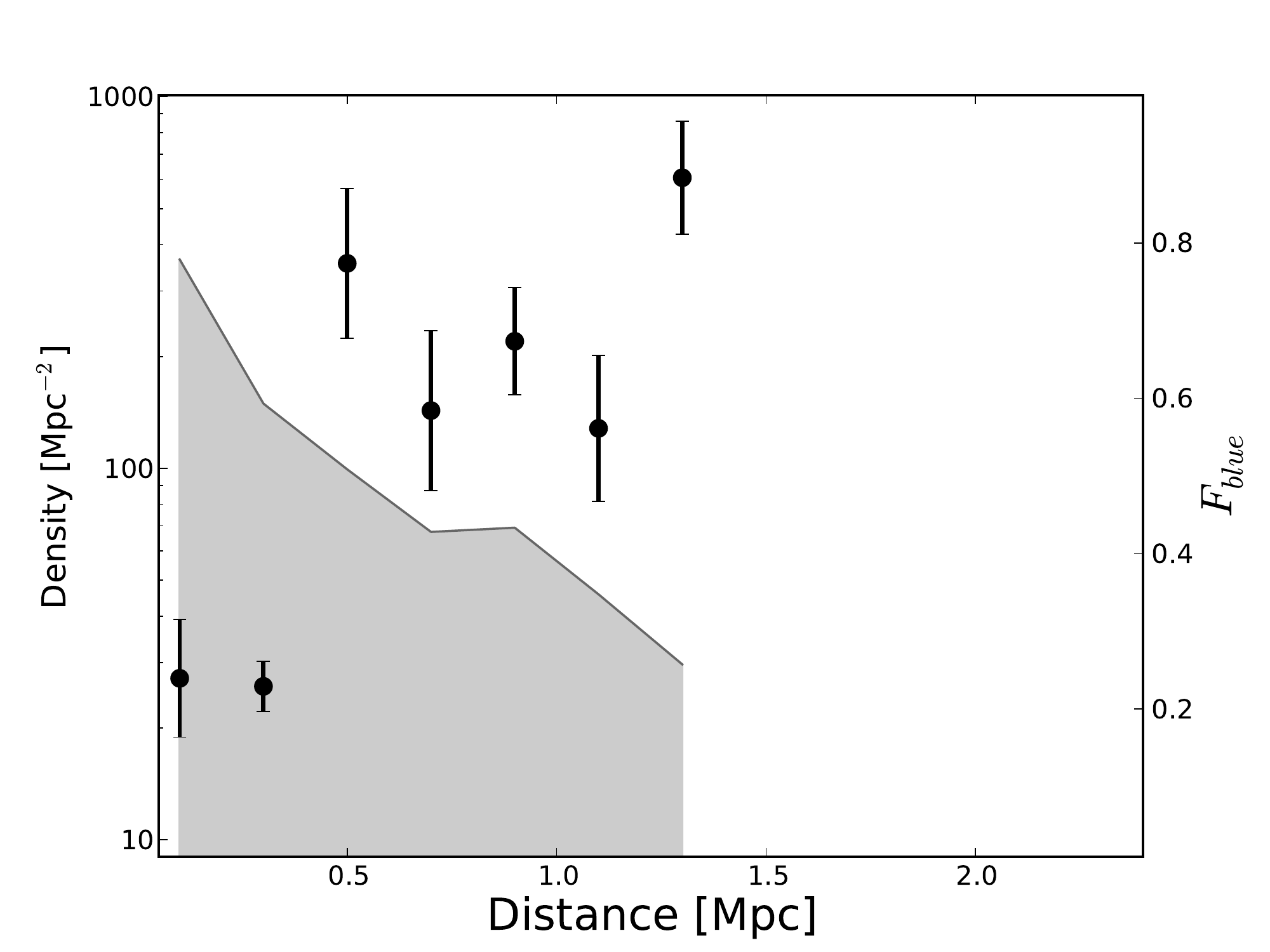}}
\subfloat[NE-Clump]{\includegraphics[width=0.33\textwidth,clip,viewport=0 5 570 395]{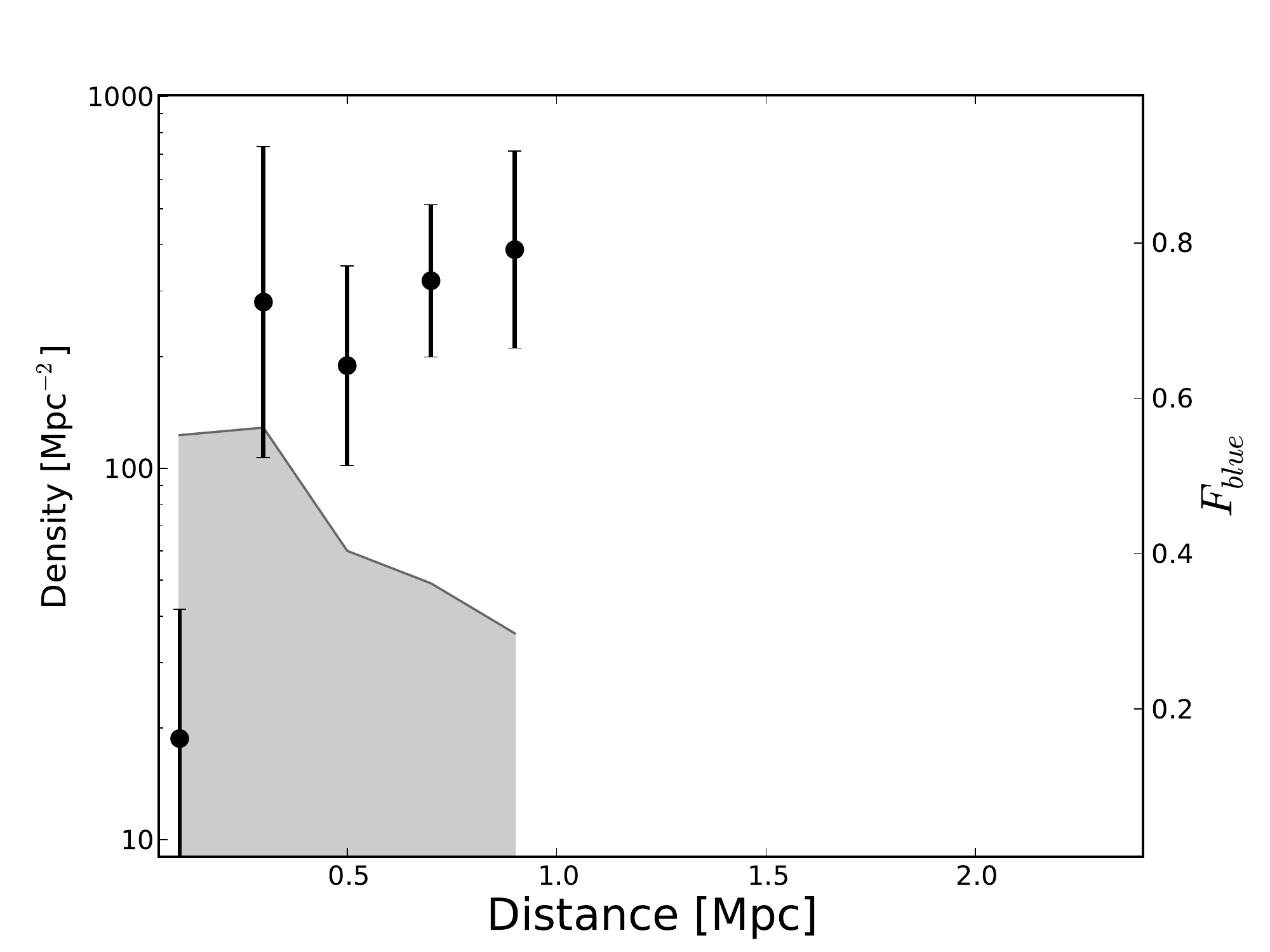}}
\caption{Blue fraction profiles for RXJ1347, LCDS0825 and the 
NE-Clump  in bins of 200\,kpc out to a distance of $R=1 R_{200}^{gal}$. 
The shaded areas in the background show the density profile calculated in rings of the same size.
Note the large difference in central densities and differences in the blue fraction profile.  
}
\label{F:PlotsClusters}
\end{figure*}
 
\subsection{LCDCS\,0825} 
\label{SS:lcdcs0825}
This structure has been previously identified as a tentative cluster by \citealt{Gonzalez2001} as part of the Las Campanas 
Distant Cluster Survey. 
\citet{Lu2010} obtained a redshift $z\sim 0.47$ (RXJ1347-SW in their analysis), which we confirm. This implies a  
comoving distance of $\sim$62\,Mpc to  RXJ1347.  

Although the concentration of galaxies is evident between  
RXJ1347 and LCDCS0825, \citet{Lu2010} were unable to confirm  
whether both structures are physically connected or are drifting away with cosmic expansion. \citet{Lu2010} reported a 
velocity dispersion of $\sigma=780\pm100$\,km\,s$^{-1}$ and a mass of $M_{200}=3.4^{+1.4}_{-1.1}\times10^{14}$\,M$_\odot$ 
for LCDCS0825, which would make it a relatively massive cluster. This mass is somewhat higher than the derived from the 
X-ray luminosity from the RASS data. However, we note that our X-ray detection has low significance and thus a large 
associated error (see Section\,\ref{SS:Xray}). Alternatively, the mass estimates from 
the cluster dynamics and weak lensing might be overestimated by the presence of substructures along the line of sight. 
Both hypotheses are plausible given the complex optical morphology of this cluster.

The total rest-frame $i$-band luminosity for this system within 1.22\,Mpc ($R_{200}$ as reported by \citealt{Lu2010}) 
is $L=2.72 \pm 0.12 \times 10^{12} L_\odot$, which implies a mass-to-light ratio of $M/L=125\pm 46 $\,$M_\odot/L_\odot$.

We also estimate the mass from optical traces  using the \citet{Reyes2008} procedure.  The centre of this object is taken 
from our group detection algorithm, but it is only 30\arcsec\ ($\sim$170\,kpc) away from the 
brightest cluster galaxy.

We obtain the following parameters: $N_{200}=40.7 \pm 2.1 $  galaxies, $R_{200}^{gal}=1.44$\,Mpc, which
correct the richness to $N_{200}=54 \pm 4$ yielding finally a mass of  
$M_{200}^{N_{200}}=6.4^{+2}_{-1.9} \times  10^{14} M_\odot$, somewhat higher than the \citet{Lu2010} estimate 
from weak lensing and dynamics but well within the combined error bars.

We have at the moment few redshifts for this object, so we cannot improve the  \citet{Lu2010} analysis, however we 
find a secondary (albeit with low significance) 
peak at $z\sim0.45$ (see Fig.\,\ref{F:histoZgroups}) that may indicate a complex system. \citealt{Lu2010} with a 
larger sample do not report the lower redshift feature, although  they targeted preferentially red galaxies. The 
redshift distribution of the associated large structure  shows, however, a bimodal distribution with one peak 
at $z\sim0.45$ and another at $z\sim0.47$. It is possible that we are observing a superposition of structures
 in this field.

\subsection{North-East clump} This object has not previously reported by studies of this field, although density 
and weak lensing maps by \citet{Lu2010} show a detection of this structure.  We spectroscopically confirm that it is 
at a similar redshift with $\langle z\rangle=0.456$ and thus likely belonging to the large scale structure associated 
with RXJ1347. It forms, however, a sparse association without a clearly defined centre. Many bright red galaxies are 
associated with this structure.

The redshift distribution from the 11 available spectra for this system  is quite broad, indicating that it may be still in process of assembling (see Fig.\,\ref{F:histoZgroups}). Using the biweight gapper algorithm of \citet{Beers1990} 
(with bootstrapping), we obtain a velocity dispersion of $\sigma= 1260 \pm 260$\,km\,s$^{-1}$, which would indicate 
a mass of $M_{200}\approx1.9 \times 10^{15}$\,M$_\odot$ following \citet{Girardi1998} formula (corrected for the adopted 
cosmology). This mass estimate would make this system almost as massive as the central cluster and is surely a gross 
overestimate. The redshift distribution of the associated large structure is also very broad but contrary to LCDCS0825 
does not show a secondary peak at $z=0.47$. Instead, it appears centred at $z\sim 0.45$ with an extended tail towards larger redshifts. 

We also use the optical mass estimates, with the centre  taken from our group detection  algorithm, obtaining a 
$N_{200}=14.9 \pm 01.3 $  galaxies, $R_{200}^{gal}=1.12$\,Mpc, a corrected richness of $N_{200}=20.5 \pm 4$ and an 
optical mass of  $M_{200}^{N_{200}}=2.1^{+1.1}_{-1} \times  10^{14} M_\odot$, which is likely  closer  to the real value 
and  concordant with the upper limits for the X-ray luminosity (see Section\,\ref{SS:Xray}).

\subsubsection{LCDCS\,0836} This object is a small, compact and isolated association of galaxies at $\zphot=0.45$. 
No spectroscopic redshifts are available for this object.

\subsubsection{BGV2006-042} This source was detected as a candidate cluster by \citet{Barkhouse2006} in their 
serendipitous X-ray cluster search using Chandra archive observations. Its position coincides with a possibly infalling 
bright elliptical galaxy located in the middle of a filament extending towards the  South-West. 
Five spectroscopic members are closely associated with this system (groups 15 and 16) with a median redshift 
of $z=0.4502$.

\subsection{Blue fraction profiles for individual clusters}

In Fig\,\ref{F:PlotsClusters} we plot the density profile and blue fractions for the three clusters
presented in the field (RXJ1347, LCDS0825 and the NE-Clump). These quantities have been calculated 
in radial bins of 200\,kpc centred in the clusters. Error bars are shown only for the blue fraction 
and are obtained from the statistics of individual measurements of the 100 Monte Carlo catalogues.

The blue fraction profile for RXJ1347 is apparently smoother than for the other clusters, which seem to 
``jump'' from low to high fractions in just a bin of radius. This may be, however, an statistical artefact given the fixed bins used and the limited number of members for the lower mass systems. In fact, the radius where LCDCS0825 and the NE-Clump reach a relatively constant value is around $R\sim0.3 R_{200}^{gal}$, which is also similar in RXJ1347. 
This radius is much lower than found in other studies, which  is typically  $R\sim1-2R_{200}$ 
(\eg \citealt{Ellingson2001,Andreon2006,Verdugo2008}). 

First, it is important to note that we are referring to $R_{200}^{gal}$, which is typically larger than $R_{200}$ 
(2.9\,Mpc versus 2.3\,Mpc for RXJ1347). Second, we are using a photometrically selected sample, reaching fainter 
magnitudes than typical spectroscopic studies. As shown in Section\,\ref{S:env}, less luminous 
galaxies start to change their properties at higher densities than brighter ones, thus these trends are affected by this 
behaviour as faint galaxies dominate the number counts.

Differences on  the radial trends of galaxy population for individual clusters have been, however, detected before 
(\eg\ \citealt{Verdugo2008,Mahajan2010}) and may be related to the large scale structure, dynamical properties 
or nature of the intracluster medium (\citealt{Urquhart2010}).

The density profiles for each cluster are provided to highlight the colour-density relation for these individual 
objects. Note the difference in the central density for each object, which also tends to be lower in the lower mass systems.

%%%%%%%%%%%%%%%%%%%%%%%%%%%%%%%%%%%%%%%%%
\section{The cluster environment and galaxy evolution}
\label{S:env}

\subsection{Blue fraction vs environment}
\label{SS:Bluefrac}

To further quantify the dependence of the galaxy mix with. environment, we plot in Fig.\,\ref{F:FracEnv} the fraction of 
blue galaxies ($F_{blue}$) versus galaxy  density ($\Sigma_{10}$) calculated in each 100 background subtracted catalogue.  
We calculated this statistic in logarithmic spaced bins and weighted it by the total number of sources in each bin. 
Errors bars are the 1-$\sigma$ standard deviation weighted in a similar manner.

The decline of  $F_{blue}$ towards higher densities found by many authors 
(\eg\ \citealt{Kodama2001,Pimbblet2002,Tanaka2005}) 
is also clearly appreciable around RXJ1347. At low densities almost 90\% of the galaxies are blue, whereas in the 
highest density bin only 10\% are part of the blue population. 

We then explore the dependence of this behaviour with luminosity, so we split the sample into bright 
($r<22$\,mag) and faint ($22\leq r<23.5$\,mag) galaxies. These two populations display a somewhat distinct behaviour. 
The blue fraction in the bright galaxy population displays a steady decline with increasing galaxy density reaching a 
saturation point at $\Sigma_{10}\approx 100$\,Mpc$^{-2}$.  On the other hand, $F_{blue}$ for fainter galaxies starts  shows a slow decline until  $\Sigma_{10}\approx 40$\,Mpc$^{-2}$ after which it starts to decrease at faster pace. 

The fraction of blue galaxies at the lowest density bin is compatible with the values found to the field population.
They are calculated using the same redshifts and magnitude cuts in the CFHTLS-Deep fields from the public available 
catalogues from \citet{Coupon2009} (see Section\,\ref{SS:Field}).

We also constructed a map of the distribution of the galaxy types traced by the fraction of blue galaxies, using a
similar method as for the density map, \ie\ for each point in a grid we calculate instead the fraction of blue 
galaxies using the 10th nearest neighbours. As in the density map, this procedure is done for each of the 100 
randomly drawn catalogues and averaged afterwards. This map is presented in Fig.\,\ref{F:MapFrac}. 
Comparing with the density map (Fig.\,\ref{F:MapDens}), it is possible to visualise the relation 
between environment. Regions with higher fraction of red galaxies tend to located in denser environments. 
Despite this clear trend, the map allows to appreciate the complexities of galaxy mix with environment. There are few 
regions with low fraction of red galaxies without the corresponding increase in density. The opposite results also true, a 
few dense regions contain a high fraction of blue galaxies. 

Similar maps, either of the fraction of blue galaxies or mean colour have been shown by \eg\ \citet{Haines2006a} for the
Shapley supercluster ($z=0.05$), \citet{Schirmer2011} for the supercluster SCL2243-0935 at $z=0.45$ and 
\citet{Fassbender2008} for a cluster complex at $z=0.95$. All these works have highlighted the complexities of the 
distribution of galaxy types against mean galaxy density over large scales.

\subsection{Star formation activy around RXJ1347}
\label{SS:SFfrac}

We also quantify the distribution of the star-forming population by comparing the fraction of GALEX NUV sources as a 
function of galaxy number density. The results are presented in Fig.\,\ref{F:UVfraction}. Due to the smaller sample, we 
are unable to cover the same range of densities as in the case of the general blue population. The decline of the fraction 
of star-forming galaxies is a factor two between 10\,Mpc$^{-2}$ and 80\,Mpc$^{-2}$, which is similar for the blue fraction.

Comparing the number of the unobscured starbursts detected by GALEX to the general blue population, we find that this 
fraction is similar across the full range of densities traced here. It is also similar to  to the fraction found for COSMOS  field (see more in Section\,\ref{SS:Field}). The large error bars in the highest density bin are an indication  of the very low number  of star-forming galaxies found in the central  regions of  clusters. This results indicates that the GALEX NUV sources at our flux limits are a constant fraction of the blue population. 

To test whether the properties star-forming population change with environment, we calculate the rest-frame UV  
luminosity function in three environments: The low density environment with $\Sigma_{10}<20$\,Mpc$^{-2}$,  the medium with 
$20\leq\Sigma_{10}<40$\,Mpc$^{-2}$  and high density one with  $\Sigma_{10}\geq 40$\,Mpc$^{-2}$.  

\begin{figure}
\centering
\includegraphics[width=0.85\columnwidth,clip,viewport=0 7 540 395]{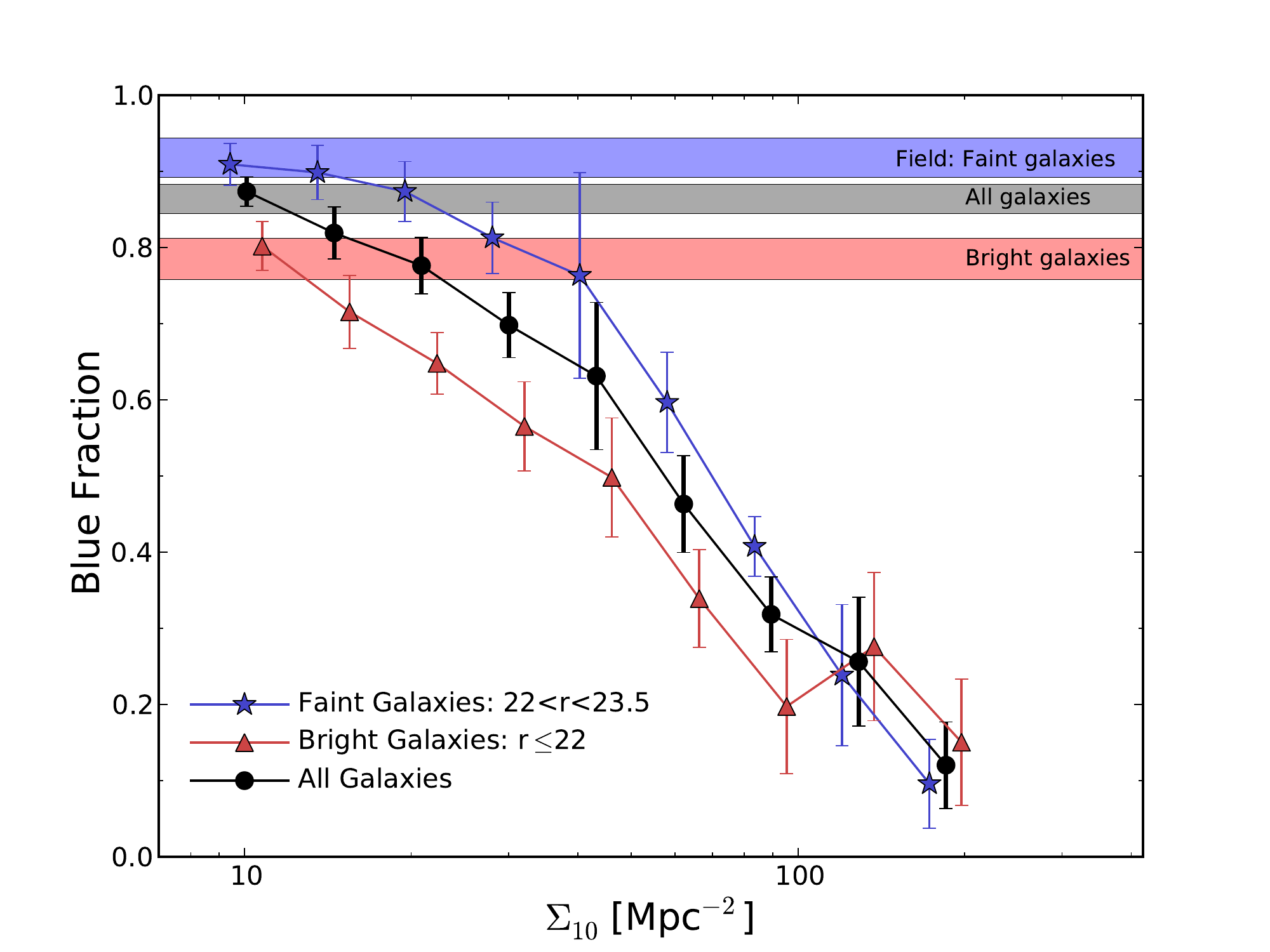}
\caption{Fraction of blue galaxies as a function of the galaxy number density, for all galaxies (black circles), 
faint galaxies (blue stars) and bright ones (red triangles) as indicated in the figure. The horizontal regions mark the 
field fraction for the same magnitude cuts, obtained by averaging the four CFHTLS-Deep fields (\citealt{Coupon2009}). }
\label{F:FracEnv}
\end{figure}

Absolute magnitudes were calculated directly from the apparent NUV magnitudes without applying k-corrections. As the GALEX NUV filter at $z=0.45$ has  a $\lambda_{eff}\approx1500$\,\AA\ (similar to the rest-frame FUV filter), we expect that the introduced errors  are  small.

The luminosity functions are plotted in Fig.\,\ref{F:OptNUV}, together with the best-fit Schechter function 
(\citealt{Schechter1976}, Eq.\,\ref{eq:lfeq}), obtained from a $\chi^2$ minimisation algorithm. The inset show 
the  1-$\sigma$ confidence limits. The results for the characteristic magnitude $M^\star_{UV}$ 
and the faint-end slope $\alpha$ are the following: 

\begin{itemize}
\item Low density:   $M^\star_{UV}$ = -19.47 $\pm$ 0.24,   $\alpha$=-0.77 $\pm$ 0.29 
\item Medium density: $M^\star_{UV}$ = -19.46 $\pm$ 0.45,   $\alpha$=-0.65 $\pm$ 0.50
\item High density: $M^\star_{UV}$ = -19.33 $\pm$ 0.93,   $\alpha$=-0.43 $\pm$ 1.25
\end{itemize}

The values of $M^\star_{UV}$ are about 1.3\,mag brighter than in $z\sim 0$ clusters 
(\eg\ \citealt{Haines2011a}), but consistent with field values at $z\sim 0.5$ 
(\citet{Arnouts2005}).  The faint-end slopes are however in our case somewhat shallower. This might 
be due to our relatively bright magnitude limit which does not allow us to sample adequately the faint-end 
of the luminosity function.

We also show in  Fig.\,\ref{F:OptNUV} the median $(g-r)$ colours for the star-forming population
as a function of the environment, together with the respective 25 and 75 percentiles of
the distribution.   The density bins are the same as in Fig.\,\ref{F:UVfraction}. 
The colours of the star-forming population is very similar across all galaxy densities.

Finally, we calculate the ultraviolet spectral slope $\beta$  for NUV sources. This parameter has 
been shown to be sensitive to the extinction in normal star-forming galaxies (\eg\ \citealt{Kong2004}). 
We use the  $u$-band and NUV magnitudes 
($\lambda_{eff}\approx 1560$ and $2580$\,\AA\ at  $z=0.45$ respectively) and a variation of formula of 
\citet{Kong2004} for local galaxies:

\begin{figure}
\centering
\includegraphics[width=0.85\columnwidth,clip,viewport=0 7 540 395]{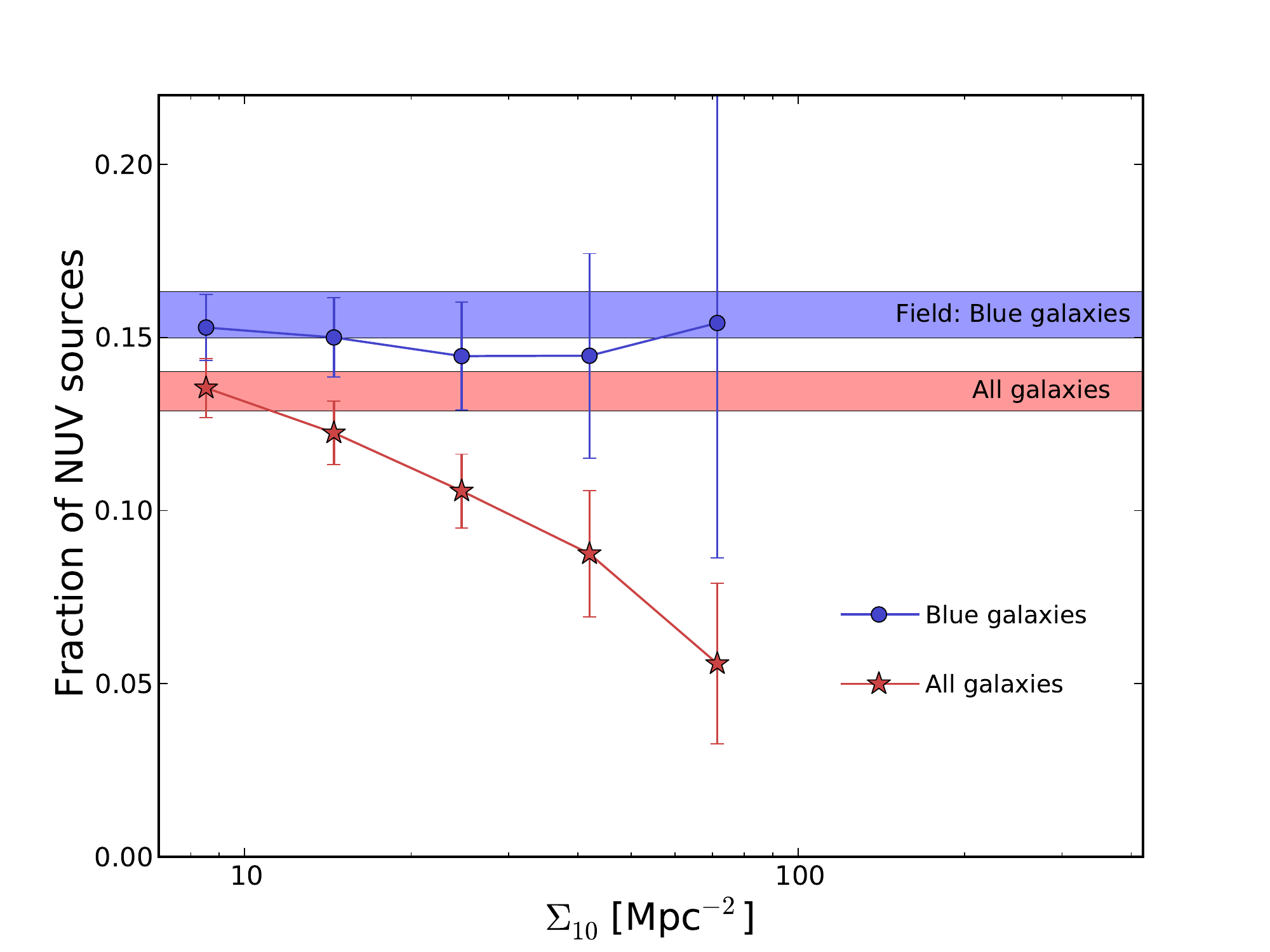}
\caption{Fraction of detected GALEX NUV sources compared with all galaxies and only blue ones, plotted as a function of 
galaxy number density. The number of the strong starburst detected by GALEX decreases towards the high density regions, 
but they represent a constant fraction of the blue population. The field fraction obtained for the COSMOS field is shown 
by the horizontal regions.}
\label{F:UVfraction}
\end{figure}

\begin{equation}
 \beta = \frac{ \log(f_{NUV}) - \log(f_{u})}{\log(\lambda_{NUV}) - \log(\lambda_{u})}
\end{equation}

where $\lambda_{NUV}$ and $\lambda_{u}$ are the respective effective wavelengths for those bands. 

The median of the values as a function of galaxy density is shown in the middle in the right panel of  
Fig.\,\ref{F:OptNUV}, together with the 25 and 75 percentiles of the distribution. These values are 
typical for UV-selected star forming galaxies  (\citealt{Schiminovich2005}) and do not differ
much with the population in low redshift clusters (\citealt{Haines2011b}).
They imply typical extinctions between  $0.5<A_{UV}< 2.5$\,mag at $\lambda=1500$\,\AA\ for 50\% of the sample
with a median of $A_{UV}\sim 1.5$\,mag. These values are also practically independent on environment.

A slight increase of the extinction is detected at intermediate densities (the top 75 percentile), however
it is still within the expectations for the star forming population.

\begin{figure*}
\centering
\includegraphics[width=0.33\textwidth,clip,viewport=15 5 520 400]{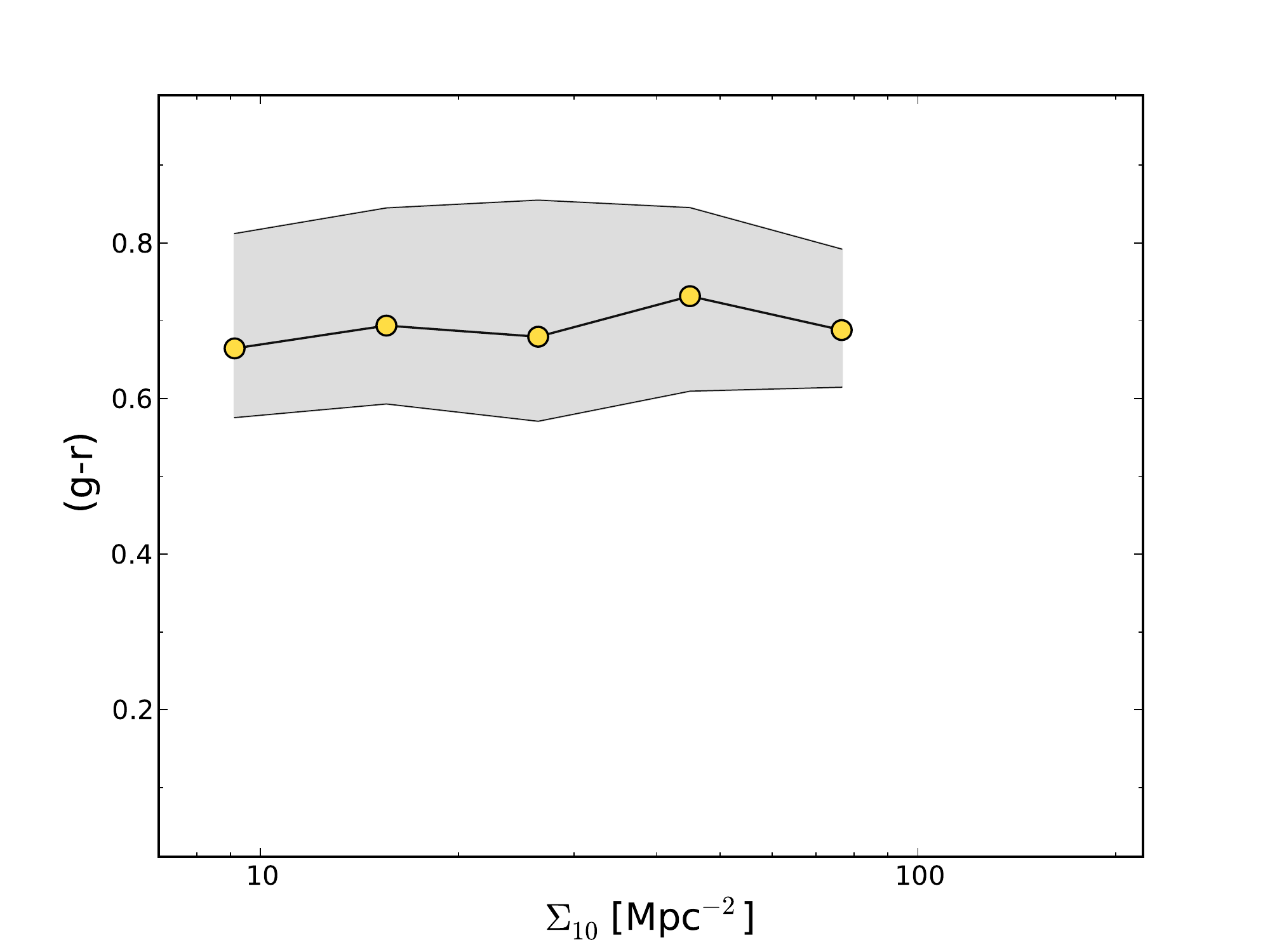}
\includegraphics[width=0.33\textwidth,clip,viewport=15 5 520 400]{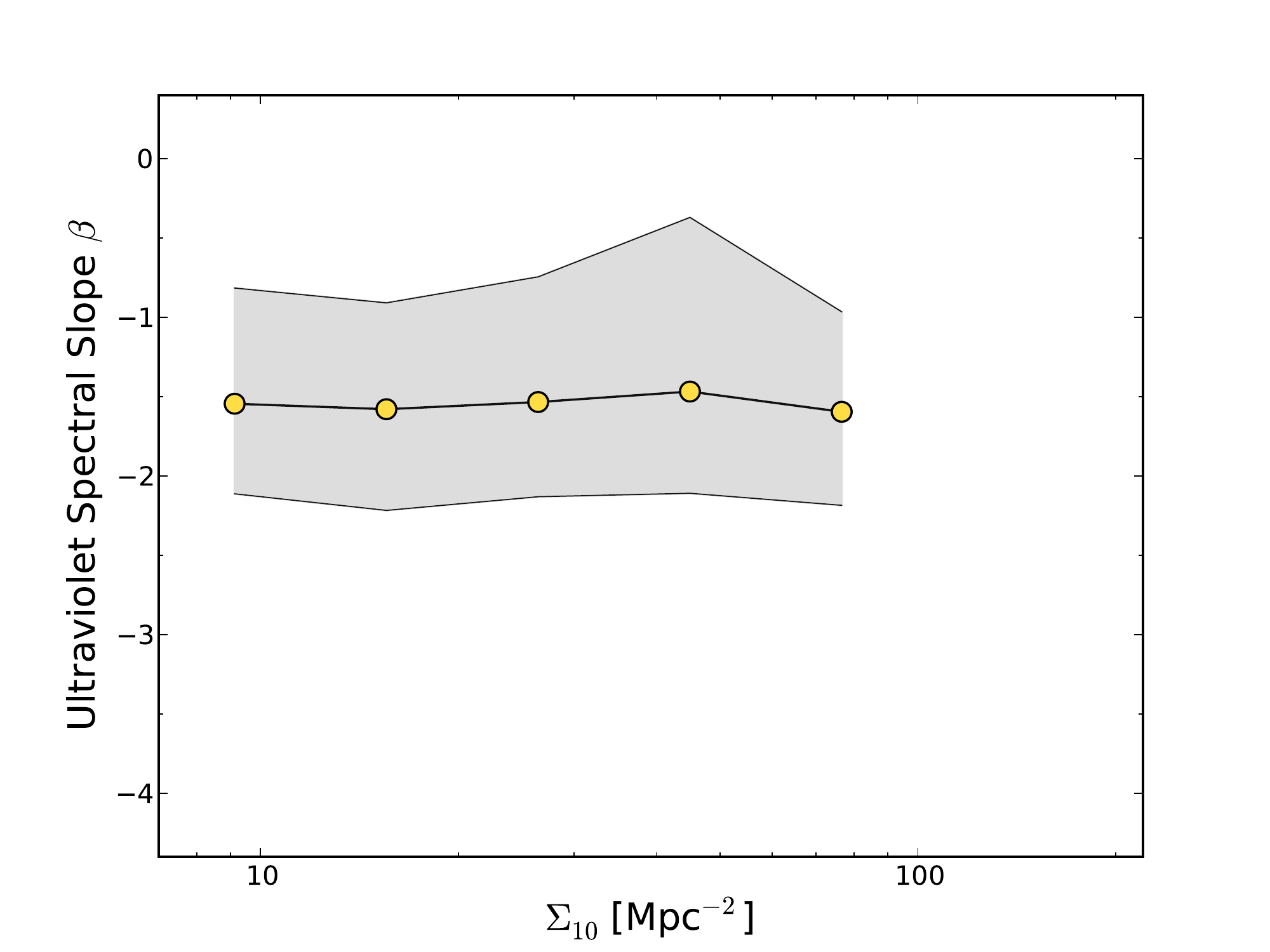}
\includegraphics[width=0.33\textwidth,clip,viewport=15 5 520 400]{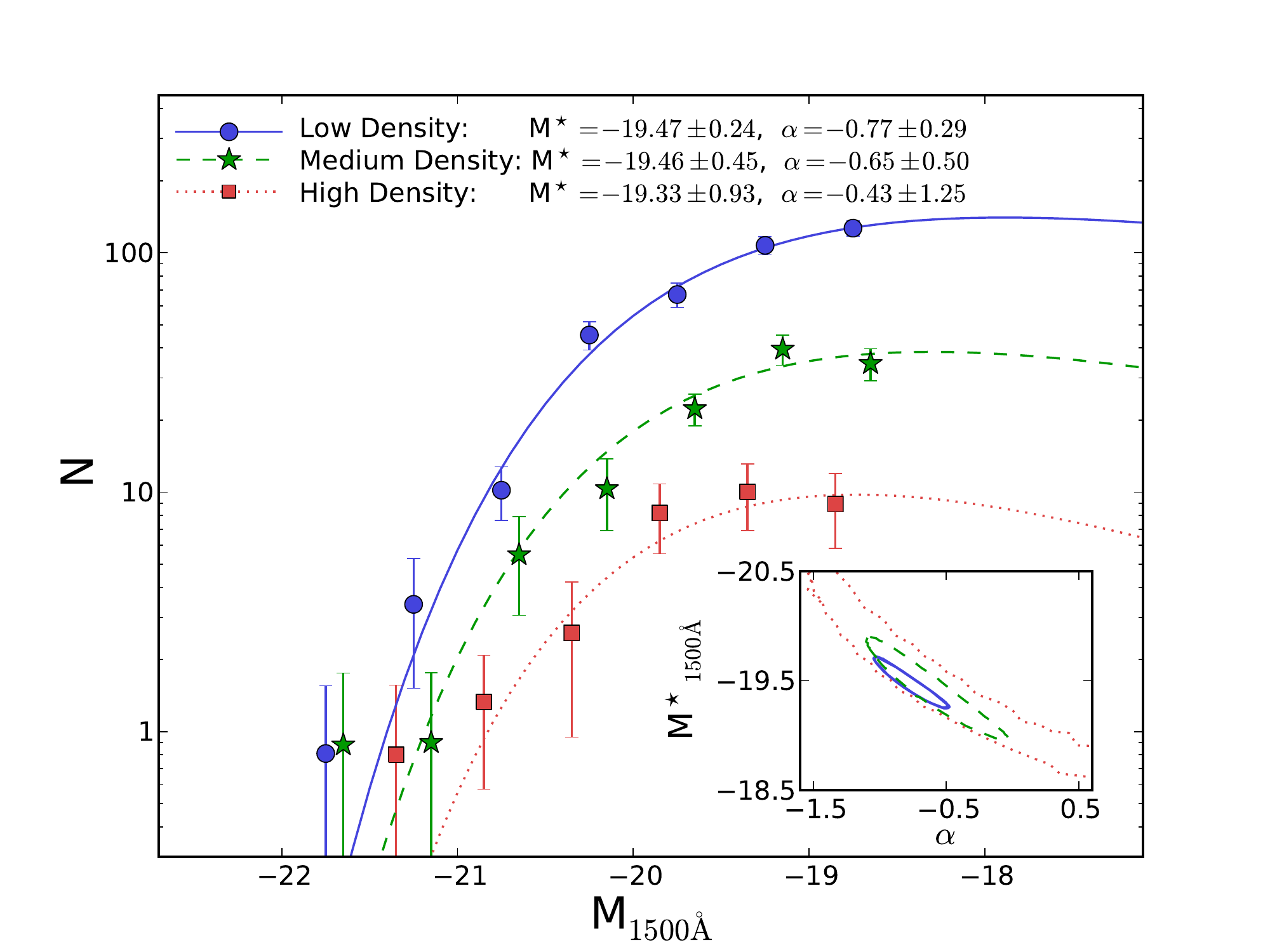}
\caption{Properties of the NUV emitters as a function
of environment. Left: The median $(g-r)$ colour. Middle: The median of the 
extinction sensitive ultraviolet spectral slope $\beta$. In both cases, the lower and upper lines 
that enclose the shaded areas represent the 25 and 75 percentiles of the distribution respectively. 
Right: Rest-frame UV (1450\,\AA) luminosity function for the GALEX sources in three different environments: 
The lines show the best-fit Schechter functions for three different environments: Low $\Sigma_{10}<20$\,Mpc$^{-2}$,  
Medium with $20\leq\Sigma_{10}<40$\,Mpc$^{-2}$  and high density  with  $\Sigma_{10}\geq 40$\,Mpc$^{-2}$. 
 The inset displays the  1-$\sigma$ uncertainty for  $\alpha$ and M$^\star_{UV}$ parameters.  }
\label{F:OptNUV}
\end{figure*}

These tests show that the star-forming population across the supercluster environment 
is notably homogeneous and only their relative abundance is decreasing with increasing density. 
This argues against a slow  decline of the star formation with environment and favours a rapid change 
between galaxy types, as transition objects appear to be statistically insignificant from the UV perspective.

It is possible, however, that the transformation is hidden by dust as some studies suggest (\eg\ \citealt{Wolf2009}). 
In that case, the nature of the star-formation should be different than in normal galaxies, probably more centrally 
concentrated, making difficult to probe the  UV emission due to the larger dust columns 
(\citealt{Geach2009}). However, this transition of star formation modes should also occur in short time-scales, 
otherwise we  would detect it, either as a reddening in the colour distribution of NUV sources, as a  dimming in the NUV 
luminosity function or as a change in the UV spectral slopes.

In Fig.\,\ref{F:MapGalex} we show the distribution of the NUV emitters. They are sparsely with very few obvious 
concentrations. Some of these concentration are coincident with detected group candidates. When we construct the 
density map  of NUV sources, using the same method described in Section\,\ref{S:env}, additional concentrations become 
appreciable. This is  due to the different probabilities for individual galaxies assigned during the statistical 
background subtraction. We plot  only concentrations  with at least 3-$\sigma$ significance over the mean of the field. 
It is interesting to note that some of  these concentrations are coincident with the large scale structure and candidate 
galaxy groups. In particular, the cluster  LCDS0825 features a relatively large population of star-forming galaxies.

\subsubsection{Comparison with the general field population}
\label{SS:Field}

Our investigation is focused in the galaxy properties in the large scale structures around RXJ1347. However, 
the lowest density bin traced in Figs.\,\ref{F:FracEnv} and \ref{F:UVfraction} can be considered as representative of the 
field population  as its density is lower than the mean of the CFHTLS fields. To test this and to provide proper comparison with the general field population, we use available data from current large area surveys.

In the case of Fig.\,\ref{F:FracEnv} we used the latest public  reduction of the CFHTLS-Deep fields (\citealt{Coupon2009}), which provides accurate photometric redshifts using a very similar methodology as our study. We use the same photometric redshifts and magnitude cuts than in our case. Since the CFHTLS-Deep fields are deeper than the data use in our investigation, their \photz\ are also more accurate, however we do not expect that this introduce any bias, as the effects of contamination from different redshift bins should be averaged out. In fact, it is reassuring that the blue fraction averaged for the four fields is very similar to the one measured in our lowest density bin, as can be seen in Fig.\,\ref{F:FracEnv}.

In the case of the NUV fraction comparison, we use data taken as part of the COSMOS survey centred on the CFHTLS-D2 field. 
Deep GALEX data has been taken over 2$\times$2\,deg$^2$ on this field (PI: Schiminovich) as part of the Deep Imaging Survey (DIS) and advanced data products have been made available (\eg\ \citealt{Zamojski2007}). However, to make our study fully comparable, we prefer to use the data from {\tt COSMOS\_00} stack, which is the shallowest stack available at MAST in this field. The  total exposure time in the NUV band of 8128\,s, similar to the data we are using for the cluster. We also use the catalogues  produced by the GALEX pipeline. We then match the CFHTLS-D2 field (in which the COSMOS field is centred) using the same procedure as described in \ref{SS:Galex}. NUV fractions are then calculated relative to the total population and the blue one. Like in the previous case, the general field star-forming fractions are very similar those 
calculated for the lowest density bin.

\subsection{Galaxy properties in groups near clusters}
\label{SS:groupevol}

In Fig.\,\ref{F:MapGroups} we have plotted the distribution of the groups detected by the Voronoi-Delaunay 
technique (Section\,\ref{SS:groups}). Blue fractions have been calculated within an aperture
given by the radius $R_{group}$ for each group for  each of the individual 
100 Monte Carlo catalogues. We consider  ``red groups'',  those with $F_{blue}<0.5$; ``green'', 
 those with $0.5<F_{blue}<0.75$ and ``blue'',  groups with $F_{blue}>0.75$. We have marked them with the 
corresponding colour in Fig.\,\ref{F:MapGroups}.
With some exceptions, we note that more evolved groups are located in the large scale structure and tend to be 
closer to larger overdensities. 

To test how the evolutionary status of galaxies depends on the environmental conditions within each group, we 
plot in Fig.\,\ref{F:GroupFrac} the fraction of blue galaxies as a function of the mean galaxy density for each group. 
Despite the large scatter, a trend is observed, in which denser groups (and clusters) have lower $F_{blue}$.
However, we do find few relatively dense groups (25--30\,Mpc$^{-2}$ with high $F_{blue}$.
Most of these groups are located at the South-West extreme of the large scale structure (see Fig.\,\ref{F:MapGroups}).
A number of groups with similar mean densities contain very low $F_{blue}$, at the same level of 
the very massive  central cluster.

Note that the fraction of blue galaxies for RXJ1347 is within the expectations from the Butcher-Oemler effect for 
clusters at those redshifts (\eg\ \citealt{Ellingson2001,Zenteno2011}).

Comparing $F_{blue}$ with the total rest-frame $i$-band luminosity for each group results 
in a similarly noisy correlation.  More luminous systems have indeed lower $F_{blue}$, whereas the fainter ones have 
high fractions.  In between, a high dispersion can be noted, including some relatively faint groups with very low 
fraction of blue galaxies.

\begin{figure}
\centering
\includegraphics[width=0.85\columnwidth,clip,viewport=20 20 520 520]{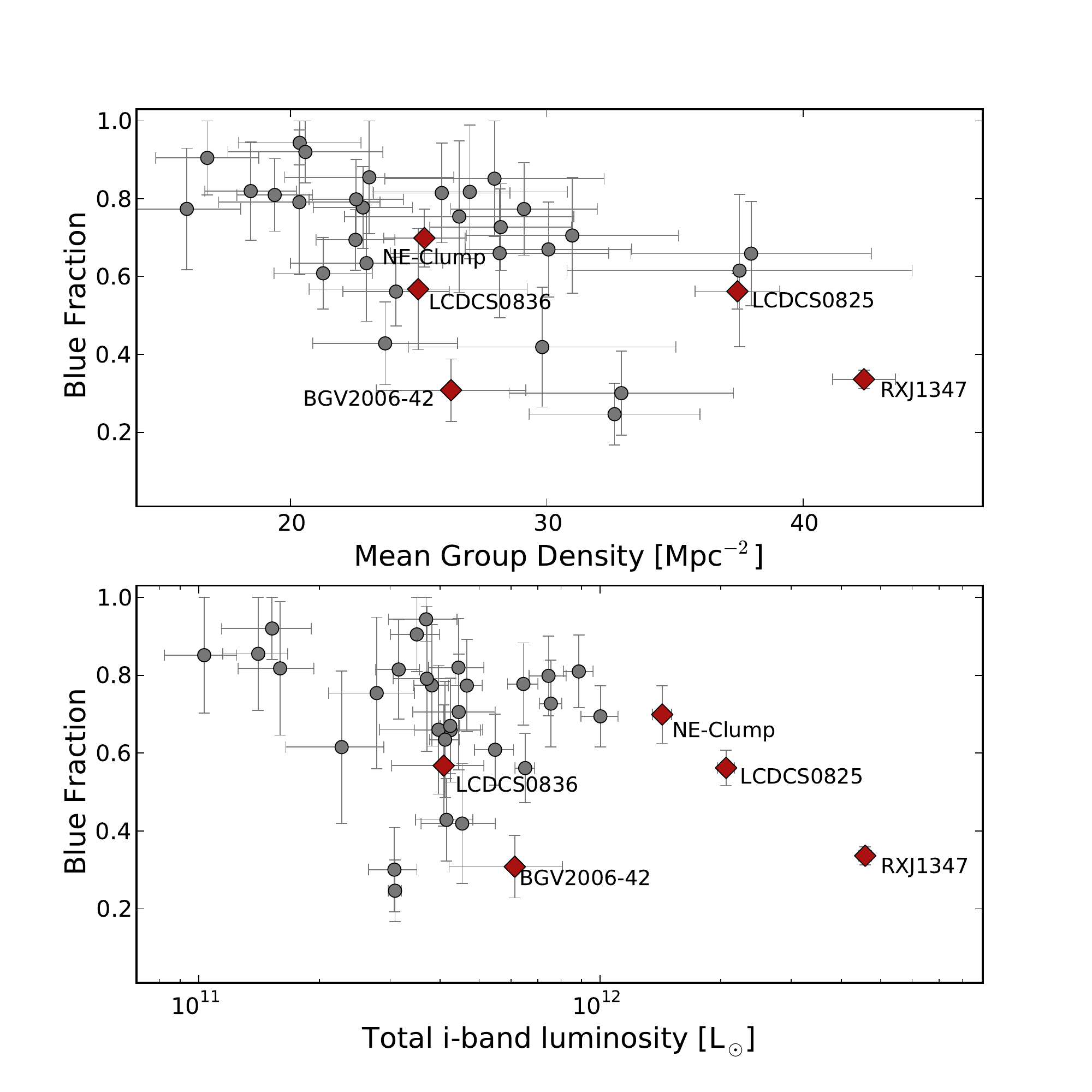}
\caption{Fraction of blue galaxies versus the mean group density and rest-frame $i$-band luminosity for each group. Error 
bars are the 1-$\sigma$ standard deviations obtained from the 100 Monte Carlo catalogues. Previously detected groups and 
clusters are marked by red diamonds and individual identifications. Note the large scatter in galaxy content for groups of 
similar 
properties.}
\label{F:GroupFrac}
\end{figure}

Although, it is difficult to assess the reality of these galaxy associations, as well as their dynamical state from 
photometry alone,  we find these trends interesting as the evolutionary state of groups may have a stronger impact on 
their galaxy content than their total masses.

We finally test the overall impact of groups in colour-density relation 
(Fig.\,\ref{F:FracEnv}). For that we separate galaxies according to their membership. Cluster galaxies 
are those  belonging to the three clusters in our field (namely RXJ1347, LCDS0825 and the NE-Clump). Group galaxies 
are those belonging to groups detected by the Voronoi technique. We include in each case all galaxies within 
twice the radius obtained from the group detection algorithm ($R_{group}$. This means that groups within twice 
the radius of the clusters are excluded to avoid duplicity. Similarly, we checked for duplicated galaxies in the 
composite group catalogue and eliminated them. We also created a catalogue with galaxies which do not belong to 
any of both categories to check the impact of the smooth density field in their properties.

Results are plotted in Fig.\,\ref{F:FracSplit}, where the fraction of blue galaxies versus galaxy density is displayed. 
Galaxies with no membership have a rather large blue population over a relatively large range of densities 
and similar to the general field. Galaxies belonging to groups display a rather modest change in $F_{blue}$ albeit 
with a large scatter. On the other hand, clusters exhibit a very strong change of $F_{blue}$ with density, indicating 
that they are very effective in transforming galaxies from one type into another.  
This also means that the bulk of the environmental signal in the colour-density relation 
(cf. Fig.\,\ref{F:FracEnv}) is likely due to these massive systems.

This does not mean that groups are not places of galaxy transformation. In fact, Fig.\ref{F:GroupFrac} shows the 
large range of properties of group candidates, however it is possible that the mechanisms acting in those places are different than in clusters and not necessarily related to the galaxy density. Furthermore, it is important 
to note that the diversity of group properties (as well as contamination effects) may wash out some of the 
environmental signal.

This result is similar to those of \citet{Li2009} where a large number of \emph{photometrically} selected groups around 
intermediate redshift clusters were analysed. They find that the environmental signal is weaker outside the virial radius of clusters. It is however in contradiction with studies at low redshift where the environmental variation 
of the galaxy population is similar regardless of the mass of the system (\citealt{lewis02,Balogh2004}).

\begin{figure}
\centering
\includegraphics[width=0.85\columnwidth,clip,viewport=0 5 540 395]{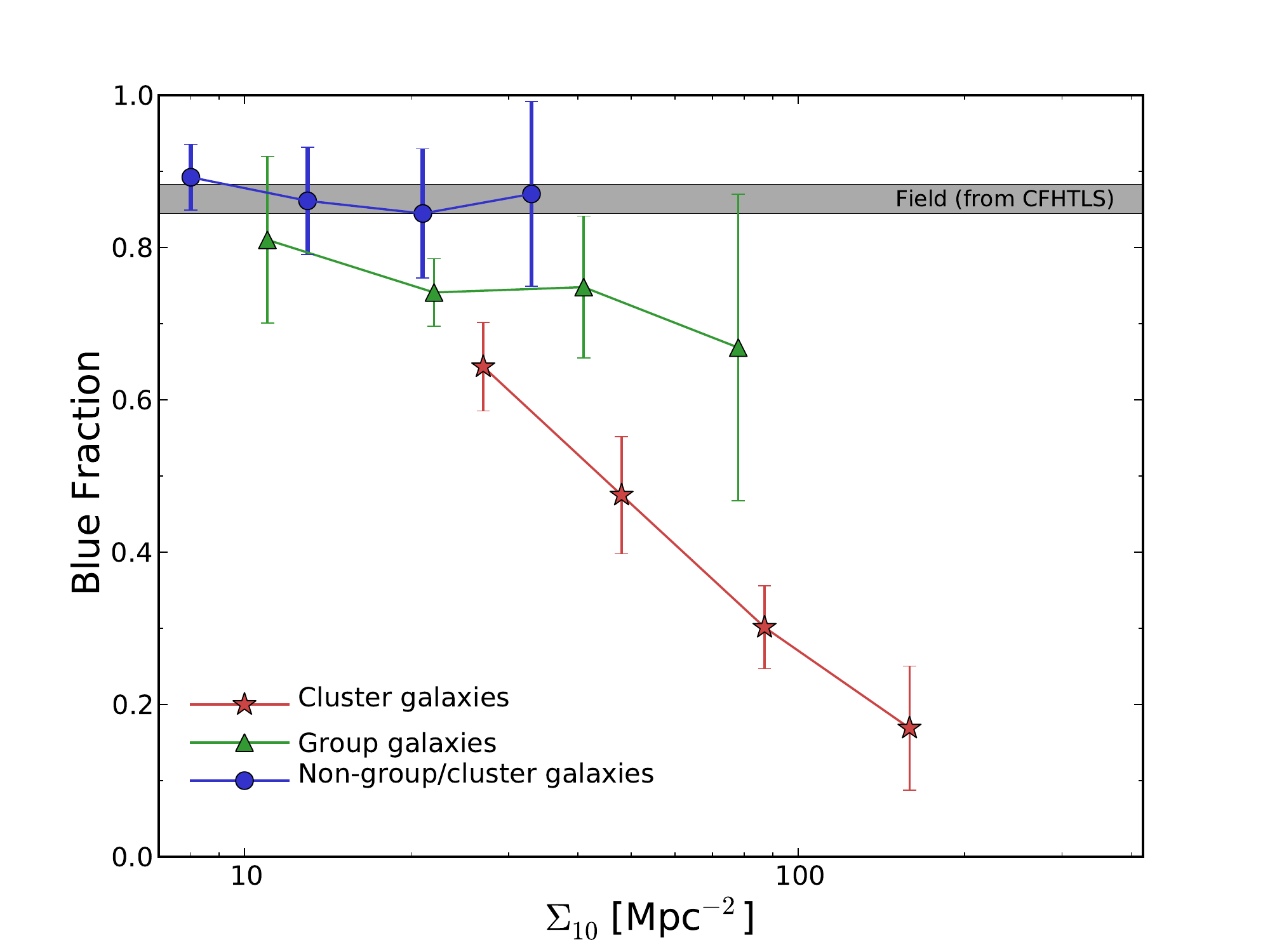}
\caption{Blue fraction as a function of galaxy density split according to galaxy membership: Clusters, groups or none.
Notice the weaker trends for non-cluster galaxies}  
\label{F:FracSplit}
\end{figure}

\subsection{Evolution along the large scale structure}
\label{SS:LSSevol}

We now test the evolution of galaxies along the large scale structure measured using an oblique area as indicated in 
Fig.\,\ref{F:MapGroups}. We trace the environment in this case using the position relative to the South-West corner. 
This may blur out the effect of small groups, but may enhance any effect purely related to the large scale structure. 

We calculate the fraction of blue galaxies and GALEX NUV sources as a function of distance. 
The results are shown in Fig.\,\ref{F:LSSfrac}. We plot in the background the mean galaxy density to appreciate its 
change along the LSS. The highest peaks associated with RXJ1347, LCDCS0825  and the North-East clump are evident.  
We can note in the plot the change of the blue fraction and how it increases when the galaxy  density decreases. In 
particular, we note the low fraction of blue galaxies associated with the position of RXJ1347, despite the contamination 
of galaxies at larger radii.

The colour-density relation is somewhat broken for the South-West extreme of the large scale structure, where few dense blue dominated groups are located. The mean density increases at that point and so does the blue fraction.

Similar increases of the star formation activity with density have been only reported at high redshifts 
(\citealt{Elbaz2007,Tran2010}) or in particular environments like galaxy pairs (\citealt{Ellison2008a,Wong2011}), but not 
often in groups at lower redshifts. We will wait for the results of our spectroscopic observations to make a complete 
assessment of the properties of these systems.

The fraction of GALEX NUV sources follows a similar trend with distance, albeit noisier due to the lower number of available sources. However, these large variations may also indicate places where the star formation is locally enhanced or suppressed depending on the local physical conditions. Unfortunately the GALEX observations do not cover the region of the  previously  mentioned blue dominated groups.

\begin{figure}
\includegraphics[width=0.9\columnwidth,clip,viewport=32 160 370 370]{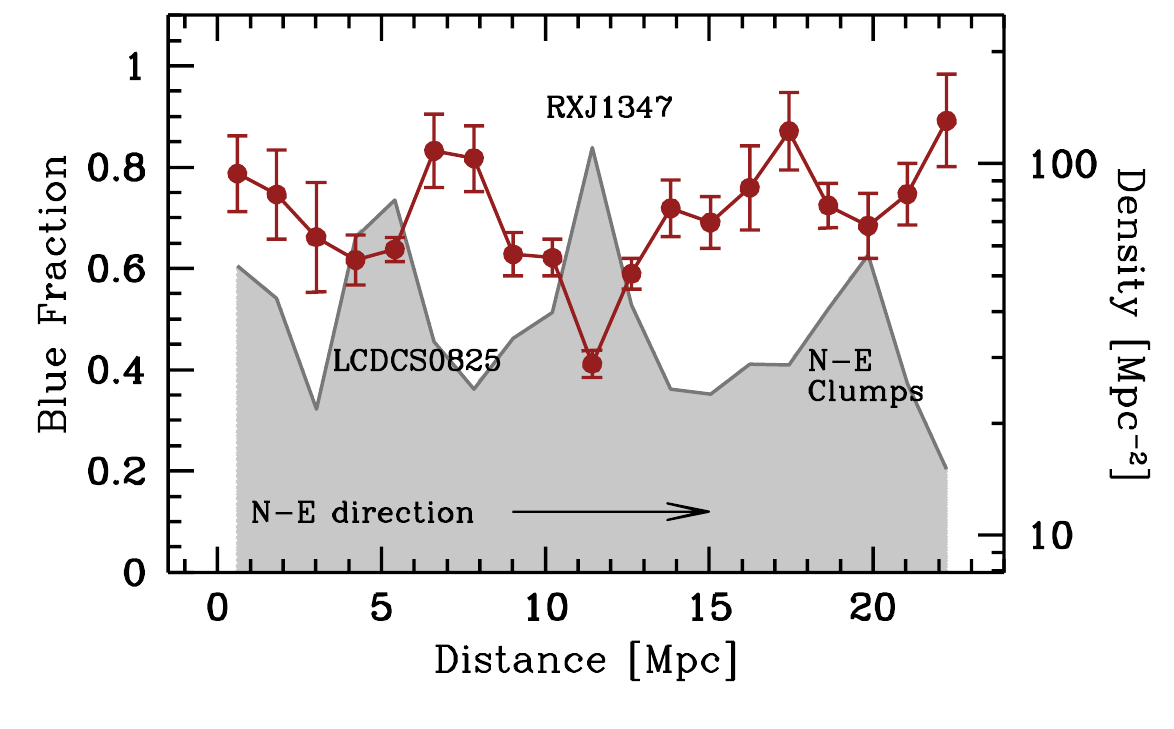}
\includegraphics[width=0.9\columnwidth,clip,viewport=32 160 370 370]{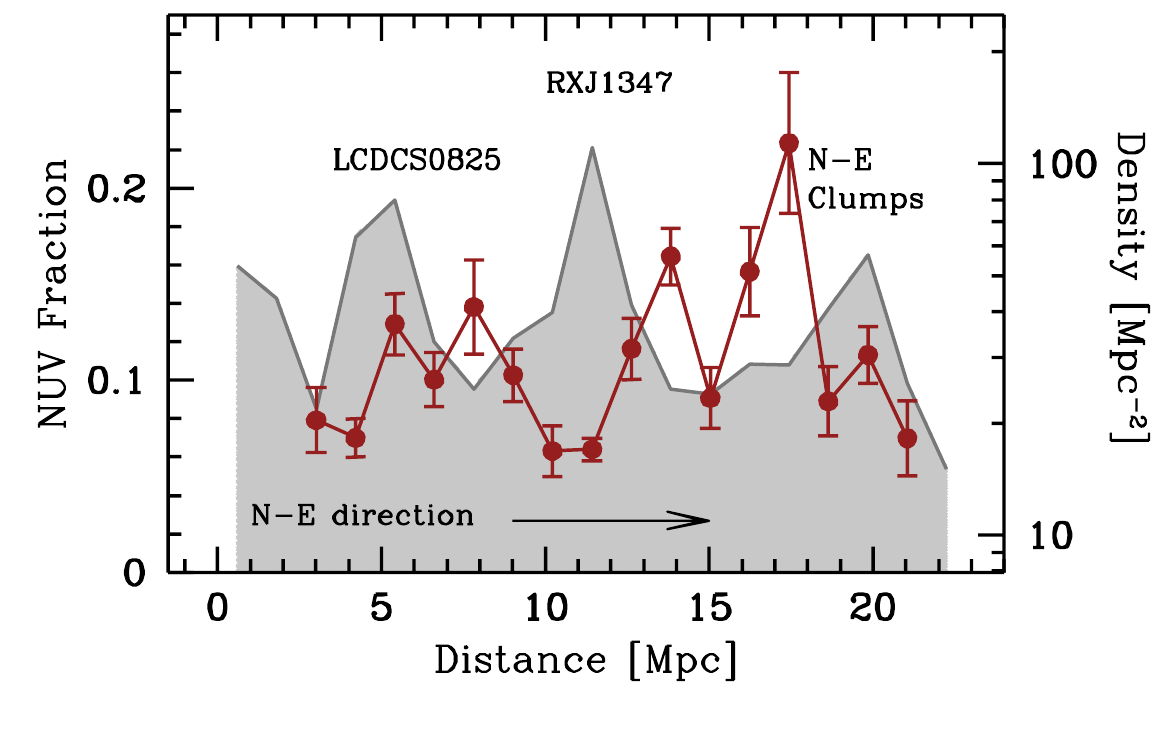}
\caption{Fraction of blue galaxies (top)  GALEX  NUV sources (bottom) as a function of position within the large scale 
structure (red points). The origin is taken from the South-West corner (see Fig.\,\ref{F:MapGroups}). In the background, we plot in grey shade the mean galaxy density of the large scale structure. The peaks associated to main clusters are clearly discernible and marked to guide the eye. An density enhancement is also observed in the South-West extreme of
the large scale structure.
}
\label{F:LSSfrac}
\end{figure}

\section{Discussion}
\label{S:Discu}

The study of effects of the large scale structure around clusters has been recently matter of intense research. This
is motivated by the evidence that galaxy transformations start at  large cluster-centric radii 
(\citealt{Rines2005,Porter2007,Verdugo2008}; etc). As simulations in $\Lambda$CDM cosmologies predict that matter is 
accreted into rich clusters through filaments and infalling smaller systems (\eg\ \citealt{Bond1996,Suhhonenko2011}),
these regions are expected to have an increase of interactions that might explain the origin of the cluster population
dichotomy.

In the local universe, \citet{Mercurio2006} and \citealt{Haines2006} have studied the optical properties of
galaxies inhabiting the Shapley supercluster ($z=0.05$). They have shown that the environmental dependence
of the global galaxy luminosity function is mostly driven by the change of the galaxy mix (colour-density relation).
The luminosity function for blue galaxies is similar at all environments.
Similar results were reported by \citet{Gavazzi2010} for galaxies belonging to the Coma supercluster.

In subsequent works, \citet{Haines2011a,Haines2011b} have completed a census of the star formation around 
the Shapley supercluster by using GALEX UV and Spitzer MIR observations. They found that the respective 
luminosity functions for cluster star-forming galaxies do not differ significantly from 
comparison fields.  \citet{Bai2008} and \citet{Biviano2011} have also found similar behaviour.

Using a very large catalogue of optically selected clusters, \citet{Hansen2009} have also provided
evidence of the very homogeneous nature of the blue population, which depends little on system mass
or clustercentric distance. This has also been confirmed by spectroscopic studies of intermediate 
redshift clusters. For example, \citet{Poggianti2008} and \citet{Verdugo2008}
have found the mean strengths of  emission lines ([OII] and H$\alpha$) for  the \emph{star forming} 
population show no dependence on environment.

Furthermore, detailed kinematical studies of spiral galaxies on distant clusters ($z=0.3 - 0.5$)
have failed to find important differences in their velocity fields compared to the field population
(\citealt{Kutdemir2010}). Larger samples are however needed to probe this in more detail.

In this paper, we have showed that the UV-selected star forming galaxies have similar properties at 
$z\sim 0.5$, regardless of their position in the large scale structure (Fig.\,\ref{F:OptNUV}).
Furthermore, we provide evidence that the cluster environment 
plays a fundamental role in shaping their own galaxy populations and trends due to less massive
galaxy systems are  much weaker (Fig\,\ref{F:FracSplit}, see also \citealt{Li2009}).  

Probably, the most viable transformation process to explain the homogeneity of the active population with environment
and its rapid transformation in massive systems, is ram-pressure stripping  (\eg\ \citealt{Fujita2004}). 
This mechanism provides the necessary time-scales to make transition between galaxy types fast enough to not easily 
detect transition galaxy types. 

Recent simulations (\eg\ \citealt{Book2010,Tecce2011}) have shown that the effects
of ram-pressure stripping are able to reproduce the 
star formation profiles in clusters  out to large clustercentric distances. 
In \citet{Tecce2011} models, the fraction of galaxies with cold gas within $R_{200}$ depends 
both on cluster mass and redshift. The predicted fraction at $z=0$ is very low for clusters with masses $M_{200}>10^{14}$\,M$_\odot$ and much higher for clusters with smaller masses. 
Interestingly, the relation between the blue fraction withi $R_{200}$ and cluster mass found by \citet{Hansen2009}
at $z\approx 0$ remains practically flat for clusters with masses   $M_{200}>10^{14}$\,M$_\odot$ and changes rapidly below that point, indicating a change in the effectivity of the transformation processes at that mass threshold
in concordance with the previous simulations.   

Regarding the effects of the large scale structure in the galaxy population,
we have shown that galaxy groups have a large diversity in their galaxy content 
(Fig\,\ref{F:GroupFrac}), with little dependence on their global galaxy density or total stellar mass content.
This might be signature that other effects, perhaps related to their dynamical state, alter the overall trends.

Finally, we should stress that we have found no 
evidence of enhanced star formation activity in the LSS as recent works  have claimed 
(\citealt{Braglia2007,Marcillac2007,Fadda2008,Porter2008}). 
Instead we find regions with larger fractions of blue and/or star forming galaxies at moderate densities,
indicating that some preprocessing is occurring in the large scale structure.  Those pockets of concentrated but rather 
normal star formation have been also recently found  around Cl\,0016+16 ($z=0.55$) by \citet{Geach2011}. 

\citet{Tran2009}, however, have found a clear increase of the star formation activity
around a super group at $z\sim0.4$.  Whether such groups are abundant at intermediate redshifts
is unclear. If such systems exists around RXJ1347, it might be possible that our optical-UV 
study are missing them due to the effects of dust attenuation. 
However, as it has been argued before, we should be able to detect the change of star formation modes 
(unobscured to obscured), if it occurs in relatively long time scales and it is caused by environmental 
effects.

\section{Summary and prospects}
\label{S:Conclusions}

We have presented a study on the large scale structures around the most luminous 
X-ray cluster RX\,J1347.5-1145. As expected from the large scale growth theory, we found that 
RXJ1347, one of the most massive known galaxy clusters, is embedded in a rich filamentary network, which
extends for at least 20\,Mpc.  We have identified a number of galaxy groups in the filaments, which are likely to be 
accreted by the main cluster at later times. Two other massive systems are associated to the large scale structure, making  RXJ1347 a good candidate for a supercluster. 

We calculated the fraction of blue galaxies as a function of environment characterised by the local 
galaxy density. Similar to other studies we confirm the low number of blue galaxies in the high densities regions.
This fraction drop from $\sim$85\% in the low density environment to $\sim$10\% at the highest densities in the cluster cores. This relation is also luminosity (mass) dependent. 

The relation of the galaxy population with environment is also studied by using 9\,ks GALEX NUV exposures, which trace 
the unobscured star-forming population. As a function of environment, its fraction 
follows closely the colour-density relation. The properties of the star-forming population
is also remarkably similar at all environments. We interpretate this as evidence of 
a rapid transformation of galaxy types.

This is further confirmed when we analyse different environments independently. We find that the cluster environment
carries most of the signal of the colour-density relation. Galaxy group candidates display, however, a large 
variation in their properties, signal that some transformations are also occurring in those environments, but this
might not be related directly to galaxy density.

Our results are compatible with a scenario of rapid suppresion of the star-formation activity in the vicinity
of clusters, likely due to specific processes. Preprocessing is also occurring in the large scale structure but 
its nature is apparently heterogeneous and probably depend on particular conditions. 

The recently completed  VIMOS campaign (30\,h) over the full 1\,deg$^2$ will allow us to confirm the trends 
studied here, via additional star formation indicators. 
We can identify optical AGNs and assess the properties of passive galaxies by using strong absorption lines. 
We would also be able to verify and investigate the properties of groups and clusters 
identified in this study (mostly) by photometric means.

\section*{Acknowledgments}

MV thanks the  anonymous referee for constructive comments that helped to clarify the 
focus of the paper. Encouraging discussions with the members of the Cluster group at MPE are
also acknowledged.

MV acknowledges support by the Universe Cluster of Excellence and the MPE. 
HH is supported by the Marie Curie International outgoing Fellowship 252760 and by a CITA National Fellowship.

This investigation is based on observations obtained with MegaPrime/MegaCam, a joint project of CFHT and CEA/DAPNIA, at the
Canada-France-Hawaii Telescope (CFHT) which is operated by the National Research Council (NRC) of Canada, the Institut
National des Science de l'Univers of the Centre National de la Recherche Scientifique (CNRS) of France, and the
University of Hawaii. 

We have made use of observations taken with ESO Telescopes at the Paranal Observatory under programme 
169.A-0595 and 381.A-0823.  

Based on observations made with the NASA Galaxy Evolution Explorer. GALEX is operated for 
NASA by the California Institute of Technology  under NASA contract NAS5-98034.

%%%%%%%%%%%%%%%%%%%%%%%%%%%%%%%%%%%%%%%%%%%%%%%%%%%%%%%%
%%%%%%%%%%%%%%%%%%%%%%%%%%%%%%%%%%%%%%%%%%%%%%%%%%%%%%%
%%%%%%%%%%%%%%%%%%%%%%%%%%%%%%%%%%%%%%%%%%%%%%%%%%%5

\bibliographystyle{mn2e} %

 \bibliography{mverdugo.mnras}

\begin{thebibliography}{}

\bibitem[\protect\citeauthoryear{{Andreon}, {Quintana}, {Tajer}, {Galaz} \&
  {Surdej}}{{Andreon} et~al.}{2006}]{Andreon2006}
{Andreon} S.,  {Quintana} H.,  {Tajer} M.,  {Galaz} G.,    {Surdej} J.,  2006,
  \mnras, 365, 915

\bibitem[\protect\citeauthoryear{{Arnouts}, {Schiminovich}, {Ilbert}, {Tresse},
  {Milliard}, {Treyer} \& et al.}{{Arnouts} et~al.}{2005}]{Arnouts2005}
{Arnouts} S.,  {Schiminovich} D.,  {Ilbert} O.,  {Tresse} L.,  {Milliard} B.,
  {Treyer} M.,    et al. 2005, \apjl, 619, L43

\bibitem[\protect\citeauthoryear{{Bai}, {Rieke}, {Rieke}, {Christlein} \&
  {Zabludoff}}{{Bai} et~al.}{2009}]{Bai2008}
{Bai} L.,  {Rieke} G.~H.,  {Rieke} M.~J.,  {Christlein} D.,    {Zabludoff}
  A.~I.,  2009, \apj, 693, 1840

\bibitem[\protect\citeauthoryear{{Baldry}, {Balogh}, {Bower}, {Glazebrook} \&
  et al.}{{Baldry} et~al.}{2006}]{Baldry2006}
{Baldry} I.~K.,  {Balogh} M.~L.,  {Bower} R.~G.,  {Glazebrook} K.,    et al.
  2006, \mnras, 373, 469

\bibitem[\protect\citeauthoryear{{Balogh}, {Eke}, {Miller}, {Lewis}, {Bower} \&
  et al.}{{Balogh} et~al.}{2004}]{Balogh2004}
{Balogh} M.,  {Eke} V.,  {Miller} C.,  {Lewis} I.,  {Bower} R.,    et al. 2004,
  \mnras, 348, 1355

\bibitem[\protect\citeauthoryear{{Balogh}, {Morris}, {Yee}, {Carlberg} \&
  {Ellingson}}{{Balogh} et~al.}{1999}]{balogh99}
{Balogh} M.~L.,  {Morris} S.~L.,  {Yee} H.~K.~C.,  {Carlberg} R.~G.,
  {Ellingson} E.,  1999, \apj, 527, 54

\bibitem[\protect\citeauthoryear{{Barkhouse}, {Green}, {Vikhlinin}, {Kim},
  {Perley} \& et al.}{{Barkhouse} et~al.}{2006}]{Barkhouse2006}
{Barkhouse} W.~A.,  {Green} P.~J.,  {Vikhlinin} A.,  {Kim} D.,  {Perley} D.,
  et al. 2006, \apj, 645, 955

\bibitem[\protect\citeauthoryear{{Bauer}, {Gr{\"u}tzbauch}, {J{\o}rgensen},
  {Varela} \& {Bergmann}}{{Bauer} et~al.}{2011}]{Bauer2011}
{Bauer} A.~E.,  {Gr{\"u}tzbauch} R.,  {J{\o}rgensen} I.,  {Varela} J.,
  {Bergmann} M.,  2011, \mnras, 411, 2009

\bibitem[\protect\citeauthoryear{{Baum}}{{Baum}}{1959}]{Baum1959}
{Baum} W.~A.,  1959, \pasp, 71, 106

\bibitem[\protect\citeauthoryear{{Beers}, {Flynn} \& {Gebhardt}}{{Beers}
  et~al.}{1990}]{Beers1990}
{Beers} T.~C.,  {Flynn} K.,    {Gebhardt} K.,  1990, \aj, 100, 32

\bibitem[\protect\citeauthoryear{{Bender}, {Appenzeller}, {B{\"o}hm} \& et
  al.}{{Bender} et~al.}{2001}]{Bender2001}
{Bender} R.,  {Appenzeller} I.,  {B{\"o}hm} A.,    et al. 2001, pp 96--+

\bibitem[\protect\citeauthoryear{{Ben{\'{\i}}tez}}{{Ben{\'{\i}}tez}}{2000}]{Be%
nitez2000}
{Ben{\'{\i}}tez} N.,  2000, \apj, 536, 571

\bibitem[\protect\citeauthoryear{{Bertin} \& {Arnouts}}{{Bertin} \&
  {Arnouts}}{1996}]{Bertin1996}
{Bertin} E.,  {Arnouts} S.,  1996, \aaps, 117, 393

\bibitem[\protect\citeauthoryear{{Bielby}, {Finoguenov}, {Tanaka}, {McCracken},
  {Daddi}, {Hudelot},  \& et al.}{{Bielby} et~al.}{2010}]{Bielby2010}
{Bielby} R.~M.,  {Finoguenov} A.,  {Tanaka} M.,  {McCracken} H.~J.,  {Daddi}
  E.,  {Hudelot} P.,     et al. 2010, \aap, 523, A66+

\bibitem[\protect\citeauthoryear{{Biviano}, {Fadda}, {Durret}, {Edwards} \&
  {Marleau}}{{Biviano} et~al.}{2011}]{Biviano2011}
{Biviano} A.,  {Fadda} D.,  {Durret} F.,  {Edwards} L.~O.~V.,    {Marleau} F.,
  2011, \aap, 532, A77

\bibitem[\protect\citeauthoryear{{B{\"o}hringer}, {Schuecker}, {Guzzo},
  {Collins}, {Voges} \& et al.}{{B{\"o}hringer} et~al.}{2004}]{Bohringer2004}
{B{\"o}hringer} H.,  {Schuecker} P.,  {Guzzo} L.,  {Collins} C.~A.,  {Voges}
  W.,    et al. 2004, \aap, 425, 367

\bibitem[\protect\citeauthoryear{{B{\"o}hringer}, {Schuecker}, {Guzzo},
  {Collins}, {Voges}, {Schindler} \& et al.}{{B{\"o}hringer}
  et~al.}{2001}]{Bohringer2001}
{B{\"o}hringer} H.,  {Schuecker} P.,  {Guzzo} L.,  {Collins} C.~A.,  {Voges}
  W.,  {Schindler} S.,    et al. 2001, \aap, 369, 826

\bibitem[\protect\citeauthoryear{{B{\"o}hringer}, {Voges}, {Huchra}, {McLean},
  {Giacconi}, {Rosati} \& et al.}{{B{\"o}hringer} et~al.}{2000}]{Bohringer2000}
{B{\"o}hringer} H.,  {Voges} W.,  {Huchra} J.~P.,  {McLean} B.,  {Giacconi} R.,
   {Rosati} P.,    et al. 2000, \apjs, 129, 435

\bibitem[\protect\citeauthoryear{{Bond}, {Kofman} \& {Pogosyan}}{{Bond}
  et~al.}{1996}]{Bond1996}
{Bond} J.~R.,  {Kofman} L.,    {Pogosyan} D.,  1996, \nat, 380, 603

\bibitem[\protect\citeauthoryear{{Book} \& {Benson}}{{Book} \&
  {Benson}}{2010}]{Book2010}
{Book} L.~G.,  {Benson} A.~J.,  2010, \apj, 716, 810

\bibitem[\protect\citeauthoryear{{Brada{\v c}}, {Erben}, {Schneider},
  {Hildebrandt}, {Lombardi}, {Schirmer} \& et al.}{{Brada{\v c}}
  et~al.}{2005}]{Bradac2005}
{Brada{\v c}} M.,  {Erben} T.,  {Schneider} P.,  {Hildebrandt} H.,  {Lombardi}
  M.,  {Schirmer} M.,    et al. 2005, \aap, 437, 49

\bibitem[\protect\citeauthoryear{{Brada{\v c}}, {Schrabback}, {Erben},
  {McCourt}, {Million}, {Mantz}, {Allen} \& et al.}{{Brada{\v c}}
  et~al.}{2008}]{Bradac2008b}
{Brada{\v c}} M.,  {Schrabback} T.,  {Erben} T.,  {McCourt} M.,  {Million} E.,
  {Mantz} A.,  {Allen}   et al. 2008, \apj, 681, 187

\bibitem[\protect\citeauthoryear{{Braglia}, {Pierini} \&
  {B{\"o}hringer}}{{Braglia} et~al.}{2007}]{Braglia2007}
{Braglia} F.,  {Pierini} D.,    {B{\"o}hringer} H.,  2007, \aap, 470, 425

\bibitem[\protect\citeauthoryear{{Braglia}, {Pierini}, {Biviano} \&
  {B{\"o}hringer}}{{Braglia} et~al.}{2009}]{Braglia2009}
{Braglia} F.~G.,  {Pierini} D.,  {Biviano} A.,    {B{\"o}hringer} H.,  2009,
  \aap, 500, 947

\bibitem[\protect\citeauthoryear{{Brimioulle}, {Lerchster}, {Seitz}, {Bender}
  \& {Snigula}}{{Brimioulle} et~al.}{2008}]{Brimioulle2008}
{Brimioulle} F.,  {Lerchster} M.,  {Seitz} S.,  {Bender} R.,    {Snigula} J.,
  2008, ArXiv e-prints

\bibitem[\protect\citeauthoryear{{Butcher} \& {Oemler} Jr.}{{Butcher} \&
  {Oemler}}{1978}]{BO78}
{Butcher} H.,  {Oemler} Jr. A.,  1978, \apj, 226, 559

\bibitem[\protect\citeauthoryear{{Chilingarian}, {Melchior} \&
  {Zolotukhin}}{{Chilingarian} et~al.}{2010}]{Chilingarian2010}
{Chilingarian} I.~V.,  {Melchior} A.-L.,    {Zolotukhin} I.~Y.,  2010, \mnras,
  405, 1409

\bibitem[\protect\citeauthoryear{{Christlein} \& {Zabludoff}}{{Christlein} \&
  {Zabludoff}}{2005}]{Christlein2005}
{Christlein} D.,  {Zabludoff} A.~I.,  2005, \apj, 621, 201

\bibitem[\protect\citeauthoryear{{Cohen} \& {Kneib}}{{Cohen} \&
  {Kneib}}{2002}]{Cohen2002}
{Cohen} J.~G.,  {Kneib} J.-P.,  2002, \apj, 573, 524

\bibitem[\protect\citeauthoryear{{Coupon}, {Ilbert}, {Kilbinger}, {McCracken},
  {Mellier}, {Arnouts} \& et al.}{{Coupon} et~al.}{2009}]{Coupon2009}
{Coupon} J.,  {Ilbert} O.,  {Kilbinger} M.,  {McCracken} H.~J.,  {Mellier} Y.,
  {Arnouts} S.,    et al. 2009, \aap, 500, 981

\bibitem[\protect\citeauthoryear{{Dressler}}{{Dressler}}{1980}]{Dressler1980}
{Dressler} A.,  1980, \apj, 236, 351

\bibitem[\protect\citeauthoryear{{Dressler}, {Thompson} \&
  {Shectman}}{{Dressler} et~al.}{1985}]{Dressler1985}
{Dressler} A.,  {Thompson} I.~B.,    {Shectman} S.~A.,  1985, \apj, 288, 481

\bibitem[\protect\citeauthoryear{{Elbaz}, {Daddi}, {Le Borgne}, {Dickinson},
  {Alexander}, {Chary} \& et al.}{{Elbaz} et~al.}{2007}]{Elbaz2007}
{Elbaz} D.,  {Daddi} E.,  {Le Borgne} D.,  {Dickinson} M.,  {Alexander} D.~M.,
  {Chary} R.-R.,    et al. 2007, \aap, 468, 33

\bibitem[\protect\citeauthoryear{{Ellingson}, {Lin}, {Yee} \&
  {Carlberg}}{{Ellingson} et~al.}{2001}]{Ellingson2001}
{Ellingson} E.,  {Lin} H.,  {Yee} H.~K.~C.,    {Carlberg} R.~G.,  2001, \apj,
  547, 609

\bibitem[\protect\citeauthoryear{{Ellison}, {Patton}, {Simard} \&
  {McConnachie}}{{Ellison} et~al.}{2008}]{Ellison2008a}
{Ellison} S.~L.,  {Patton} D.~R.,  {Simard} L.,    {McConnachie} A.~W.,  2008,
  \aj, 135, 1877

\bibitem[\protect\citeauthoryear{{Erben}, {Hildebrandt}, {Lerchster},
  {Hudelot}, {Benjamin} \& et al.}{{Erben} et~al.}{2009}]{Erben2009}
{Erben} T.,  {Hildebrandt} H.,  {Lerchster} M.,  {Hudelot} P.,  {Benjamin} J.,
    et al. 2009, \aap, 493, 1197

\bibitem[\protect\citeauthoryear{{Ettori}, {Allen} \& {Fabian}}{{Ettori}
  et~al.}{2001}]{Ettori2001}
{Ettori} S.,  {Allen} S.~W.,    {Fabian} A.~C.,  2001, \mnras, 322, 187

\bibitem[\protect\citeauthoryear{{Fadda}, {Biviano}, {Marleau},
  {Storrie-Lombardi} \& {Durret}}{{Fadda} et~al.}{2008}]{Fadda2008}
{Fadda} D.,  {Biviano} A.,  {Marleau} F.~R.,  {Storrie-Lombardi} L.~J.,
  {Durret} F.,  2008, \apjl, 672, L9

\bibitem[\protect\citeauthoryear{{Fassbender}, {B{\"o}hringer}, {Lamer},
  {Mullis}, {Rosati} \& et al.}{{Fassbender} et~al.}{2008}]{Fassbender2008}
{Fassbender} R.,  {B{\"o}hringer} H.,  {Lamer} G.,  {Mullis} C.~R.,  {Rosati}
  P.,    et al. 2008, \aap, 481, L73

\bibitem[\protect\citeauthoryear{{Finn}, {Desai}, {Rudnick}, {Poggianti},
  {Bell}, {Hinz} \& et al.}{{Finn} et~al.}{2010}]{Finn2010}
{Finn} R.~A.,  {Desai} V.,  {Rudnick} G.,  {Poggianti} B.,  {Bell} E.~F.,
  {Hinz} J.,    et al. 2010, \apj, 720, 87

\bibitem[\protect\citeauthoryear{{Finoguenov}, {Watson}, {Tanaka}, {Simpson} \&
  et al.}{{Finoguenov} et~al.}{2010}]{Finoguenov2010}
{Finoguenov} A.,  {Watson} M.~G.,  {Tanaka} M.,  {Simpson} C.,    et al. 2010,
  \mnras, 403, 2063

\bibitem[\protect\citeauthoryear{{Fujita}}{{Fujita}}{2004}]{Fujita2004}
{Fujita} Y.,  2004, \pasj, 56, 29

\bibitem[\protect\citeauthoryear{{Gallazzi}, {Bell}, {Wolf}, {Gray},
  {Papovich}, {Barden}, {Peng} \& et al.}{{Gallazzi}
  et~al.}{2009}]{Gallazzi2009}
{Gallazzi} A.,  {Bell} E.~F.,  {Wolf} C.,  {Gray} M.~E.,  {Papovich} C.,
  {Barden} M.,  {Peng} C.~Y.,    et al. 2009, \apj, 690, 1883

\bibitem[\protect\citeauthoryear{{Garilli}, {Fumana}, {Franzetti}, {Paioro},
  {Scodeggio}, {Le F{\`e}vre} \& et al.}{{Garilli} et~al.}{2010}]{Garilli2010}
{Garilli} B.,  {Fumana} M.,  {Franzetti} P.,  {Paioro} L.,  {Scodeggio} M.,
  {Le F{\`e}vre} O.,    et al. 2010, \pasp, 122, 827

\bibitem[\protect\citeauthoryear{{Gavazzi}, {Fumagalli}, {Cucciati} \&
  {Boselli}}{{Gavazzi} et~al.}{2010}]{Gavazzi2010}
{Gavazzi} G.,  {Fumagalli} M.,  {Cucciati} O.,    {Boselli} A.,  2010, \aap,
  517, A73+

\bibitem[\protect\citeauthoryear{{Geach}, {Ellis}, {Smail}, {Rawle} \&
  {Moran}}{{Geach} et~al.}{2011}]{Geach2011}
{Geach} J.~E.,  {Ellis} R.~S.,  {Smail} I.,  {Rawle} T.~D.,    {Moran} S.~M.,
  2011, \mnras, 413, 177

\bibitem[\protect\citeauthoryear{{Geach}, {Murphy} \& {Bower}}{{Geach}
  et~al.}{2011}]{Geach2011a}
{Geach} J.~E.,  {Murphy} D.~N.~A.,    {Bower} R.~G.,  2011, \mnras, 413, 3059

\bibitem[\protect\citeauthoryear{{Geach}, {Smail}, {Moran}, {Treu} \&
  {Ellis}}{{Geach} et~al.}{2009}]{Geach2009}
{Geach} J.~E.,  {Smail} I.,  {Moran} S.~M.,  {Treu} T.,    {Ellis} R.~S.,
  2009, \apj, 691, 783

\bibitem[\protect\citeauthoryear{{Girardi}, {Giuricin}, {Mardirossian},
  {Mezzetti} \& {Boschin}}{{Girardi} et~al.}{1998}]{Girardi1998}
{Girardi} M.,  {Giuricin} G.,  {Mardirossian} F.,  {Mezzetti} M.,    {Boschin}
  W.,  1998, \apj, 505, 74

\bibitem[\protect\citeauthoryear{{Girardi}, {Manzato}, {Mezzetti}, {Giuricin}
  \& {Limboz}}{{Girardi} et~al.}{2002}]{Girardi2002}
{Girardi} M.,  {Manzato} P.,  {Mezzetti} M.,  {Giuricin} G.,    {Limboz} F.,
  2002, \apj, 569, 720

\bibitem[\protect\citeauthoryear{{Gitti}, {Ferrari}, {Domainko}, {Feretti} \&
  {Schindler}}{{Gitti} et~al.}{2007}]{Gitti2007a}
{Gitti} M.,  {Ferrari} C.,  {Domainko} W.,  {Feretti} L.,    {Schindler} S.,
  2007, \aap, 470, L25

\bibitem[\protect\citeauthoryear{{Gitti}, {Piffaretti} \& {Schindler}}{{Gitti}
  et~al.}{2007}]{Gitti2007}
{Gitti} M.,  {Piffaretti} R.,    {Schindler} S.,  2007, \aap, 472, 383

\bibitem[\protect\citeauthoryear{{Gitti} \& {Schindler}}{{Gitti} \&
  {Schindler}}{2004}]{Gitti2004}
{Gitti} M.,  {Schindler} S.,  2004, \aap, 427, L9

\bibitem[\protect\citeauthoryear{{Gladders}, {Lopez-Cruz}, {Yee} \&
  {Kodama}}{{Gladders} et~al.}{1998}]{Gladders1998}
{Gladders} M.~D.,  {Lopez-Cruz} O.,  {Yee} H.~K.~C.,    {Kodama} T.,  1998,
  \apj, 501, 571

\bibitem[\protect\citeauthoryear{{Gonzalez}, {Zaritsky}, {Dalcanton} \&
  {Nelson}}{{Gonzalez} et~al.}{2001}]{Gonzalez2001}
{Gonzalez} A.~H.,  {Zaritsky} D.,  {Dalcanton} J.~J.,    {Nelson} A.,  2001,
  \apjs, 137, 117

\bibitem[\protect\citeauthoryear{{Haines}, {Busarello}, {Merluzzi}, {Smith},
  {Raychaudhury} \& et al.}{{Haines} et~al.}{2011a}]{Haines2011a}
{Haines} C.~P.,  {Busarello} G.,  {Merluzzi} P.,  {Smith} R.~J.,
  {Raychaudhury} S.,    et al. 2011a, \mnras, 412, 127

\bibitem[\protect\citeauthoryear{{Haines}, {Busarello}, {Merluzzi}, {Smith},
  {Raychaudhury} \& et al.}{{Haines} et~al.}{2011b}]{Haines2011b}
{Haines} C.~P.,  {Busarello} G.,  {Merluzzi} P.,  {Smith} R.~J.,
  {Raychaudhury} S.,    et al. 2011b, \mnras, 412, 145

\bibitem[\protect\citeauthoryear{{Haines}, {La Barbera}, {Mercurio}, {Merluzzi}
  \& {Busarello}}{{Haines} et~al.}{2006}]{Haines2006}
{Haines} C.~P.,  {La Barbera} F.,  {Mercurio} A.,  {Merluzzi} P.,
  {Busarello} G.,  2006, \apjl, 647, L21

\bibitem[\protect\citeauthoryear{{Haines}, {Merluzzi}, {Mercurio}, {Gargiulo},
  {Krusanova}, {Busarello} \& et al.}{{Haines} et~al.}{2006}]{Haines2006a}
{Haines} C.~P.,  {Merluzzi} P.,  {Mercurio} A.,  {Gargiulo} A.,  {Krusanova}
  N.,  {Busarello} G.,    et al. 2006, \mnras, 371, 55

\bibitem[\protect\citeauthoryear{{Haines}, {Smith}, {Egami}, {Ellis}, {Moran},
  {Sanderson} \& et al.}{{Haines} et~al.}{2009}]{Haines2009a}
{Haines} C.~P.,  {Smith} G.~P.,  {Egami} E.,  {Ellis} R.~S.,  {Moran} S.~M.,
  {Sanderson} A.~J.~R.,    et al. 2009, \apj, 704, 126

\bibitem[\protect\citeauthoryear{{Haines}, {Smith}, {Egami}, {Okabe}, {Takada},
  {Ellis} \& et al.}{{Haines} et~al.}{2009}]{Haines2009}
{Haines} C.~P.,  {Smith} G.~P.,  {Egami} E.,  {Okabe} N.,  {Takada} M.,
  {Ellis} R.~S.,    et al. 2009, \mnras, 396, 1297

\bibitem[\protect\citeauthoryear{{Halkola}, {Hildebrandt}, {Schrabback},
  {Lombardi}, {Brada{\v c}}, {Erben} \& et al.}{{Halkola}
  et~al.}{2008}]{Halkola2008}
{Halkola} A.,  {Hildebrandt} H.,  {Schrabback} T.,  {Lombardi} M.,  {Brada{\v
  c}} M.,  {Erben} T.,    et al. 2008, \aap, 481, 65

\bibitem[\protect\citeauthoryear{{Hansen}, {Sheldon}, {Wechsler} \&
  {Koester}}{{Hansen} et~al.}{2009}]{Hansen2009}
{Hansen} S.~M.,  {Sheldon} E.~S.,  {Wechsler} R.~H.,    {Koester} B.~P.,  2009,
  \apj, 699, 1333

\bibitem[\protect\citeauthoryear{{Hicks}, {Mushotzky} \& {Donahue}}{{Hicks}
  et~al.}{2010}]{Hicks2010}
{Hicks} A.~K.,  {Mushotzky} R.,    {Donahue} M.,  2010, \apj, 719, 1844

\bibitem[\protect\citeauthoryear{{Hildebrandt}, {Pielorz}, {Erben}, {van
  Waerbeke}, {Simon} \& {Capak}}{{Hildebrandt} et~al.}{2009}]{Hildebrandt2009}
{Hildebrandt} H.,  {Pielorz} J.,  {Erben} T.,  {van Waerbeke} L.,  {Simon} P.,
    {Capak} P.,  2009, \aap, 498, 725

\bibitem[\protect\citeauthoryear{{Ilbert}, {Arnouts}, {McCracken},
  {Bolzonella}, {Bertin}, {Le F{\`e}vre} \& et al.}{{Ilbert}
  et~al.}{2006}]{Ilbert2006}
{Ilbert} O.,  {Arnouts} S.,  {McCracken} H.~J.,  {Bolzonella} M.,  {Bertin} E.,
   {Le F{\`e}vre} O.,    et al. 2006, \aap, 457, 841

\bibitem[\protect\citeauthoryear{{Kiang}}{{Kiang}}{1966}]{Kiang1966}
{Kiang} T.,  1966, \zap, 64, 433

\bibitem[\protect\citeauthoryear{{Kim}, {Kepner}, {Postman}, {Strauss},
  {Bahcall}, {Gunn} \& et al.}{{Kim} et~al.}{2002}]{Kim2002}
{Kim} R.~S.~J.,  {Kepner} J.~V.,  {Postman} M.,  {Strauss} M.~A.,  {Bahcall}
  N.~A.,  {Gunn} J.~E.,    et al. 2002, \aj, 123, 20

\bibitem[\protect\citeauthoryear{{Kitayama}, {Komatsu}, {Ota}, {Kuwabara},
  {Suto}, {Yoshikawa}, {Hattori} \& {Matsuo}}{{Kitayama}
  et~al.}{2004}]{Kitayama2004}
{Kitayama} T.,  {Komatsu} E.,  {Ota} N.,  {Kuwabara} T.,  {Suto} Y.,
  {Yoshikawa} K.,  {Hattori} M.,    {Matsuo} H.,  2004, \pasj, 56, 17

\bibitem[\protect\citeauthoryear{{Kodama}, {Smail}, {Nakata}, {Okamura} \&
  {Bower}}{{Kodama} et~al.}{2001}]{Kodama2001}
{Kodama} T.,  {Smail} I.,  {Nakata} F.,  {Okamura} S.,    {Bower} R.~G.,  2001,
  \apjl, 562, L9

\bibitem[\protect\citeauthoryear{{Komatsu}, {Matsuo}, {Kitayama}, {Hattori},
  {Kawabe} \& et al.}{{Komatsu} et~al.}{2001}]{Komatsu2001}
{Komatsu} E.,  {Matsuo} H.,  {Kitayama} T.,  {Hattori} M.,  {Kawabe} R.,    et
  al. 2001, \pasj, 53, 57

\bibitem[\protect\citeauthoryear{{Kong}, {Charlot}, {Brinchmann} \&
  {Fall}}{{Kong} et~al.}{2004}]{Kong2004}
{Kong} X.,  {Charlot} S.,  {Brinchmann} J.,    {Fall} S.~M.,  2004, \mnras,
  349, 769

\bibitem[\protect\citeauthoryear{{Koyama}, {Kodama}, {Shimasaku}, {Okamura},
  {Tanaka}, {Lee} \& et al.}{{Koyama} et~al.}{2008}]{Koyama2008}
{Koyama} Y.,  {Kodama} T.,  {Shimasaku} K.,  {Okamura} S.,  {Tanaka} M.,  {Lee}
  H.~M.,    et al. 2008, \mnras, 391, 1758

\bibitem[\protect\citeauthoryear{{Kutdemir}, {Ziegler}, {Peletier}, {Da Rocha},
  {B{\"o}hm} \& {Verdugo}}{{Kutdemir} et~al.}{2010}]{Kutdemir2010}
{Kutdemir} E.,  {Ziegler} B.~L.,  {Peletier} R.~F.,  {Da Rocha} C.,  {B{\"o}hm}
  A.,    {Verdugo} M.,  2010, \aap, 520, A109+

\bibitem[\protect\citeauthoryear{{Leauthaud}, {Finoguenov}, {Kneib}, {Taylor},
  {Massey}, {Rhodes},  \& et al.}{{Leauthaud} et~al.}{2010}]{Leauthaud2010}
{Leauthaud} A.,  {Finoguenov} A.,  {Kneib} J.,  {Taylor} J.~E.,  {Massey} R.,
  {Rhodes} J.,     et al. 2010, \apj, 709, 97

\bibitem[\protect\citeauthoryear{{Lewis}, {Balogh}, {De Propris}, {Couch},
  {Bower}, {Offer} \& et al.}{{Lewis} et~al.}{2002}]{lewis02}
{Lewis} I.,  {Balogh} M.,  {De Propris} R.,  {Couch} W.,  {Bower} R.,  {Offer}
  A.,    et al. 2002, \mnras, 334, 673

\bibitem[\protect\citeauthoryear{{Li}, {Yee} \& {Ellingson}}{{Li}
  et~al.}{2009}]{Li2009}
{Li} I.~H.,  {Yee} H.~K.~C.,    {Ellingson} E.,  2009, \apj, 698, 83

\bibitem[\protect\citeauthoryear{{Loh}, {Ellingson}, {Yee}, {Gilbank},
  {Gladders} \& {Barrientos}}{{Loh} et~al.}{2008}]{Loh2008}
{Loh} Y.-S.,  {Ellingson} E.,  {Yee} H.~K.~C.,  {Gilbank} D.~G.,  {Gladders}
  M.~D.,    {Barrientos} L.~F.,  2008, \apj, 680, 214

\bibitem[\protect\citeauthoryear{{Lu}, {Gilbank}, {Balogh}, {Milkeraitis},
  {Hoekstra}, {van Waerbeke} \& et al.}{{Lu} et~al.}{2010}]{Lu2010}
{Lu} T.,  {Gilbank} D.~G.,  {Balogh} M.~L.,  {Milkeraitis} M.,  {Hoekstra} H.,
  {van Waerbeke} L.,    et al. 2010, \mnras, 403, 1787

\bibitem[\protect\citeauthoryear{{Mahajan}, {Haines} \&
  {Raychaudhury}}{{Mahajan} et~al.}{2010}]{Mahajan2010}
{Mahajan} S.,  {Haines} C.~P.,    {Raychaudhury} S.,  2010, \mnras, 404, 1745

\bibitem[\protect\citeauthoryear{{Marcillac}, {Rigby}, {Rieke} \&
  {Kelly}}{{Marcillac} et~al.}{2007}]{Marcillac2007}
{Marcillac} D.,  {Rigby} J.~R.,  {Rieke} G.~H.,    {Kelly} D.~M.,  2007, \apj,
  654, 825

\bibitem[\protect\citeauthoryear{{Marinoni}, {Davis}, {Newman} \&
  {Coil}}{{Marinoni} et~al.}{2002}]{Marinoni2002}
{Marinoni} C.,  {Davis} M.,  {Newman} J.~A.,    {Coil} A.~L.,  2002, \apj, 580,
  122

\bibitem[\protect\citeauthoryear{{Mason}, {Dicker}, {Korngut} \& et
  al.}{{Mason} et~al.}{2010}]{Mason2009}
{Mason} B.~S.,  {Dicker} S.~R.,  {Korngut} P.~M.,    et al. 2010, \apj, 716,
  739

\bibitem[\protect\citeauthoryear{{Medezinski}, {Broadhurst}, {Umetsu}, {Oguri},
  {Rephaeli} \& {Ben{\'{\i}}tez}}{{Medezinski} et~al.}{2010}]{Medezinski2010}
{Medezinski} E.,  {Broadhurst} T.,  {Umetsu} K.,  {Oguri} M.,  {Rephaeli} Y.,
   {Ben{\'{\i}}tez} N.,  2010, \mnras, 405, 257

\bibitem[\protect\citeauthoryear{{Mercurio}, {Merluzzi}, {Haines}, {Gargiulo},
  {Krusanova}, {Busarello} \& et al.}{{Mercurio} et~al.}{2006}]{Mercurio2006}
{Mercurio} A.,  {Merluzzi} P.,  {Haines} C.~P.,  {Gargiulo} A.,  {Krusanova}
  N.,  {Busarello} G.,    et al. 2006, \mnras, 368, 109

\bibitem[\protect\citeauthoryear{{Moran}, {Ellis}, {Treu}, {Smith}, {Rich} \&
  {Smail}}{{Moran} et~al.}{2007}]{Moran2007a}
{Moran} S.~M.,  {Ellis} R.~S.,  {Treu} T.,  {Smith} G.~P.,  {Rich} R.~M.,
  {Smail} I.,  2007, \apj, 671, 1503

\bibitem[\protect\citeauthoryear{{Morrissey}, {Conrow}, {Barlow}, {Small} \& et
  al.}{{Morrissey} et~al.}{2007}]{Morrissey2007}
{Morrissey} P.,  {Conrow} T.,  {Barlow} T.~A.,  {Small} T.,    et al. 2007,
  \apjs, 173, 682

\bibitem[\protect\citeauthoryear{{Ota}, {Murase}, {Kitayama}, {Komatsu},
  {Hattori} \& et al.}{{Ota} et~al.}{2008}]{Ota2008}
{Ota} N.,  {Murase} K.,  {Kitayama} T.,  {Komatsu} E.,  {Hattori} M.,    et al.
  2008, \aap, 491, 363

\bibitem[\protect\citeauthoryear{{Pell{\'o}}, {Rudnick}, {De Lucia}, {Simard},
  {Clowe}, {Jablonka} \& et al.}{{Pell{\'o}} et~al.}{2009}]{Pello2009}
{Pell{\'o}} R.,  {Rudnick} G.,  {De Lucia} G.,  {Simard} L.,  {Clowe} D.~I.,
  {Jablonka} P.,    et al. 2009, \aap, 508, 1173

\bibitem[\protect\citeauthoryear{{Pickles}}{{Pickles}}{1998}]{Pickles1998}
{Pickles} A.~J.,  1998, \pasp, 110, 863

\bibitem[\protect\citeauthoryear{{Pimbblet}, {Smail}, {Kodama}, {Couch},
  {Edge}, {Zabludoff} \& {O'Hely}}{{Pimbblet} et~al.}{2002}]{Pimbblet2002}
{Pimbblet} K.~A.,  {Smail} I.,  {Kodama} T.,  {Couch} W.~J.,  {Edge} A.~C.,
  {Zabludoff} A.~I.,    {O'Hely} E.,  2002, \mnras, 331, 333

\bibitem[\protect\citeauthoryear{{Poggianti}, {Desai}, {Finn}, {Bamford}, {De
  Lucia} \& et al.}{{Poggianti} et~al.}{2008}]{Poggianti2008}
{Poggianti} B.~M.,  {Desai} V.,  {Finn} R.,  {Bamford} S.,  {De Lucia} G.,
  et al. 2008, \apj, 684, 888

\bibitem[\protect\citeauthoryear{{Popesso}, {Biviano}, {Romaniello} \&
  {B{\"o}hringer}}{{Popesso} et~al.}{2007}]{Popesso2007}
{Popesso} P.,  {Biviano} A.,  {Romaniello} M.,    {B{\"o}hringer} H.,  2007,
  \aap, 461, 411

\bibitem[\protect\citeauthoryear{{Porter} \& {Raychaudhury}}{{Porter} \&
  {Raychaudhury}}{2007}]{Porter2007}
{Porter} S.~C.,  {Raychaudhury} S.,  2007, \mnras, 375, 1409

\bibitem[\protect\citeauthoryear{{Porter}, {Raychaudhury}, {Pimbblet} \&
  {Drinkwater}}{{Porter} et~al.}{2008}]{Porter2008}
{Porter} S.~C.,  {Raychaudhury} S.,  {Pimbblet} K.~A.,    {Drinkwater} M.~J.,
  2008, \mnras, 388, 1152

\bibitem[\protect\citeauthoryear{{Ramella}, {Boschin}, {Fadda} \&
  {Nonino}}{{Ramella} et~al.}{2001}]{Ramella2001}
{Ramella} M.,  {Boschin} W.,  {Fadda} D.,    {Nonino} M.,  2001, \aap, 368, 776

\bibitem[\protect\citeauthoryear{{Reyes}, {Mandelbaum}, {Hirata}, {Bahcall} \&
  {Seljak}}{{Reyes} et~al.}{2008}]{Reyes2008}
{Reyes} R.,  {Mandelbaum} R.,  {Hirata} C.,  {Bahcall} N.,    {Seljak} U.,
  2008, \mnras, 390, 1157

\bibitem[\protect\citeauthoryear{{Rines}, {Geller}, {Kurtz} \&
  {Diaferio}}{{Rines} et~al.}{2005}]{Rines2005}
{Rines} K.,  {Geller} M.~J.,  {Kurtz} M.~J.,    {Diaferio} A.,  2005, \aj, 130,
  1482

\bibitem[\protect\citeauthoryear{{Rudnick}, {von der Linden}, {Pell{\'o}},
  {Arag{\'o}n-Salamanca} \& et al.}{{Rudnick} et~al.}{2009}]{Rudnick2009}
{Rudnick} G.,  {von der Linden} A.,  {Pell{\'o}} R.,  {Arag{\'o}n-Salamanca}
  A.~.,    et al. 2009, \apj, 700, 1559

\bibitem[\protect\citeauthoryear{{Saintonge}, {Tran} \& {Holden}}{{Saintonge}
  et~al.}{2008}]{Saintonge2008}
{Saintonge} A.,  {Tran} K.,    {Holden} B.~P.,  2008, \apjl, 685, L113

\bibitem[\protect\citeauthoryear{{Schechter}}{{Schechter}}{1976}]{Schechter197%
6}
{Schechter} P.,  1976, \apj, 203, 297

\bibitem[\protect\citeauthoryear{{Schiminovich}, {Ilbert}, {Arnouts},
  {Milliard}, {Tresse}, {Le F{\`e}vre} \& et al.}{{Schiminovich}
  et~al.}{2005}]{Schiminovich2005}
{Schiminovich} D.,  {Ilbert} O.,  {Arnouts} S.,  {Milliard} B.,  {Tresse} L.,
  {Le F{\`e}vre} O.,    et al. 2005, \apjl, 619, L47

\bibitem[\protect\citeauthoryear{{Schindler}, {Guzzo}, {Ebeling}, {Boehringer},
  {Chincarini}, {Collins} \& et al.}{{Schindler} et~al.}{1995}]{Schindler1995}
{Schindler} S.,  {Guzzo} L.,  {Ebeling} H.,  {Boehringer} H.,  {Chincarini} G.,
   {Collins} C.~A.,    et al. 1995, \aap, 299, L9+

\bibitem[\protect\citeauthoryear{{Schindler}, {Hattori}, {Neumann} \&
  {Boehringer}}{{Schindler} et~al.}{1997}]{Schindler1997}
{Schindler} S.,  {Hattori} M.,  {Neumann} D.~M.,    {Boehringer} H.,  1997,
  \aap, 317, 646

\bibitem[\protect\citeauthoryear{{Schirmer}, {Hildebrandt}, {Kuijken} \&
  {Erben}}{{Schirmer} et~al.}{2011}]{Schirmer2011}
{Schirmer} M.,  {Hildebrandt} H.,  {Kuijken} K.,    {Erben} T.,  2011, \aap,
  532, A57+

\bibitem[\protect\citeauthoryear{{Schlegel}, {Finkbeiner} \&
  {Davis}}{{Schlegel} et~al.}{1998}]{schlegel98}
{Schlegel} D.~J.,  {Finkbeiner} D.~P.,    {Davis} M.,  1998, \apj, 500, 525

\bibitem[\protect\citeauthoryear{{Scodeggio}, {Franzetti}, {Garilli},
  {Zanichelli}, {Paltani} \& et al.}{{Scodeggio} et~al.}{2005}]{Scodeggio2005}
{Scodeggio} M.,  {Franzetti} P.,  {Garilli} B.,  {Zanichelli} A.,  {Paltani}
  S.,    et al. 2005, \pasp, 117, 1284

\bibitem[\protect\citeauthoryear{{Suhhonenko}, {Einasto}, {Liivam{\"a}gi},
  {Saar}, {Einasto} \& et al.}{{Suhhonenko} et~al.}{2011}]{Suhhonenko2011}
{Suhhonenko} I.,  {Einasto} J.,  {Liivam{\"a}gi} L.~J.,  {Saar} E.,  {Einasto}
  M.,    et al. 2011, \aap, 531, A149+

\bibitem[\protect\citeauthoryear{{Tanaka}, {Finoguenov}, {Kodama}, {Koyama},
  {Maughan} \& {Nakata}}{{Tanaka} et~al.}{2009}]{Tanaka2009a}
{Tanaka} M.,  {Finoguenov} A.,  {Kodama} T.,  {Koyama} Y.,  {Maughan} B.,
  {Nakata} F.,  2009, \aap, 505, L9

\bibitem[\protect\citeauthoryear{{Tanaka}, {Hoshi}, {Kodama} \&
  {Kashikawa}}{{Tanaka} et~al.}{2007}]{Tanaka2007}
{Tanaka} M.,  {Hoshi} T.,  {Kodama} T.,    {Kashikawa} N.,  2007, \mnras, 379,
  1546

\bibitem[\protect\citeauthoryear{{Tanaka}, {Kodama}, {Arimoto}, {Okamura},
  {Umetsu} \& et al.}{{Tanaka} et~al.}{2005}]{Tanaka2005}
{Tanaka} M.,  {Kodama} T.,  {Arimoto} N.,  {Okamura} S.,  {Umetsu} K.,    et
  al. 2005, \mnras, 362, 268

\bibitem[\protect\citeauthoryear{{Tecce}, {Cora} \& {Tissera}}{{Tecce}
  et~al.}{2011}]{Tecce2011}
{Tecce} T.~E.,  {Cora} S.~A.,    {Tissera} P.~B.,  2011, \mnras, 416, 3170

\bibitem[\protect\citeauthoryear{{Tran}, {Papovich}, {Saintonge} \& et
  al.}{{Tran} et~al.}{2010}]{Tran2010}
{Tran} K.,  {Papovich} C.,  {Saintonge} A.,    et al. 2010, \apjl, 719, L126

\bibitem[\protect\citeauthoryear{{Tran}, {Saintonge}, {Moustakas}, {Bai},
  {Gonzalez} \& et al.}{{Tran} et~al.}{2009}]{Tran2009}
{Tran} K.,  {Saintonge} A.,  {Moustakas} J.,  {Bai} L.,  {Gonzalez} A.~H.,
  et al. 2009, \apj, 705, 809

\bibitem[\protect\citeauthoryear{{Urquhart}, {Willis}, {Hoekstra} \&
  {Pierre}}{{Urquhart} et~al.}{2010}]{Urquhart2010}
{Urquhart} S.~A.,  {Willis} J.~P.,  {Hoekstra} H.,    {Pierre} M.,  2010,
  \mnras, 406, 368

\bibitem[\protect\citeauthoryear{{Verdugo}, {Ziegler} \& {Gerken}}{{Verdugo}
  et~al.}{2008}]{Verdugo2008}
{Verdugo} M.,  {Ziegler} B.~L.,    {Gerken} B.,  2008, \aap, 486, 9

\bibitem[\protect\citeauthoryear{{von der Linden}, {Wild}, {Kauffmann}, {White}
  \& {Weinmann}}{{von der Linden} et~al.}{2010}]{vonderLinden2009}
{von der Linden} A.,  {Wild} V.,  {Kauffmann} G.,  {White} S.~D.~M.,
  {Weinmann} S.,  2010, \mnras, 404, 1231

\bibitem[\protect\citeauthoryear{{Wolf}, {Arag{\'o}n-Salamanca}, {Balogh},
  {Barden}, {Bell}, {Gray} \& et al.}{{Wolf} et~al.}{2009}]{Wolf2009}
{Wolf} C.,  {Arag{\'o}n-Salamanca} A.,  {Balogh} M.,  {Barden} M.,  {Bell}
  E.~F.,  {Gray} M.~E.~.,    et al. 2009, \mnras, 393, 1302

\bibitem[\protect\citeauthoryear{{Wong}, {Blanton}, {Burles}, {Coil}, {Cool} \&
  et al.}{{Wong} et~al.}{2011}]{Wong2011}
{Wong} K.~C.,  {Blanton} M.~R.,  {Burles} S.~M.,  {Coil} A.~L.,  {Cool} R.~J.,
    et al. 2011, \apj, 728, 119

\bibitem[\protect\citeauthoryear{{Yee}, {Ellingson} \& {Carlberg}}{{Yee}
  et~al.}{1996}]{Yee1996}
{Yee} H.~K.~C.,  {Ellingson} E.,    {Carlberg} R.~G.,  1996, \apjs, 102, 269

\bibitem[\protect\citeauthoryear{{Zamojski}, {Schiminovich}, {Rich},
  {Mobasher}, {Koekemoer} \& et al.}{{Zamojski} et~al.}{2007}]{Zamojski2007}
{Zamojski} M.~A.,  {Schiminovich} D.,  {Rich} R.~M.,  {Mobasher} B.,
  {Koekemoer} A.~M.,    et al. 2007, \apjs, 172, 468

\bibitem[\protect\citeauthoryear{{Zenteno}, {Song}, {Desai}, {Armstrong},
  {Mohr}, {Ngeow}, {Barkhouse} \& et al.}{{Zenteno} et~al.}{2011}]{Zenteno2011}
{Zenteno} A.,  {Song} J.,  {Desai} S.,  {Armstrong} R.,  {Mohr} J.~J.,  {Ngeow}
  C.-C.,  {Barkhouse} W.~A.,    et al. 2011, \apj, 734, 3

\bibitem[\protect\citeauthoryear{{Zhang}, {B{\"o}hringer}, {Finoguenov},
  {Ikebe}, {Matsushita} \& et al.}{{Zhang} et~al.}{2006}]{Zhang2006}
{Zhang} Y.,  {B{\"o}hringer} H.,  {Finoguenov} A.,  {Ikebe} Y.,  {Matsushita}
  K.,    et al. 2006, \aap, 456, 55

\bibitem[\protect\citeauthoryear{{Zhang}, {Finoguenov}, {B{\"o}hringer},
  {Ikebe}, {Matsushita} \& {Schuecker}}{{Zhang} et~al.}{2004}]{Zhang2004}
{Zhang} Y.,  {Finoguenov} A.,  {B{\"o}hringer} H.,  {Ikebe} Y.,  {Matsushita}
  K.,    {Schuecker} P.,  2004, \aap, 413, 49

\end{thebibliography}

\appendix

\section{Comparison of densities estimates}
\label{A:DenComp}
We provide here two plots that compare estimates of the
two measures of density. 
In the first plot  (Fig.\,\ref{F:DenDen}) two densities estimates are compared  ($\Sigma_{5}$ and $\Sigma_{10}$)
for all galaxies drawn from two randomly selected realizations of the 100 Monte Carlo realizations. 
The second plot shows the mean group density measured within the aperture   $R_{group}$ versus the median 
density $\Sigma_{5}$ and $\Sigma_{10}$ for all galaxies within the same aperture.

\begin{figure}
\centering
\includegraphics[width=0.9\columnwidth,clip,viewport=32 260 550 520]{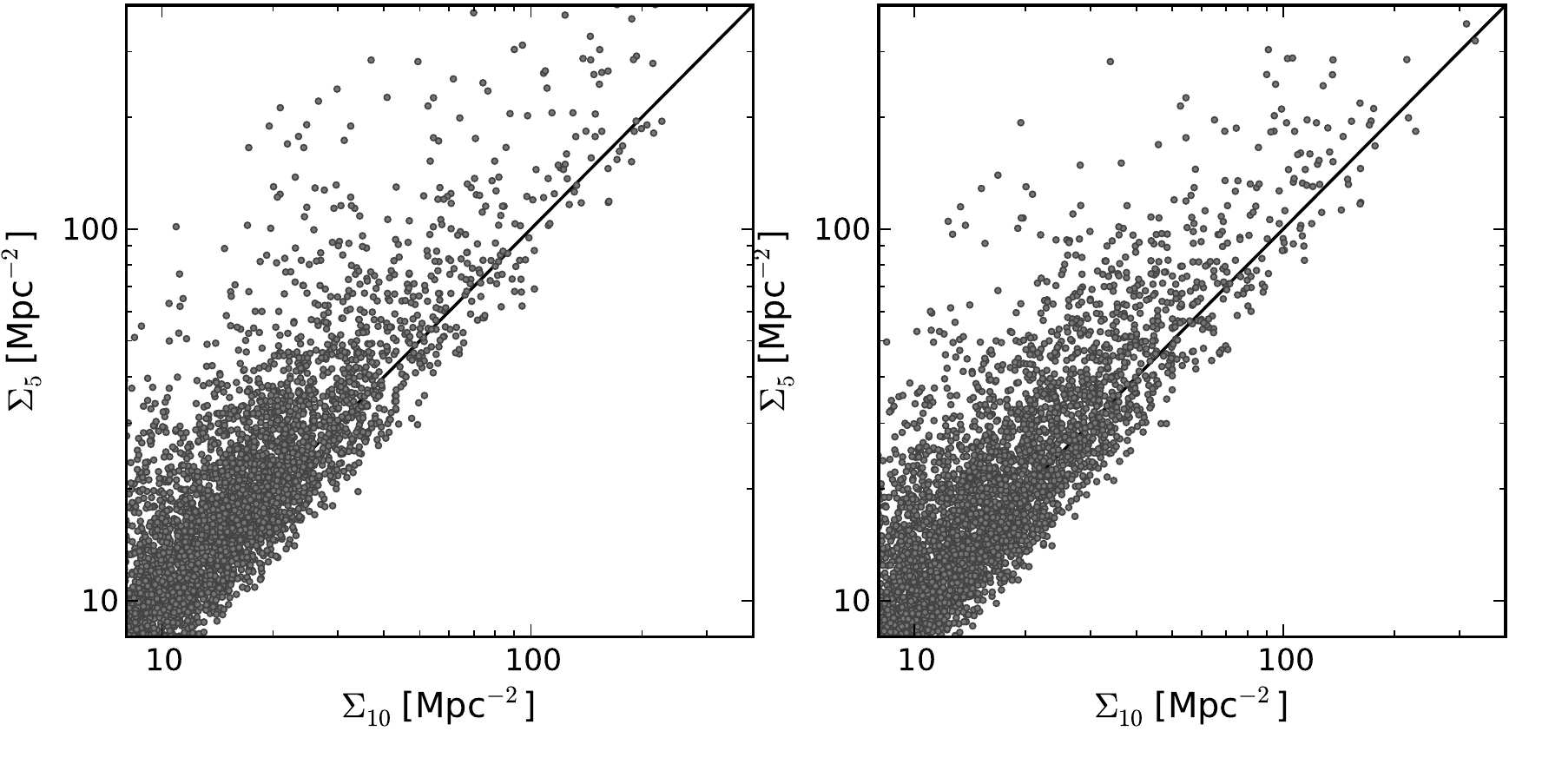}
\caption{Comparison between $\Sigma_{10}$ and the most commonly used $\Sigma_{5}$ (10 and five neigbours respectively)
for two random Monte Carlo realization of the original catalog. The line represents the one to one relation. Although both
values correlate, a large scatter is appreciated.}
\label{F:DenDen}
\end{figure}

\begin{figure}
\centering
\includegraphics[width=0.9\columnwidth,clip,viewport=32 260 550 520]{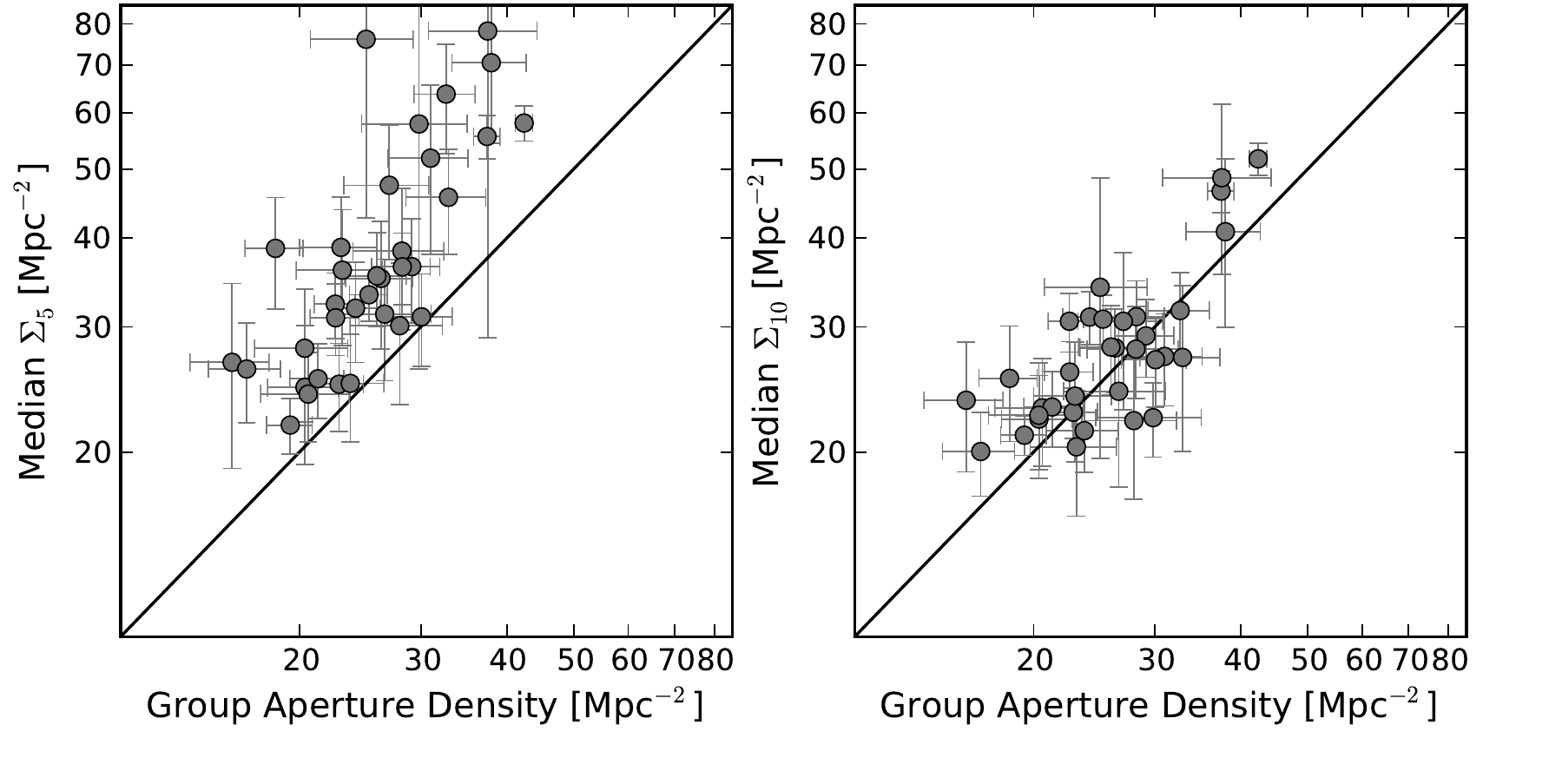}
\caption{Comparison between the group mean density, measured within the aperture $R_{group}$,
and the median  $\Sigma_{5}$ and $\Sigma_{10}$ for galaxies within the same apperture.
The line represents the one to one relation. Error bars are the standard deviation from the 100
Monte Carlo realizations. It is clear the $\Sigma_{5}$ overestimate the densiy 
for practically all groups, either due to its sensitivity to small compact association or projection effects 
in a \photz\ selected catalog.}
\label{F:DenDen}
\end{figure}

\section{Tables with properties of groups around RX\,J1347.5-1145}
\label{A:GroupTab}

In the next page, we provide the tables with properties of the optical selected groups candidates around RX\,J1347.5-1145.
Table\,\ref{T:tablegroups} contains the optical properties of these groups, whereas in Table\,\ref{T:xmmgroups} we provide
the derived X-ray properties for those groups with detectable signal within the XMM field of view. Please see Sections\,\ref{SS:groups} and \ref{SS:Xraygroups} for additional details.

\begin{table*}
\caption{List of the groups detected by the Voronoi tessellation technique. Columns: 1) Identification number, 2) and 3) Coordinates for the J2000 equinox, 4) Radius of the overdensity 5) and 6) average number of members and uncertainty, 7) and 8) Fraction of blue galaxies and associated error, 9) and 10) Total rest-frame $i$-band luminosity and error, 11) Notes for individual clusters to  provide a cross identification. }
\centering
\begin{tabular}{rcccrcccccl}
\hline\hline \\
 (1)	& (2)		& (3)		& (4)	& (5) &  (6)	& (7)  & (8)  & (9)  & (10) & (11) \\
 ID	& RA		& DEC		& $R_{group}$	& $\langle N\rangle$  &  $\sigma(N)$	& $\langle F_{blue}\rangle$	& $\sigma(F_{blue})$	& $L$	& Error($L$) & Notes\\ 
	& (J2000)	& (J2000)	& (Mpc)		& 	& 		& 		& 			& ($10^{11}L_\odot$)	& ($10^{11}L_\odot$) &	\\ \hline \\
1	& 13:47:12.2	& -12:14:09	& 0.369		& 13.28	& 1.77		& 0.71		& 0.15	   & 4.45  & 1.03		  &        \\
2	& 13:46:40.0	& -12:12:14	& 0.493		& 15.6	& 1.83		& 0.94		& 0.06	   & 3.69  & 0.72		  & 	\\
3	& 13:46:08.8	& -12:10:04	& 0.649		& 24.45	& 2.36		& 0.81		& 0.13	   & 4.44  & 0.70		  &	\\
4	& 13:46:42.0	& -12:08:49	& 0.493		& 15.77	& 2.31		& 0.92		& 0.08	   & 1.52  & 0.38		  &	\\
5	& 13:48:52.0	& -12:04:19	& 0.406		& 12.98	& 2.22		& 0.56		& 0.16	   & 4.08  & 1.06		  & LCDS0836	\\
6	& 13:45:53.2	& -12:02:38	& 0.340		& 10.17	& 1.55		& 0.85		& 0.15	   & 1.03  & 0.11		  &	\\
7	& 13:45:47.7	& -12:01:37	& 0.356		& 15.19	& 1.87		& 0.65		& 0.13	   & 4.25  & 0.79		  &	\\
8	& 13:46:03.6	& -12:00:57	& 0.587		& 17.29	& 2.26		& 0.77		& 0.16	   & 3.81  & 3.77		  &	\\
9	& 13:46:43.9	& -11:57:36	& 0.612		& 34.26	& 3.35		& 0.77		& 0.12	   & 4.66  & 4.27		  &	\\
10	& 13:45:44.4	& -11:57:54	& 0.581		& 17.77	& 2.12		& 0.90		& 0.09	   & 3.49  & 0.49		  & 	\\
11	& 13:46:26.8	& -11:54:28	& 0.987		& 114.7	& 5.06		& 0.56		& 0.05	   & 20.6  & 1.02		  & LCDS0825	\\
12	& 13:45:55.2	& -11:55:04	& 0.273		& 8.84	& 1.58		& 0.61		& 0.19	   & 2.27  & 0.62		  &	\\
13	& 13:46:51.8	& -11:53:13	& 0.894		& 48.71	& 3.69		& 0.80		& 0.09	   & 8.86  & 0.75		  &	\\
14	& 13:46:34.0	& -11:50:45	& 0.643		& 29.67	& 2.51		& 0.77		& 0.11	   & 6.45  & 0.56		  &	\\
15	& 13:47:18.4	& -11:52:19	& 0.446		& 16.42	& 1.82		& 0.31		& 0.08	   & 6.14  & 1.93		  & BGV2006-042	\\
16	& 13:47:13.9	& -11:51:32	& 0.321		& 9.15	& 1.38		& 0.66		& 0.17	   & 3.96  & 1.14		  &	\\
17	& 13:47:24.7	& -11:50:16	& 0.551		& 23.08	& 1.98		& 0.56		& 0.09	   & 6.51  & 0.37		  &	\\
18	& 13:47:09.1	& -11:47:52	& 0.512		& 23.27	& 2.28		& 0.72		& 0.11	   & 7.55  & 0.48		  &	\\
19	& 13:47:31.6	& -11:45:21	& 1.201		& 192.1	& 5.53		& 0.33		& 0.02	   & 45.8  & 1.69		  & RXJ1347.5-1145	\\
20	& 13:46:57.8	& -11:46:08	& 0.435		& 17.93	& 1.93		& 0.66		& 0.12	   & 4.24  & 0.24		  &	\\
21	& 13:47:47.0	& -11:42:21	& 0.309		& 9.88	& 1.31		& 0.30		& 0.11	   & 3.08  & 0.42		  &	\\
22	& 13:47:09.1	& -11:41:24	& 0.400		& 13.6	& 1.91		& 0.81		& 0.17	   & 1.59  & 0.34		  &	\\
23	& 13:45:56.8	& -11:40:12	& 0.466		& 15.8	& 2.25		& 0.85		& 0.14	   & 1.41  & 0.26		  &	\\
24	& 13:47:33.6	& -11:39:57	& 0.676		& 30.59	& 2.75		& 0.60		& 0.09	   & 5.48  & 0.61		  &	\\
25	& 13:47:51.6	& -11:38:49	& 0.773		& 42.41	& 2.88		& 0.69		& 0.07	   & 10.0  & 1.06		  &	\\
26	& 13:48:18.0	& -11:38:42	& 0.773		& 42.47	& 3.46		& 0.79		& 0.10	   & 7.44  & 0.79		  &	\\
27	& 13:45:54.9	& -11:35:56	& 0.377		& 10.29	& 1.33		& 0.63		& 0.15	   & 4.11  & 0.35		  &	\\
28	& 13:48:18.9	& -11:34:30	& 0.529		& 22.78	& 2.33		& 0.81		& 0.13	   & 3.15  & 0.39		  &	\\
29	& 13:47:46.5	& -11:32:52	& 0.464		& 16.08	& 1.91		& 0.42		& 0.11	   & 4.15  & 0.68		  &	\\
30	& 13:48:26.8	& -11:29:02	& 0.280		& 7.35	& 1.28		& 0.41		& 0.15	   & 4.54  & 0.95		  &	\\
31	& 13:48:30.0	& -11:27:25	& 0.302		& 9.41	& 0.96		& 0.24		& 0.08	   & 3.09  & 0.12		  &       \\
32	& 13:48:22.5	& -11:24:39	& 0.925		& 67.87	& 4.25		& 0.69		& 0.07	   & 14.3  & 0.79		  & NE-Clump		\\
33	& 13:46:48.9	& -11:23:34	& 0.379		& 12.04	& 2.02		& 0.75		& 0.19	   & 2.78  & 0.67		  &	\\
34	& 13:46:30.9	& -11:19:08	& 0.427		& 11.68	& 1.80		& 0.79		& 0.18	   & 3.71  & 0.65		  &	\\
\hline
\end{tabular}
\label{T:tablegroups}
\end{table*}

\begin{table*}
\caption{X-ray fluxes and derived parameters for optically detected groups in the area covered by XMM-$Newton$ observations. 
Columns: 1) Identification numbers (same as in Table\,\ref{T:tablegroups}), 2) and 3) Coordinates, 4) Measured X-ray flux and error,
5) X-ray luminosity, 6) Total mass within the 7) the $R_{200}$ radius, 8) Flux significance.}
\centering
\begin{tabular}{cccccccc}\hline\hline \\
(1) &  (2)        &  (3)      &   (4)	&  (5) &  (6) &  (7)   &  (8)   \\
ID &  RA        &  DEC      & $F_X$ (0.5-2 keV)  			& $L_X$				& $M_{200}$  	 	& $R_{200}$   &  $\sigma(F_X)$   \\
   &  (J2000)   & (J2000)   & ($10^{-15}$\,erg\,s$^{-1}$\,cm$^{-2}$)   	& ($10^{42}$\,erg\,s$^{-1}$)	& ($10^{13} M_\odot$ )  & (Mpc)  &                  \\ \hline \\
13 & 13:46:51.8	& -11:53:13 & $2.77 \pm 1.66 $				& $2.83\pm 2.08	$		& $ 3.22\pm 1.12 $	& 0.57	 &   1.67 \\      
15 & 13:47:18.4	& -11:52:19 & $3.65 \pm 1.57 $				& $3.71\pm 1.96	$		& $ 3.83\pm 0.98$	& 0.60	 &   2.32 \\    
16 & 13:47:13.9	& -11:51:32 & $1.12 \pm 0.79 $				& $1.21\pm 1.04	$		& $ 1.87\pm 0.75$	& 0.47	 &   1.42 \\   
17 & 13:47:24.7	& -11:50:16 & $3.26 \pm 1.26 $				& $3.32\pm 1.57	$		& $ 3.57\pm 0.82$	& 0.59	 &   2.59 \\    
18 & 13:47:09.1	& -11:47:52 & $2.14 \pm 0.94 $				& $2.20\pm 1.18	$		& $ 2.74\pm 0.71$	& 0.54	 &   2.29 \\   
22 & 13:47:09.1	& -11:41:24 & $1.13 \pm 0.73 $				& $1.22\pm 0.96	$		& $ 1.87\pm 0.70$	& 0.47	 &   1.55\\   
25 & 13:47:51.6	& -11:38:49 & $1.80 \pm 1.34 $				& $1.87\pm 1.71	$		& $ 2.47\pm 1.05$	& 0.52	 &   1.34 \\    
26 & 13:48:18.0	& -11:38:42 & $5.72 \pm 1.77 $				& $5.77\pm 2.19	$		& $ 5.08\pm 0.95$	& 0.66	 &   3.24 \\     
29 & 13:47:46.5	& -11:32:52 & $2.53 \pm 1.51 $				& $2.59\pm 1.90	$		& $ 3.04\pm 1.06$	& 0.56	 &   1.68 \\ \hline
\end{tabular}

\label{T:xmmgroups}
\end{table*}

\label{lastpage}

\end{document}